\documentclass[a4paper,14pt]{article}
\usepackage[english]{babel}
\pdfoutput=1 

\usepackage{jheppub} 

\usepackage[T1]{fontenc} 

\usepackage{comment}

\usepackage{cases}

\newcommand{\bi}{\begin{itemize}}
	\newcommand{\ei}{\end{itemize}}
\newcommand{\bea}{\begin{eqnarray}}
	\newcommand{\eea}{\end{eqnarray}}
\newcommand{\bt}{\begin{tabular}}
	\newcommand{\et}{\end{tabular}}
\newcommand{\bc}{\begin{center}}
	\newcommand{\ec}{\end{center}}

\newcommand{\be}{\begin{equation}}
	\newcommand{\ee}{\end{equation}}
\newcommand{\ba}{\begin{array}}
	\newcommand{\ea}{\end{array}}

\newcommand{\lb}[1]{\label{#1}}

\def\bbox{{\,\lower0.9pt\vbox{\hrule \hbox{\vrule height 0.2 cm
				\hskip 0.2 cm \vrule height 0.2 cm}\hrule}\,}}
\newcommand{\dsl}{\pa \kern-0.5em /}

\makeatletter \@addtoreset{equation}{section} \makeatother

\def\slashchar#1{\setbox0=\hbox{$#1$}           
	\dimen0=\wd0                                 
	\setbox1=\hbox{/} \dimen1=\wd1               
	\ifdim\dimen0>\dimen1                        
	\rlap{\hbox to \dimen0{\hfil/\hfil}}      
	#1                                        
	\else                                        
	\rlap{\hbox to \dimen1{\hfil$#1$\hfil}}   
	/                                         
	\fi}

\pdfoutput=1

\title{\boldmath Structure of $\mathcal{N} = 2$ superfield higher-spin abelian
	cubic interactions}

\author[a,b]{Nikita~Zaigraev}

\affiliation[a]{Bogoliubov Laboratory of Theoretical Physics, JINR,\\
	Joliot-Curie st. 6, 141980 Dubna, Moscow region, Russia}
\affiliation[b]{Moscow Institute of Physics and Technology,\\ Institutskiy per. 9, 141700 Dolgoprudny, Moscow region, Russia}

\emailAdd{nikita.zaigraev@phystech.edu}

\abstract{In this article we study the structure of the $\mathcal{N}=2$ abelian higher-spin cubic $(\mathbf{s_1}, \mathbf{s_2}, \mathbf{s_2})$ vertices and the corresponding $\mathcal{N}=2$ higher-spin supercurrents, introduced in \href{https://arxiv.org/abs/2408.00668}{arXiv:2408.00668}.  These interactions are possible only for $\mathbf{s_1} \geq 2 \mathbf{s_2}$. Conserved supercurrents are constructed as descendants of the \textit{principal supercurrent}, which is uniquely characterized by simple differential conditions and admits an explicit representation in terms of $\mathcal{N}=2$ higher-spin super-Weyl tensors.  We derive the analytic form of the abelian vertices and identify the corresponding analytic higher-spin $\mathcal{N}=2$ supercurrents. We show that the vertex structure is fully determined by the analytic supercurrents $J^{++}_{\alpha(s-1)\dot{\alpha}(s-1)}$, $J^+_{\alpha(s-1)\dot{\alpha}(s-2)}$, and $\bar{J}^+_{\alpha(s-2)\dot{\alpha}(s-1)}$.  The analytic form of the vertices provides a simple framework for analyzing their component structure. As an example, we explore the component content of such interactions on the Bel--Robinson diagonal. Using the superfield inverse Noether procedure, we study higher-spin gauge transformations for the $\mathcal{N}=2$ vector multiplet associated with the $(\mathbf{s}, \mathbf{1}, \mathbf{1})$ interaction. In the rigid limit, for odd $\mathbf{s}$ these transformations reduce to the $\mathcal{N}=2$ superspace generalization of zilch-type higher-spin symmetries. }

\arxivnumber{2605.27206}

\makeatletter
\gdef\@fpheader{}
\makeatother

\begin{document} 
\maketitle
\flushbottom

\section{Introduction}

The construction of cubic interaction vertices for higher-spin fields is a necessary step toward understanding consistent interacting higher-spin theories. Many important results have been obtained in this direction. Cubic vertices in Minkowski space of any dimension $d\geq 4$   were classified by Metsaev  \cite{Metsaev:2005ar,Metsaev:2007rn} in the light-cone formalism.
This classification was generalized to $\mathcal{N}=1$ supersymmetry \cite{Metsaev:2019dqt}  and to $\mathcal{N}= 4 \mathbb{N}$ supersymmetry \cite{Metsaev:2019aig} in four-dimensional flat space.
  Lorentz-covariant realizations of cubic vertices in flat space have also been extensively developed in the metric-like approach \cite{Berends:1984rq, Berends:1984wp, Berends:1985xx, Damour:1987fp, Deser:1990bk, Boulanger:2008tg, Zinoviev:2008ck, Boulanger:2006gr,  Manvelyan:2009vy, Manvelyan:2010je, Manvelyan:2010jr, Manvelyan:2010wp, Henneaux:2012wg, Henneaux:2013gba, Joung:2011ww} and the frame-like approach \cite{Zinoviev:2010cr, Khabarov:2020bgr, Zinoviev:2024eto, Fradkin:1986qy, Fradkin:1987ks, Khabarov:2020deh}.
Reviews of these results and more complete lists of references can be found in  \cite{Bekaert:2010hw, Ponomarev:2022vjb, Bekaert:2022poo}.

  The Metsaev classification implies that in four dimensions there exist only two possible numbers of derivatives for nontrivial Lorentz-covariant cubic vertices $(s_1,s_2,s_3)$\footnote{If one abandons the requirement of Lorentz covariance, new types of vertices appear, see \cite{Ponomarev:2016lrm, Ponomarev:2022vjb}.}.
  Such vertices (with integer spins $s_1\geq s_2 \geq s_3$) contain either $s_1+s_2+s_3$ derivatives or $s_1+s_2-s_3$ derivatives.
The two derivative branches identified by Metsaev admit a natural interpretation in terms of the gauge-algebra deformation properties of the corresponding vertices:

\begin{itemize}

\item[\textbf{I.}]  \textbf{Trivially gauge invariant or Born-Infeld type ($CCC$)}

These interactions exist for $s_1 \leq s_2+s_3$. They can be constructed entirely in terms of abelian curvatures and do not deform the linearized higher-spin gauge transformations.
The algebra of gauge transformations is abelian. Examples of such vertices were considered in \cite{Damour:1987fp}.  The typical form of  a Born-Infeld $(s_1,s_2,s_3)$ vertex is given by:
\begin{equation}\label{eq: type 1}
		\mathcal{L}^{\text{I}}_{int} \sim \epsilon^{\alpha_1\beta_1} \dots \epsilon^{\alpha_{2s_1}\gamma_{2s_3}}
		\;\; C_{\alpha(2s_1)} C_{\beta(2s_2)} C_{\gamma(2s_3)}.
\end{equation}
Here we assume contraction over all spinor indices.
Such vertices contain a maximal number of derivatives $k_{BBB}=s_1+s_2+s_3$, since each bosonic higher-spin Weyl-like tensor contains $s_i$ derivatives.
 While $k_{BFF} = s_1+s_2+s_3-1$, if we consider fermionic fields, since a spin $s-\frac{1}{2}$ fermion Weyl-like tensor contains $s-1$ derivatives.

\item[\textbf{II.}]  \textbf{Abelian or Chern-Simons type ($\Phi CC $)}

These interactions exist if $s_1\geq s_2+s_3$.  
Vertices of this type belong to one of the two derivative branches classified by Metsaev and therefore contain either
$k_{BBB}=s_1+s_2+s_3$
or
$k_{BBB}=s_1+s_2-s_3$
derivatives ($k_{BFF} = s_1+s_2 \pm s_3-1$ for vertices with fermions).
In this case, gauge transformations receive nontrivial corrections that mix fields of different spins, although the gauge algebra remains abelian.  Such vertices typically take the form:
\begin{equation}
	\mathcal{L}^{\text{II}}_{int} \sim \Phi_1 C_2 C_3.
\end{equation}		
The simplest examples of such vertices were constructed by  Berends, Burgers and van Dam~\cite{Berends:1985xx}. The Lagrangian for the $(s_1, s_2, s_2)$ vertex with $s_1\geq 2s_2$ has universal form and contains $k_{BBB} = s_1$ or $k_{BFF} = s_1-1$ spacetime derivatives:  
\begin{equation}\label{eq: type 2}
	\mathcal{L}^{\text{II}}_{\text{BBvD}} \sim \Phi^{\alpha(s_1)\dot{\alpha}(s_1)} J_{\alpha(s_1)\dot{\alpha}(s_1)} + \Phi^{\alpha(s_1-2)\dot{\alpha}(s_1-2)} T_{\alpha(s_1-2)\dot{\alpha}(s_1-2)}.
\end{equation}	
Here $J_{\alpha(s_1)\dot{\alpha}(s_1)}$ is the traceless gauge-invariant higher-spin current and $T_{\alpha(s_1-2)\dot{\alpha}(s_1-2)}$ is the gauge-invariant  current trace. $J$ and $T$ are constructed from spin-$s_2$ fields
and satisfy the  on-shell conservation law\footnote{We use ''weak equality'' symbol $\approx$ to denote equations, which are satisfied  only on the equations of motion.}:
\begin{equation}\label{eq: bos cons eq}
	\partial^{\alpha\dot{\alpha}} J_{\alpha(s_1)\dot{\alpha}(s_1)} + \partial_{(\alpha(\dot{\alpha}} T_{\alpha(s-2))\dot{\alpha}(s-2))} \approx 0.
\end{equation}	
These conserved currents are higher-spin generalizations of the linearized gravitational Bel--Robinson tensor \cite{Bel}, which is known to be conserved. 
By the inverse Noether procedure,  the conservation law \eqref{eq: bos cons eq} corresponds to global spin-$s_1$ symmetry of the free Fronsdal spin-$s_2$ action. This global symmetry determines the corresponding spin-$s_1$ gauge transformations for spin-$s_2$ fields.  

\item[\textbf{III}.]  \textbf{Non-abelian or Yang-Mills type ($\Phi\Phi C$ or $\Phi\Phi\Phi$)}

If $s_1 < s_2 +s_3$,  cubic vertices require deformation of gauge transformations, which leads to non-abelian  gauge algebra. The structure of such vertices is considerably more involved, see, e.g., examples \cite{Berends:1984wp, Fradkin:1986qy, Fradkin:1987ks, Boulanger:2006gr, Boulanger:2008tg, Zinoviev:2008ck}. The number of derivatives in such vertices is $k_{BBB} = s_1+s_2-s_3$, $k_{BFF} = s_1+s_2-s_3-1$.
\end{itemize}

\noindent Type-II and type-III vertices are consistent only at the leading order.
The inconsistency of type-II vertices at higher orders is discussed, e.g., in \cite{Deser:1990bk}, type-III vertices in \cite{Berends:1984rq, Bekaert:2010hp}.

\medskip

The connection of these vertices with the nonlinear higher‑spin theory \cite{Vasiliev:1990en, Vasiliev:1992av} is being actively investigated.  In particular, in Ref. \cite{Misuna:2017bjb} (using  the local form of second-order Vasiliev equations developed 
 in  \cite{Gelfond:2017wrh}) an explicit expression is obtained for the quadratic corrections to the bosonic Fronsdal equations that are generated by gauge‑invariant higher‑spin currents, consistent with the  known parity‑invariant cubic vertices. 
A recent analysis \cite{Tatarenko:2024csa} confirms that for $d = 4$, the vertices obtained from Metsaev’s classification match exactly the current deformation of the free higher‑spin equations coming from the non‑linear $4D$ higher-spin theory  \cite{Vasiliev:1992av}.

\medskip

Supersymmetry unifies diverse vertices into a single, supertransformation-invariant structure — a property systematically explored for 
$\mathcal{N}=1$  in Refs. \cite{Metsaev:2019dqt, Khabarov:2020deh}. The most efficient way to build actions that respect this invariance from the outset is the superspace approach \cite{Gates:1983nr, Wess:1992cp, BK,18}, which guarantees manifest supersymmetry. Therefore, it is natural to ask whether these distinct bosonic vertices admit a unified superspace description. However, the explicit construction of superfield actions that generate these cubic vertices in a manifestly supersymmetric way remains an open problem. Addressing this challenge is particularly timely in light of the recent construction of nonlinear field equations for the supersymmetric higher-spin gauge theory  presented in \cite{Vasiliev:2025hfh}. The construction of such superfield actions is expected to reveal new underlying (super)geometric structures, ultimately contributing to a deeper understanding of the full nonlinear theory.

\smallskip

There are some notable results about the superfield structure of cubic interactions of massless $\mathcal{N}=1$ supermultiplets. Off-shell abelian cubic vertices (type-II) of $\mathcal{N}=1$ higher-spin multiplets with $\mathcal{N}=1$ matter via the supercurrent multiplets were considered in \cite{Kuzenko:2017ujh, Buchbinder:2017nuc, Hutomo:2017nce, Hutomo:2017phh, Koutrolikos:2017qkx, Buchbinder:2018wwg}. Construction of type-II interactions and gauge-invariant supercurrents for massless $\mathcal{N}=1$ higher-spin supermultiplets was elaborated in \cite{Buchbinder:2018wzq}. 
In the work \cite{Gates:2019cnl}, a class of vertices of the type-I  was constructed using gauge-invariant supercurrents. 
In both cases, supercurrents were constructed in terms of  $\mathcal{N}=1$ superfield strengths, which  were first introduced in \cite{Kuzenko:1993jp, Kuzenko:1993jq} (see also \cite{BK}).

\smallskip
 
Berends-Burgers-van Dam (type-II) cubic vertices for the massless higher-spin $\mathcal{N}=2$ off-shell supermultiplets were constructed in \cite{Zaigraev:2024ryg} (see also \cite{Zaigraev:2025xqo}) using the formulation of free $4D \,\mathcal{N}=2$ higher-spin theory \cite{Buchbinder:2021ite}\footnote{Recently, this construction was generalized to 5D massless higher-spin multiplets \cite{Buchbinder:2025yef}.} in the harmonic superspace \cite{Galperin:1984av, 18}. In harmonic superspace $\mathcal{N}=2$ higher-spin supermultiplets (with integer highest spin) are naturally described by the unconstrained analytic prepotentials. Using these prepotentials one can construct cubic interactions with the fundamental multiplet of $\mathcal{N}=2$ matter, the hypermultiplet  \cite{Buchbinder:2022kzl, Buchbinder:2022vra} (see also for review \cite{Zaigraev:2023ogo, Zaigraev:2024xve}). Interactions with the hypermultiplet possess a natural geometric structure and are obtained by  covariantizing the harmonic derivative $\mathcal{D}^{++}$ in  direct analogy with the gauge-covariantization procedure in ordinary gauge theory:
\begin{equation}\label{eq: hyper int}
	\mathcal{L}_{free} = - \frac{1}{2} q^{+a} \mathcal{D}^{++} q^+_a
	\quad
	\Rightarrow
	\quad 
		\mathcal{L}_{free+int} = - \frac{1}{2} q^{+a} \left(\mathcal{D}^{++} +
		\kappa_s \hat{\mathcal{H}}^{++}_{(s)} (J)^{P(s)} \right)q^+_a,
\end{equation}
where $\hat{\mathcal{H}}^{++}_{(s)}$ is a higher-derivative differential operator, which will be defined in \eqref{eq: hat H}. The form of the interaction makes it possible to identify the corresponding higher-spin hypermultiplet supercurrents directly, which were explored in \cite{Buchbinder:2022vra}. 
Although the $(\mathbf{s}, \mathbf{\frac{1}{2}}, \mathbf{\frac{1}{2}})$\footnote{We will denote as $\mathbf{s}$ $\mathcal{N}=2$ supermultiplet with highest spin $s$. So spin-$\mathbf{s}$ supermultiplet corresponds to off-shell supermultiplet with physical spins $\{s, s -\frac{1}{2}, s-\frac{1}{2}, s-1\}$, spin-$\mathbf{\frac{1}{2}}$ supermultiplet corresponds to hypermultiplet with physical spins $\{\frac{1}{2}, \frac{1}{2}, 0, 0\}$.} vertices admit a natural geometric interpretation, the resulting $\mathcal{N}=2$ supercurrents have a rather special form:
\begin{equation}\label{eq: geom supercurrent}
	J^{++}_{\alpha(s)\dot{\alpha}(s)} = -\frac{1}{2} q^{+a} \partial^{s} (J)^{P(s)} q^+_a,
	\qquad 
	\mathcal{D}^{++} 	J^{++}_{\alpha(s)\dot{\alpha}(s)} \approx 0.
\end{equation}	 
Interestingly due to the special form of the free hypermultiplet equations of motion $\mathcal{D}^{++} q^+_a \approx 0$, it is possible to construct a whole family of conserved higher-spin hypermultiplet supercurrents that are not fully fixed by the gauge transformations of the $\mathcal{N}=2$ higher-spin prepotentials, e.g.,
\begin{equation}
	J^{++}_{\alpha(s)\dot{\alpha}(s)} = \sum_{k=0}^{s}  a_k \; \partial^k q^{+a} \partial^{s-k} (J)^{P(s)} q_a^+,
	\qquad
	\mathcal{D}^{++} J^{++}_{\alpha(s)\dot{\alpha}(s)}  \approx 0.
\end{equation}	 
Here $a_k$ are arbitrary coefficients, constrained only by the reality condition $\widetilde{J^{++}_{\alpha(s)\dot{\alpha}(s)} } = J^{++}_{\alpha(s)\dot{\alpha}(s)} $.
On shell, all these supercurrents appear to be equivalent at the level of cubic interactions. The supercurrent \eqref{eq: geom supercurrent} is distinguished by its direct geometric origin from the interaction \eqref{eq: hyper int}.
For arbitrary coefficients $a_k$ such supercurrents contain only one conserved component \cite{Buchbinder:2022vra}. 

\smallskip

In this regard, Kuzenko and Raptakis \cite{Kuzenko:2023vgf} examined the $\mathcal{N}=2$ supercurrent structure and showed that the special conserved hypermultiplet supercurrents can be obtained as descendants of the $\mathcal{N}=2$ \textit{primary higher-spin supercurrent}, introduced in \cite{Kuzenko:2021pqm}. This approach uniquely fixes the structure of the supercurrent. It also ensures that all component currents contained in the supermultiplet satisfy conservation equations. From the point of view of harmonic superspace, the origin of these additional conservation laws is less transparent, since they are not determined by the gauge transformations of the higher-spin analytic prepotentials.

\smallskip

 Construction of $\mathcal{N}=2$ Berends-Burgers-van Dam  interactions  \cite{Zaigraev:2024ryg} uses higher-spin  Mezincescu prepotentials (introduced in \cite{Buchbinder:2022vra}) and gauge-invariant \textit{$\mathcal{N}=2$ principal supercurrents}, which satisfy equations analogous to those defining the $\mathcal{N}=2$ primary supercurrents. $\mathcal{N}=2$ principal supercurrents were constructed in terms of gauge-invariant $\mathcal{N}=2$ Weyl-like higher-spin supertensor (superfield strength). In this article, we analyze the structure of such cubic vertices.  We will focus on three directions:

\smallskip

\textbf{1.} The analytic structure of $\mathcal{N}=2$ higher-spin supercurrents;

\textbf{2.} Higher-spin transformations for $\mathcal{N}=2$ vector multiplet;

\textbf{3.} The component structure of interactions of the form
$(\mathbf{s_1},\mathbf{s_2},\mathbf{s_2})$, which we refer to as the Bel--Robinson diagonal.

\smallskip

The paper is organized as follows. In section \ref{sec: 2}, we review the $4D,\, \mathcal{N}=2$ higher-spin theory in harmonic superspace, its superfield equations of motion and introduce Weyl-like gauge-invariant superfield strengths. In section \ref{sec: 3}, we recall Mezincescu-type higher-spin prepotentials and explore their (pre-)gauge freedom.   Section \ref{sec: N=2 higher-spin supercurrents} is devoted to the structure of $\mathcal{N}=2$ gauge-invariant supercurrents for arbitrary spin gauge $\mathcal{N}=2$ supermultiplets. We find explicit  relations between analytic supercurrents and non-analytic ones and represent all these supercurrents as descendants of the principal $\mathcal{N}=2$ supercurrent. We also present simple examples of $\mathcal{N}=2$ supercurrents corresponding to $(\mathbf{2}, \mathbf{1}, \mathbf{1})$ and $(\mathbf{3}, \mathbf{1}, \mathbf{1})$ vertices and their component structure. 
Section \ref{sec: 5} is devoted to derivation of higher-spin gauge transformation of $\mathcal{N}=2$ vector multiplet. In section \ref{sec: 6}, we elaborate component structure of interactions on Bel--Robinson diagonal $(\mathbf{s}= 2\mathbf{s_2}, \mathbf{s_2}, \mathbf{s_2})$   The last section \ref{sec: dis} summarizes our results and presents some related open problems.  

We also include a few technical appendices. In appendix \ref{app A}, we review Fronsdal and Fang-Fronsdal higher-spin fields in spinor notation and introduce linearized field strengths, which are used throughout the text. In appendix \ref{app: conserved gauge-invariant higher-spin currents}, we study the structure of gauge-invariant higher-spin  currents, which are required for component reduction. 
In appendix \ref{sec: ZILCH}, we discuss in detail the zilch pseudotensor, the symmetries that are responsible for its conservation and corresponding higher-spin generalizations.
Appendix \ref{app: structure of strength} is devoted to the structure of $\mathcal{N}=2$ higher-spin superfield Weyl-like tensors. In appendix  \ref{eq: cons N=2 str}, we discuss component content of superfield strengths and derive on-shell equations, which we use in our analysis of conserved $\mathcal{N}=2$ supercurrents.

\section*{Harmonic superspace preliminaries}

We use the standard harmonic superspace conventions, see Ref. \cite{18} for a pedagogical introduction.  Here, we briefly summarize the main definitions and features of harmonic superspace that we will use in this article.

The essential feature of $\mathcal{N}=2$ harmonic superspace is the presence of additional $SU(2)$ harmonic variables $u_i^\pm$:
\begin{equation}
	\mathbb{HR}^{4|8} = \{x^{\alpha\dot{\alpha}}, \theta^{\alpha i}, \bar{\theta}^{\dot{\alpha}i}, u^\pm_i\}.
\end{equation}	
Harmonic variables satisfy $ \overline{u^{+i}}=u^-_i  $ and $u^{+i}u^-_i=1$. 
The introduction of harmonics
allows one to define a supersymmetry-invariant \textit{analytic superspace}, containing only half of the Grassmann coordinates:
\begin{equation}
	\mathbb{HA}^{4|4} = \{x_A^{\alpha\dot{\alpha}}, \theta_A^{\alpha +}, \bar{\theta}_A^{\dot{\alpha}+}, u^\pm_i\},
\end{equation}	
where we use the following notations:
\begin{equation}\label{eq: analit basis}
	x^{\alpha\dot{\alpha}}_A := x^{\alpha\dot{\alpha}} - 4i \theta^{\alpha(i} \bar{\theta}^{\dot{\alpha}j)} u^+_i u^-_j,
	\qquad
	\theta^{ \alpha \pm}_{A} :=  \theta^{\alpha i}u^\pm_i,
	\qquad
	\bar{\theta}^{\dot{\alpha}\pm}_{A} :=  \bar{\theta}^{\dot{\alpha}i}u_i^\pm.
\end{equation}	

In the case of $\mathcal{N} = 2$ supersymmetry with central charges it is convenient to introduce fifth (auxiliary) coordinate $x^5$, which takes the following form in the analytic basis:
\begin{equation}	
	x^5_A := x^5 + i\left( \theta^{\alpha i} \theta^j_\alpha -  \bar{\theta}^i_{\dot{\alpha}} \bar{\theta}^{\dot{\alpha}j} \right) u^+_{(i} u^-_{j)}.
\end{equation}	
The analytic superspace is preserved by rigid $\mathcal{N}=2$ supersymmetry transformations.
All known $\mathcal{N}=2$ supermultiplets admit a description in terms of analytic superfields.
Throughout this paper, we will always work in an analytic basis, so we omit the A indices.

Rigid $\mathcal{N}=2$ supersymmetry acts on coordinates of harmonic superspace according to:
\begin{equation}\label{eq: rig N=2 susy}
	\delta_\epsilon x^{\alpha\dot{\alpha}} = - 4i \left( \epsilon^{-\alpha} \bar{\theta}^{+\dot{\alpha}} + \theta^{+\alpha}  \bar{\epsilon}^{-\dot{\alpha}} \right),
	\qquad
		\delta_\epsilon \theta^{\pm}_\alpha  = \epsilon^{\pm}_\alpha,
	\qquad
		\delta_\epsilon\bar{\theta}^{\pm}_{\dot{\alpha}} = \bar{\epsilon}^{\pm}_{\dot{\alpha}}, 
		\qquad
			\delta_\epsilon u^\pm_i = 0,
\end{equation}	
where we introduced notations $\epsilon^{\pm}_\alpha : = \epsilon_{\alpha}^{i} u^\pm_i$ and $\bar{\epsilon}^{\pm}_{\dot{\alpha}} := \bar{\epsilon}^{i}_{\dot{\alpha}} u^\pm_i $.

Analytic space is real with respect to \textit{tilde-conjugation}, which is the combination of complex conjugation and antipodal map on $S^3\simeq SU(2)$: 
\begin{equation}\label{eq: analit space}
	\widetilde{x^{\alpha\dot{\alpha}}} = x^{\alpha\dot{\alpha}},
	\quad
	\widetilde{\theta^\pm_\alpha} = \bar{\theta}^\pm_{\dot{\alpha}},
	\quad
	\widetilde{\bar{\theta}^\pm_{\dot{\alpha}}} = - \theta^\pm_\alpha,
	\quad
	\widetilde{u^{\pm i}} = - u_i^\pm,
	\quad
	\widetilde{u^\pm_i} = u^{\pm i},
	\quad
	\widetilde{x^5} = x^5.
\end{equation}

An important feature of harmonic superspace is the existence of harmonic derivatives. It is convenient to select harmonic derivatives in a form that preserves the ratio $u^{+i}u^-_i =1$:
\begin{equation}
	\partial^{++} = u^{+i} \frac{\partial}{\partial u^{-i}},
	\qquad
	\partial^{--} = u^{-i} \frac{\partial}{\partial u^{+i}},
	\qquad
	\partial^0 = u^{+i} \frac{\partial}{\partial u^{+i}} - u^{-i} \frac{\partial}{\partial u^{-i}}.
\end{equation}
In the analytic basis, harmonic derivatives take the form:	
\begin{subequations}
	\begin{equation}
		\mathcal{D}^{++} = \partial^{++} - 4i \theta^{+\alpha}\bar{\theta}^{+\dot{\alpha}}\partial_{\alpha\dot{\alpha}} + \theta^{+\hat{\alpha}}\partial^+_{\hat{\alpha}} + i(\theta^{\hat{+}})^2 \partial_5, 
	\end{equation}	
	\begin{equation}
		\mathcal{D}^{--} = \partial^{--} - 4i \theta^{-\alpha}\bar{\theta}^{-\dot{\alpha}}\partial_{\alpha\dot{\alpha}} + \theta^{-\hat{\alpha}}\partial^-_{\hat{\alpha}}
		+
		i(\theta^{\hat{-}})^2 \partial_5, 
	\end{equation}	
	\begin{equation}
		\mathcal{D}^0 = \partial^0 + \theta^{+\hat{\alpha}}\partial^-_{\hat{\alpha}}
		-
		\theta^{-\hat{\alpha}}\partial^+_{\hat{\alpha}}.
	\end{equation}	
\end{subequations}
Here $\hat{\alpha} := (\alpha, \dot{\alpha})$ and $(\theta^{\hat{\pm}})^2 := \theta^{\pm\hat{\alpha}} \theta^{\pm}_{\hat{\alpha}} $. Harmonic derivatives are supersymmetry-invariant and play an important  role in the harmonic superspace formulations of $\mathcal{N}=2$ theories.

In the harmonic superspace, superfields with a fixed harmonic charge $\Phi^{(n)}$ are considered:
\begin{equation}
	\mathcal{D}^0 \Phi^{(n)} = n \Phi^{(n)}.
\end{equation}	
In the present work, the superscripts $++$, $--$, etc.,  indicate the harmonic charges carried by the superfield.

Covariant spinor derivatives in the analytic basis are given by:
\begin{equation}\label{eq: cov der}
	\begin{split}
&\mathcal{D}^+_{\hat{\alpha}} = \partial^+_{\hat{\alpha}},
\\
&\mathcal{D}^-_\alpha = - \partial^-_\alpha + 4i \bar{\theta}^{-\dot{\alpha}}\partial_{\alpha\dot{\alpha}}
-
2i \theta^-_\alpha \partial_5,
\\
&\bar{\mathcal{D}}^-_{\dot{\alpha}} = - \partial^-_{\dot{\alpha}} - 4i \theta^{-\alpha}\partial_{\alpha\dot{\alpha}}
-
2i \bar{\theta}^-_{\dot{\alpha}} \partial_5
\end{split}
\end{equation}
and satisfy relations
\begin{equation}
	\begin{split} &\mathcal{D}^\pm_{\hat{\alpha}} = [\mathcal{D}^{\pm\pm}, \mathcal{D}^{\mp}_{\hat{\alpha}}],
		\\
		 &[\mathcal{D}^{\pm\pm}, \mathcal{D}^\pm_{\hat{\alpha}}] = 0,
		\\
		& \{\mathcal{D}^+_{\alpha}, \bar{\mathcal{D}}^-_{\dot{\alpha}}\} =- \{\mathcal{D}^-_{\alpha}, \bar{\mathcal{D}}^+_{\dot{\alpha}}\}   = -4i \partial_{\alpha\dot{\alpha}}.
		\end{split}
	\end{equation}

\section{$\mathcal{N}=2$ higher-spin theories in harmonic superspace}\label{sec: 2}

In this section, we discuss $\mathcal{N}=2$ analytic higher-spin prepotentials and harmonic superspace formulations of $\mathcal{N}=2$ higher-spin theory.
We also introduce gauge-invariant higher-spin super-Weyl tensors, which serve as building blocks for consistent abelian interactions.

\subsection{$\mathcal{N}=2$ off-shell higher-spin supermultiplets in terms of the analytic prepotentials}

$\mathcal{N}=2$ integer spin-$\mathbf{s}$ supermultiplet ($\mathbf{s}\geq2$) is described by the set of unconstrained analytic prepotentials \cite{Buchbinder:2021ite}:
\begin{equation}\label{eq: HS analytic prepotentials}
	h^{++\alpha(s-1)\dot{\alpha}(s-1)},
	\quad
	h^{++\alpha(s-1)\dot{\alpha}(s-2)+},
	\quad
	h^{++\alpha(s-2)\dot{\alpha}(s-1)+},
	\quad
	h^{++\alpha(s-2)\dot{\alpha}(s-2)},
\end{equation}	
which satisfy reality conditions under tilde-conjugation \eqref{eq: analit space}:
\begin{equation}
	\begin{split}
	&\widetilde{h^{++\alpha(s-1)\dot{\alpha}(s-1)}} = h^{++\alpha(s-1)\dot{\alpha}(s-1)},
	\\
	&\widetilde{h^{++\alpha(s-1)\dot{\alpha}(s-2)+}}
	=
	h^{++\alpha(s-2)\dot{\alpha}(s-1)+},
	\\
	&\widetilde{h^{++\alpha(s-2)\dot{\alpha}(s-2)}}
	=
	h^{++\alpha(s-2)\dot{\alpha}(s-2)}
	\end{split}
\end{equation}	
and are
defined modulo the gauge freedom
 \begin{equation}\label{eq: GF}
	\begin{split}
		& \delta_\lambda h^{++\alpha(s-1)\dot\alpha(s-1)} = \mathcal{D}^{++} \lambda^{\alpha(s-1)\dot\alpha(s-1)} 
		\\&
		\qquad\qquad\qquad\qquad\qquad+
		4i \big[\lambda^{+\alpha(s-1)(\dot\alpha(s-2)}\bar\theta^{+\dot\alpha_{s-1})} +\theta^{+(\alpha_{s-1}} \bar\lambda^{+\alpha(s-2))\dot\alpha(s-1)} \big], 
		\\
		&\delta_\lambda h^{++\alpha(s-2)\dot\alpha(s-2)} =  \mathcal{D}^{++} \lambda^{\alpha(s-2)\dot\alpha(s-2)}
		\\&
		\qquad\qquad\qquad\qquad\qquad -
		2i\,\big[\lambda^{+(\alpha(s-2)\alpha_{s-1})\dot\alpha(s-2)} \theta^+_{\alpha_{s-1}} 
		+
		\bar\lambda^{+(\dot\alpha(s-2)\dot\alpha_{s-1})\alpha(s-2)} \bar\theta^+_{\dot\alpha_{s-1}} \big], 
		\\
		& \delta_\lambda  h^{++\alpha(s-1)\dot\alpha(s-2)+} =  \mathcal{D}^{++} \lambda^{+\alpha(s-1)\dot\alpha(s-2)}\,,
		\\
		& \delta_\lambda h^{++\dot\alpha(s-1)\alpha(s-2)+} =
		\mathcal{D}^{++} \bar\lambda^{+\dot\alpha(s-1)\alpha(s-2)}. 
	\end{split}
\end{equation}
The gauge parameters are unconstrained analytic superfields, which satisfy the following reality conditions under tilde-conjugation:
\begin{equation}
		\begin{split}
			&\widetilde{\lambda^{\alpha(s-1)\dot{\alpha}(s-1)}} = \lambda^{\alpha(s-1)\dot{\alpha}(s-1)},
			\\
			&\widetilde{\lambda^{+\alpha(s-1)\dot{\alpha}(s-2)}}
			=
			\lambda^{+\alpha(s-2)\dot{\alpha}(s-1)},
			\\
			&\widetilde{\lambda^{\alpha(s-2)\dot{\alpha}(s-2)}}
			=
			\lambda^{\alpha(s-2)\dot{\alpha}(s-2)}.
		\end{split}
\end{equation}

 Realization of rigid $\mathcal{N}=2$ supersymmetry transformations on analytic  prepotentials is given by equations ($\epsilon^{\hat{\alpha}\pm} := \epsilon^{\hat{\alpha}i}u_i^\pm$): 
 \begin{equation}\label{eq: susy prepotentials}
	\begin{split}
		&\delta_\epsilon h^{++\alpha(s-1)\dot\alpha(s-1)} = -4i\big[h^{++\alpha(s-1)(\dot\alpha(s-2)+}\bar\epsilon^{-\dot\alpha_{s-1})}-
		h^{++\dot\alpha(s-1)(\alpha(s-2)+}\,\epsilon^{-\alpha_{s-1})}
		\big]\,, 
		\\
		&
		\delta_\epsilon h^{++\alpha(s-2)\dot\alpha(s-2)} =2i\big[h^{++(\alpha(s-2)\alpha_{s-1})
			\dot\alpha(s-2)+}\epsilon^{-}_{\alpha_{s-1}} +
		h^{++\alpha(s-2)(\dot\alpha(s-2)\dot\alpha_{s-1})+}\,\bar\epsilon^{-}_{\dot{\alpha}_{s-1}}
		\big]\,,
		\\
		& \delta_\epsilon h^{++\alpha(s-1)\dot{\alpha}(s-2)+} = 0,
		\\
		&\delta_\epsilon h^{++\dot{\alpha}(s-1)\alpha(s-2) +} = 0.
	\end{split}
\end{equation}

Using the gauge freedom \eqref{eq: GF}, one may impose the Wess-Zumino-type gauge. In the Wess–Zumino gauge, the analytic prepotentials contain precisely the physical and auxiliary component fields of the off-shell higher-spin multiplet:
\begin{equation}\label{eq: WZ gauge}
	\begin{split}
		&  h_{WZ}^{++\alpha(s-1)\dot{\alpha}(s-1)}
		=
		-4i \theta^{+}_{\beta} \bar{\theta}^{+}_{\dot{\beta}} \Phi^{(\beta\alpha(s-1))(\dot{\beta}\dot{\alpha}(s-1))}_{}
		-
		4i \left( \frac{s-1}{s} \right)^2  \theta^{+(\alpha} \bar{\theta}^{+(\dot{\alpha}}
		\Phi^{\alpha(s-2))\dot{\alpha}(s-2))}
		\\
		&\qquad\qquad\qquad\qquad
		+  (\bar{\theta}^+)^2 \theta^{+\beta} \psi_{\beta}^{\alpha(s-1)\dot{\alpha}(s-1)i}u^-_i
		+\, (\theta^+)^2 \bar{\theta}^{+\dot{\beta}} \bar{\psi}_{\dot{\beta}}^{\alpha(s-1)\dot{\alpha}(s-1)i}u_i^-
		\\&\qquad\qquad\qquad\qquad
		+  (\theta^+)^2 (\bar{\theta}^+)^2 V^{\alpha(s-1)\dot{\alpha}(s-1)(ij)}u^-_iu^-_j\,,
		\\
		&  h_{WZ}^{++\alpha(s-2)\dot{\alpha}(s-2)} =
		-4i \theta^{+}_{\beta} \bar{\theta}^{+}_{\dot{\beta}} C^{(\beta\alpha(s-2))(\dot{\beta}\dot{\alpha}(s-2))}_{}
		-4i \theta^{+(\alpha} \bar{\theta}^{+(\dot{\alpha}} C^{\alpha(s-3))\dot{\alpha}(s-3))}
		\\&\qquad\qquad\qquad\qquad+ (\bar{\theta}^+)^2 \theta^{+}_{\beta} \rho^{(\beta\alpha(s-2))\dot{\alpha}(s-2)i}u^-_i
		+ (\bar{\theta}^+)^2 \theta^{+(\alpha} \bar{\psi}^{\alpha(s-3))\dot{\alpha}(s-2)i}u^-_i
		\\&\qquad\qquad\qquad\qquad
		+ (\theta^+)^2 \bar{\theta}^{+}_{\dot{\beta}} \bar{\rho}^{\alpha(s-2)(\dot{\beta}\dot{\alpha}(s-2))i}u_i^-
		+ (\theta^+)^2 \bar{\theta}^{+(\dot{\alpha}} \psi^{\alpha(s-2)\dot{\alpha}(s-3))i}u_i^-
		\\&\qquad\qquad\qquad\qquad+ (\theta^+)^2 (\bar{\theta}^+)^2 S^{\alpha(s-2)\dot{\alpha}(s-2)(ij)}u^-_iu^-_j\,,
		\\
		&  h_{WZ}^{++\alpha(s-1)\dot{\alpha}(s-2)+} = (\theta^+)^2 \bar{\theta}^+_{\dot{\beta}} P^{\alpha(s-1)\dot{\alpha}(s-2)\dot{\beta}}
		+  \left(\bar{\theta}^+\right)^2 \theta^+_\beta T^{\dot\alpha(s-2)\alpha(s-1)\beta}
		\\&\qquad\qquad\qquad\qquad+  (\theta^+)^2 (\bar{\theta}^+)^2 \chi^{\alpha(s-1)\dot{\alpha}(s-2)i}u^-_i\,,
		\\
		&  h_{WZ}^{++\dot{\alpha}(s-1)\alpha(s-2)+} = \widetilde{\left(h_{WZ}^{++\alpha(s-1)\dot{\alpha}(s-2)+}\right)}\,.
	\end{split}
\end{equation}
The first line contains the highest-spin Fronsdal field and its trace-like companion, the second line fermionic fields, while the last line contains auxiliary components and so on.
Finally, after eliminating pure-gauge components, the remaining fields are interpreted as  the fields of the $\mathcal{N}=2$ spin-$\mathbf{s}$ off-shell supermultiplet:
\begin{equation}
	\begin{split}
		\text{\textit{Physical fields}}&: \qquad \left(\Phi^{\alpha(s)\dot{\alpha}(s)},  \Phi^{\alpha(s-1)\dot{\alpha}(s-1)} \right), \left(\psi_\rho^{\alpha(s-1)\dot{\alpha}(s-1))i}, \psi^{\alpha(s-2)\dot{\alpha}(s-3)i} \right),
		\\&\quad \quad\;\;\;
		\left(C^{\alpha(s-1)\dot{\alpha}(s-1)}, C^{\alpha(s-3)\dot{\alpha}(s-3)}\right),
		\\
		\text{\textit{Auxiliary fields}}&:
		\qquad V^{\alpha(s-1)\dot{\alpha}(s-1)(ij)}, \rho^{\alpha(s-1)\dot{\alpha}(s-2)\,i}, 
		S^{\alpha(s-2)\dot{\alpha}(s-2)(ij)}, \\&\quad \quad\;\;\; T^{\alpha(s-1)\beta\dot{\alpha}(s-2)}, P^{\alpha(s-1)\dot{\alpha}(s-2)\dot{\beta}}, \chi^{\alpha(s-1)\dot{\alpha}(s-2) i}.
	\end{split}
\end{equation} 
Physical fields correspond to the massless Fronsdal spin $s$, the doublet of spin $s-\frac{1}{2}$ Fang-Fronsdal fields and the spin $s-1$ Fronsdal field (see appendix \ref{app A} for a review of the formulation of free higher-spin theories in spinor notation, which we use in this article). It is important to note that some of the fields require field redefinition, see, e.g.,  eq. \eqref{eq: P fields} for redefinition in the spin-$s$ sector and section 3.2 in  Ref. \cite{Ivanov:2024gjo} for the spin-$\mathbf{2}$ case.

It is worth emphasizing that the gauge parameters $\lambda$ admit a clear geometric meaning, e.g.:
$$
 \lambda^{\alpha(s-1)\dot{\alpha}(s-1)} (\zeta) = a^{\alpha(s-1)\dot{\alpha}(s-1)}(x) +\dots
 \quad
 \Rightarrow
 \quad
 \delta_\lambda \Phi_{\alpha(s)\dot{\alpha}(s)}(x)
 =
 \partial_{(\alpha(\dot{\alpha}}a_{\alpha(s-1))\dot{\alpha}(s-1))}(x).
$$
Thus, the lower components of the $\lambda$-superparameters correspond to the gauge transformation parameters for free higher-spin fields.

\subsection{$G^{++}$-superfields and superfield action principle}

The analytic prepotentials provide the off-shell formulation, while the non-analytic covariant superfields introduced below are better suited for constructing gauge-invariant quantities.

Starting from the analytic spin-$\mathbf{s}$  prepotentials, one constructs supersymmetry invariant analytic differential operator \cite{Buchbinder:2022kzl}:
\begin{equation}\label{eq: hat H}
	\hat{\mathcal{H}}^{++}_{(s)} = h^{++\alpha(s-2)\dot{\alpha}(s-2)M} \partial_M \partial^{s-2}_{\alpha(s-2)\dot{\alpha}(s-2)},
	\qquad
	\partial_M := (\partial_{\alpha\dot{\alpha}}, \partial^-_{\alpha}, \partial^-_{\dot{\alpha}}, \partial_5).
\end{equation}	
The index $M$ collectively labels vector, spinor and central-charge derivatives.
$	\hat{\mathcal{H}}^{++}_{(s)} $ has a simple gauge transformation law:
\begin{equation}
	\delta_\lambda 	\hat{\mathcal{H}}^{++}_{(s)} = [\mathcal{D}^{++}, \hat{\Lambda}_{(s)}],
	\qquad
	\hat{\Lambda}_{(s)} : = \lambda^{\alpha(s-2)\dot{\alpha}(s-2)M} \partial_M \partial^{s-2}_{\alpha(s-2)\dot{\alpha}(s-2)}.
\end{equation}	
This operator naturally defines cubic interactions of $\mathcal{N}=2$ higher-spin supermultiplets with hypermultiplet:
\begin{equation}\label{eq: hyper int 2}
	\mathcal{L}_{int} = - \frac{1}{2} q^{+a} \left(\mathcal{D}^{++} + \hat{\mathcal{H}}^{++}_{(s)} (J)^{P(s)} \right)q^+_a.
\end{equation}
Since $\hat{\mathcal{H}}^{++}_{(s)}$ is supersymmetry invariant, it naturally defines supersymmetry-invariant  superfields.
We introduce supersymmetry-invariant superfields $G^{++\dots}$ through the decomposition in the basis of covariant derivatives \eqref{eq: cov der}:
\begin{equation}\label{eq: DO cov G}
\hat{\mathcal{H}}^{++}_{(s)} = G^{++\alpha(s-2)\dot{\alpha}(s-2)M} \mathcal{D}_M \partial^{s-2}_{\alpha(s-2)\dot{\alpha}(s-2)},
\qquad
\delta_\epsilon G^{++\alpha(s-2)\dot{\alpha}(s-2)M} = 0.
\end{equation}	
Explicit relations of covariant $\mathcal{N}=2$ superfields to analytic prepotentials are given by the relations:
\begin{equation}\label{eq: G++ def}
	\begin{split}
		& G^{++\alpha(s-1)\dot\alpha(s-1)} =
		h^{++\alpha(s-1)\dot\alpha(s-1)} + 4i
		\big[h^{++\alpha(s-1)(\dot\alpha(s-2)+}\bar\theta^{-\dot\alpha_{s-1})}
		\\
		&\qquad\qquad\qquad\qquad\qquad\qquad\qquad\qquad\qquad\quad-
		h^{++\dot\alpha(s-1)(\alpha(s-2)+}\,\theta^{-\alpha_{s-1})}
		\big], \\ & G^{++\alpha(s-2)\dot\alpha(s-2)} =
		h^{++\alpha(s-2)\dot\alpha(s-2)} 
	- 2i
		\big[h^{++(\alpha(s-2)\alpha_{s-1})
			\dot\alpha(s-2)+}\theta^{-}_{\alpha_{s-1}} 
				\\&\qquad\qquad\qquad\qquad\qquad\qquad\qquad\qquad\qquad\quad+
		h^{++\alpha(s-2)(\dot\alpha(s-2)\dot\alpha_{s-1})+}\,\bar\theta^{-}_{\dot{\alpha}_{s-1}}
		\big],
		\\
		& G^{++\alpha(s-1)\dot\alpha(s-2)+} = - h^{++\alpha(s-1)\dot\alpha(s-2)+}, 
		\\
		& G^{++\alpha(s-2)\dot\alpha(s-1)+}
		=
		-h^{++\alpha(s-2)\dot\alpha(s-1)+}.
	\end{split}
\end{equation}
The superfields $G^{++}$
are convenient because supersymmetry acts trivially on them, although they are no longer analytic:
\begin{equation}
	\begin{split}
	&\mathcal{D}^+_\beta G^{++\alpha(s-1)\dot\alpha(s-1)} 
	=
	+4i \, h^{++\dot\alpha(s-1)(\alpha(s-2)+}\,\delta^{\alpha_{s-1})}_\beta,
	\\
	&\bar{\mathcal{D}}^+_{\dot{\beta}} G^{++\alpha(s-1)\dot\alpha(s-1)} 
	=
	-
	4i \,
	h^{++\alpha(s-1)(\dot\alpha(s-2)+}\delta^{\dot\alpha_{s-1})}_{\dot{\beta}},
	\\
	&
	\mathcal{D}^{+\beta}G^{++\alpha(s-2)\dot\alpha(s-2)} 
	=
		- 2i\,
	h^{++(\alpha(s-2)\beta)
		\dot\alpha(s-2)+},
		\\
		&
		\bar{\mathcal{D}}^{+\dot{\beta}}G^{++\alpha(s-2)\dot\alpha(s-2)} 
		=
		+
		2i \,	h^{++\alpha(s-2)(\dot\alpha(s-2)\dot\beta)+}. 
	\end{split}
\end{equation}	
Gauge transformations of the $G^{++}$ superfields take the simple form:
  \begin{equation}\label{eq:G++ gauge}
  \delta_\lambda G^{++\alpha(s-2)\dot\alpha(s-2)M} =
  \mathcal{D}^{++}\Lambda^{\alpha(s-2)\dot\alpha(s-2)M},
  \end{equation}
 where the gauge $\Lambda$-parameters are defined as coefficients in expansion  of operator $\hat{\Lambda}_{(s)}$ in covariant derivative basis:
 \begin{equation}\hat{\Lambda}_{(s)} : = \Lambda^{\alpha(s-2)\dot{\alpha}(s-2)M} \mathcal{D}_M \partial^{s-2}_{\alpha(s-2)\dot{\alpha}(s-2)}.
 \end{equation}

 Although the analytic prepotentials provide the natural off-shell description,
 gauge-invariant quantities cannot be constructed locally solely in terms of analytic prepotentials.  For this reason, employing the zero-curvature conditions, we introduce the non-analytic harmonic superfield $G^{--}$:
  \begin{equation}\label{zero-curv}
  	\begin{split}
  	 \mathcal{D}^{++}G^{--\alpha(s-2)\dot\alpha(s-2)M} &= \mathcal{D}^{--}G^{++\alpha(s-2)\dot\alpha(s-2)M}\,, 
  	 \\
  	 \mathcal{D}^{++}G^{--\alpha(s-2)\dot{\alpha}(s-2)\hat{\beta}-}
  	 &+
  	 G^{--\alpha(s-2)\dot{\alpha}(s-2)\hat{\beta}+} = 0.
  	\end{split} 
  \end{equation}
  Gauge transformations of these potentials take the form:
  \be\label{Lambda gauge}
  \begin{split}
  \delta_\lambda G^{--\alpha(s-1)\dot\alpha(s-1)} &=
  \mathcal{D}^{--}\Lambda^{\alpha(s-1)\dot\alpha(s-1)}\,, 
  \\
  \delta_\lambda G^{--\alpha(s-2)\dot\alpha(s-2)} &=
  \mathcal{D}^{--}\Lambda^{\alpha(s-2)\dot\alpha(s-2)},
  \\
   \delta_\lambda G^{--\alpha(s-2)\dot\alpha(s-2)\hat{\beta}+} &=
  \mathcal{D}^{--}\Lambda^{\alpha(s-2)\dot\alpha(s-2)\hat{\beta}+} + \Lambda^{\alpha(s-2)\dot\alpha(s-2)\hat{\beta}-},
  \\
  \delta_\lambda G^{--\alpha(s-2)\dot\alpha(s-2)\hat{\beta}-} &=
  \mathcal{D}^{--}\Lambda^{\alpha(s-2)\dot\alpha(s-2)\hat{\beta}-}. 
  \end{split}
  \ee
  Here parameter $\Lambda^{\alpha(s-2)\dot\alpha(s-2)\hat{\beta}-}$ is defined through the equation $$\Lambda^{\alpha(s-2)\dot\alpha(s-2)\hat{\beta}+} = \mathcal{D}^{++} \Lambda^{\alpha(s-2)\dot\alpha(s-2)\hat{\beta}-}.$$
  In what follows, we will show that negative potentials $G^{--}$ are the main objects for constructing gauge-invariant $\mathcal{N}=2$ superfield strengths and conserved $\mathcal{N}=2$ supercurrents.
  \\

 Gauge-invariant action has universal form for all $\mathcal{N}=2$ spin-$\mathbf{s}$  supermultiplets \cite{Buchbinder:2021ite}:
  \begin{equation}\lb{ActionsGen}
  	\begin{split}
  		& S_{(s)} = (-1)^{s+1} \int d^4x
  		d^8\theta du \,\Big\{G^{++
  			\alpha(s-1)\dot\alpha(s-1)}G^{--}_{\alpha(s-1)\dot\alpha(s-1)} \\
  		&\;\;\;\;\;  \qquad\qquad \qquad\qquad \qquad\qquad+\,
  		4G^{++\alpha(s-2)\dot\alpha(s-2)}G^{--}_{\alpha(s-2)\dot\alpha(s-2)}
  		\Big\}\,. 
  	\end{split}
  \end{equation}
  
  Varying this action with respect to analytic prepotentials, one can derive  superfield equations of motion \cite{Buchbinder:2022vra}, which can be written in a compact (but not analytic) form:
  \begin{equation}\label{eq: super EOM}
  \mathcal{F}^+_{\alpha(s-2)\dot{\alpha}(s-1)}
  :=
  	(\bar{\mathcal{D}}^+)^2 \mathcal{D}^{+\alpha_{s-1}} G^{--}_{\alpha(s-1) \dot{\alpha}(s-1)}
  	-2
  	(\mathcal{D}^+)^2 \bar{\mathcal{D}}^+_{(\dot{\alpha}} G^{--}_{\alpha(s-2) \dot{\alpha}(s-2))} \;
  	\approx 0.
  \end{equation}
  These equations are gauge invariant and harmonic analytic, i.e. $\mathcal{D}^{++} \mathcal{F}^+_{\alpha(s-2)\dot{\alpha}(s-1)} = 0 $. The superfield $ \mathcal{F}^+_{\alpha(s-2)\dot{\alpha}(s-1)}$ is an $\mathcal{N}=2$ higher-spin generalization of the linearized  Einstein tensor and vanishes on-shell.
  In components, these equations lead to free equations for physical gauge fields (see \eqref{eom s} and \eqref{eom s-1/2} in appendix~\ref{app A}). Auxiliary fields vanish on-shell (for $\mathbf{s}\geq3$).

  \subsection{$\mathcal{N}=2$ higher-spin Weyl supertensor}
  
 The propagating degrees of freedom are encoded in a gauge invariant higher-spin super-Weyl tensor\footnote{ $\mathcal{N}=2$ higher-spin super-Weyl  tensor has an interesting geometrical structure. Technical details of the construction are presented in appendix \ref{app: structure of strength}.}. 
 The superfield strength is constructed as the simplest local gauge-invariant chiral combination of $G^{--}$ potentials containing $2s-2$ undotted spinor indices. Its explicit form is:
   \begin{equation}\label{eq: SF strenghts}
  	\begin{split}
  		\mathcal{W}_{\alpha(2s-2)}
  		:=
  		&\left(\bar{\mathcal{D}}^+\right)^2 \partial_{(\alpha_1}^{\;\;\dot{\beta}_1} \dots  	\partial_{\alpha_{s-2}}^{\;\;\dot{\beta}_{s-2}} \Bigr\{
  		\mathcal{D}^+_{\alpha_{s-1}} G^{---}_{\alpha_s\dots\alpha_{2s-2})(\dot{\beta}_1\dots\dot{\beta}_{s-2})}
  		\\&	\qquad\qquad +
  		\mathcal{D}^-_{\alpha_{s-1}}
  		G^{--+}_{\alpha_s\dots\alpha_{2s-2})(\dot{\beta}_1\dots\dot{\beta}_{s-2})}
  		+
  		\partial_{\alpha_{s-1}}^{\;\;\dot{\beta}_{s-1}} G^{--}_{\alpha_s\dots\alpha_{2s-2})(\dot{\beta}_1\dots\dot{\beta}_{s-1})}\Bigr\}.
  	\end{split}
  \end{equation}	
  Similarly,  one can constructs  conjugated superfield strength $	\bar{\mathcal{W}}_{\dot{\alpha}(2s-2)} := \widetilde{\mathcal{W}}_{\alpha(2s-2)}$. This superfield provides the natural higher-spin generalization of linearized $\mathcal{N}=2$ super-Weyl tensor, introduced in harmonic approach in \cite{Ivanov:2024gjo}.

  Higher-spin super-Weyl tensor $\mathcal{W}_{\alpha(2s-2)}$ is the chiral $\mathcal{N}=2$ superfield, the tilde-conjugate superfield strength $\bar{\mathcal{W}}_{\dot{\alpha}(2s-2)}$ is the anti-chiral superfield:
  \begin{equation}
  	\bar{\mathcal{D}}^\pm_{\dot{\beta}}\mathcal{W}_{\alpha(2s-2)} = 0,
  	\qquad
  	\mathcal{D}^\pm_{\beta} \bar{\mathcal{W}}_{\dot{\alpha}(2s-2)} = 0.
  \end{equation}
  Also higher-spin Weyl superfields $\mathcal{W}_{\alpha(2s-2)}$ and $\bar{\mathcal{W}}_{\dot{\alpha}(2s-2)}$ are covariantly harmonic independent, i.e., they satisfy
  \begin{equation}\label{eq: harm indep}
  	\mathcal{D}^{\pm\pm} \mathcal{W}_{\alpha(2s-2)} = 0,
  	\qquad
  		\mathcal{D}^{\pm\pm} \bar{\mathcal{W}}_{\dot{\alpha}(2s-2)} = 0.
  \end{equation}	
  
   On-shell superfield strengths satisfy the identities (derivation of this result is presented in the appendix \ref{eq: cons N=2 str}):
  \begin{equation}\label{eq: SEOM}
  	\mathcal{D}^{+\alpha}	\mathcal{W}_{\alpha(2s-2)} \approx 0,
  	\qquad
  	\bar{\mathcal{D}}^{+\dot{\alpha}}	\bar{\mathcal{W}}_{\dot{\alpha}(2s-2)}  \approx 0.
  \end{equation}	
  A number of other useful on-shell relations which follow from \eqref{eq: harm indep} and \eqref{eq: SEOM} are presented in the formula \eqref{eq: on-shell W cond}.

  In components, superfield strengths contain the spin-$s$  Weyl-like tensor $C_{\alpha(2s)}$ (see eq. \eqref{eq: bos C}):
  \begin{equation}
  	\mathcal{W}_{\alpha(2s-2)} \sim \theta^{+\beta} \theta^{-\gamma}  C_{(\alpha(2s-2)\beta\gamma)} + \dots
  \end{equation}	
  The highest-spin bosonic Weyl tensor appears at the $\theta^+\theta^-$ level of the superfield expansion, thereby increasing the number of spinor indices by two.
  Thus, one can naturally consider $\mathcal{W}_{\alpha(2s-2)}$ as a generalization of the linearized $\mathcal{N}=2$ super Weyl tensor $\mathcal{W}_{(\alpha\beta)}$ to higher-spin case. Full component content of superstrength $\mathcal{W}_{\alpha(2s-2)}$ is discussed in appendix \ref{eq: cons N=2 str}.

  \subsection{$\mathcal{N}=2$ vector multiplet as the limiting case of $\mathcal{N}=2$ higher-spin theories}
  
  The $\mathcal{N}=2$ \textit{vector multiplet} may be viewed as the spin-$1$ member of the $\mathcal{N}=2$ higher-spin family.
   $\mathcal{N}=2$ vector multiplet is described by the real analytic superfield $V^{++}$ defined up to gauge freedom 
  \begin{equation}\label{eq: spin 1 gauge}
  	\delta_\lambda V^{++} = \mathcal{D}^{++} \lambda.
  \end{equation}	
  Using the gauge freedom, we fix Wess-Zumino-type gauge:
  \begin{equation}\label{eq: vector WZ gauge}
  	\begin{split}
	V_{WZ}^{++}  = &\;
	i (\bar{\theta}^+)^2 \phi
	- 
		i (\theta^+)^2 \bar{\phi} 
	-4i \theta^{+\beta} \bar{\theta}^{+\dot{\beta}} A_{\beta\dot{\beta}} 
	\\&+ (\bar{\theta}^+)^2 \theta^{+\beta}\psi^i_\beta u^-_i + 
	(\theta^+)^2 \bar{\theta}^{+\dot{\beta}} \bar{\psi}^i_{\dot{\beta}} u^-_i
	+
	(\theta^+)^2 (\bar{\theta}^+)^2  D^{(ij)} u^-_i u^-_j.
	\end{split}
\end{equation}  
In this gauge, we obtain the content of $\mathcal{N}=2$ spin-$\mathbf{1}$ off-shell supermultiplet:
\begin{equation}\label{eq: N=2 vector}
	\begin{split}
		\text{Physical fields}&: \qquad 
		A_{\alpha\dot{\alpha}}, \psi^i_\alpha, \phi,
		\\
		\text{Auxiliary fields}&:
		\qquad D^{(ij)}.
	\end{split}
\end{equation} 

  Dynamical $\mathcal{N}=2$ superfield action is given by:
  \begin{equation}\label{eq: spin 1 action}
  	S_{(s=1)} = \int d^4x d^8\theta du \; V^{++} V^{--},
  \end{equation}
  where $V^{--}$ is defined as the solution of zero-curvature equation
  \begin{equation}
  	\mathcal{D}^{++} V^{--} = \mathcal{D}^{--} V^{++}.
  \end{equation}	 
  Dynamical equations of motion can be easily derived from the action:
  \begin{equation}\label{eq: EOM spin 1}
  	(\mathcal{D}^+)^4 V^{--} \approx 0.
  \end{equation}

  \textit{Gauge invariant superfield strengths}  have  a simpler structure in contrast to higher-spin case \eqref{eq: SF strenghts}:
  \begin{equation}\label{eq: W spin 1}
  	\mathcal{W} := (\bar{\mathcal{D}}^+)^2 V^{--},
  	\qquad
  	\bar{\mathcal{W}} := (\mathcal{D}^+)^2 V^{--} 
  \end{equation}
  and  also satisfy  harmonic-independence and chirality/antichirality respectively.
  Also $\mathcal{W}$ and $\bar{\mathcal{W}}$ satisfy the Bianchi identities:
  \begin{equation}
  	(\mathcal{D}^+)^2 \mathcal{W} = (\bar{\mathcal{D}}^+)^2 \bar{\mathcal{W}},
  	\qquad
  	(\mathcal{D}^+ \mathcal{D}^-) \mathcal{W} 
  	=
  	(\bar{\mathcal{D}}^+ \bar{\mathcal{D}}^-)  \bar{\mathcal{W}},
  	\qquad
  	 (\mathcal{D}^-)^2 \mathcal{W} = (\bar{\mathcal{D}}^-)^2 \bar{\mathcal{W}}.
  \end{equation}	
  
   On-shell $\mathcal{W}$ and $\bar{\mathcal{W}}$ satisfy the relations:
  \begin{equation}\label{eq: spin 1 on-shell}
  		\left(\mathcal{D}^+\right)^2\mathcal{W}  \approx 0,
  		\qquad
  		\left(\bar{\mathcal{D}}^+\right)^2 	\bar{\mathcal{W}} \approx 0.
  \end{equation}	
These equations are equivalent to the equation of motion \eqref{eq: EOM spin 1}, formulated in terms of superfield strengths. 
  Acting on the first and second equations by the harmonic derivative $\mathcal{D}^{--}$, we derive useful on-shell equations:
  \begin{equation}
  		\left( \mathcal{D}^+ \mathcal{D}^- \right) \mathcal{W} \approx 0,
  		\qquad
  		\left( \bar{\mathcal{D}}^+ \bar{\mathcal{D}}^- \right) \bar{\mathcal{W}} \approx 0. 
  \end{equation}	
  Acting on these equations by derivatives $\bar{\mathcal{D}}^+$ and $\mathcal{D}^+$ gives:
  \begin{equation}
  			\partial_{\dot{\alpha}}^{\;\alpha} \mathcal{D}^{\pm}_\alpha  \mathcal{W}  \approx 0,
  		\qquad
  		\partial_{\alpha}^{\;\dot{\alpha}} \bar{\mathcal{D}}^\pm_{\dot{\alpha}} \bar{\mathcal{W}} \approx 0.
  \end{equation}	
  Acting on equations by $\bar{\mathcal{D}}^+$ and $\mathcal{D}^+$ correspondingly, one finds
  \begin{equation}
  	\Box \mathcal{W} \approx 0 ,
  	\qquad
  	  	\Box \bar{\mathcal{W}} \approx 0.
  \end{equation}	
  
  In what follows, we will need the expression for the superfield strength on-shell:
  \begin{equation}
  	\mathcal{W} \sim \phi + \theta^{+\beta} \psi^i_\beta u^-_i
  	-
  	\theta^{-\beta} \psi_\beta^i u^+_i
  	+
  \theta^{+\alpha}	\theta^{-\beta} C_{\alpha\beta} 
  	+
  	 4i \theta^{-\beta}  \bar{\theta}^{+\dot{\beta}}  \partial_{\beta\dot{\beta}} \phi 
  	 -
  	 4i \theta^{-(\beta} \theta^{+\gamma)} \bar{\theta}^{+\dot{\gamma}} \partial_{\gamma\dot{\gamma}}\psi^i_\beta u^-_i.
  \end{equation}	
  This expansion can be viewed as a degenerate case of the general integer-spin $\mathcal{N}=2$ Weyl supertensor, presented in eq. \eqref{eq: W on-shell}.

  \newpage
  
  \section{Higher-spin Mezincescu-type prepotentials}\label{sec: 3}

 An alternative representation of $\mathcal{N}=2$ higher-spin prepotentials is given by  Mezincescu-type higher-spin non-analytic prepotentials, first introduced in \cite{Buchbinder:2022vra}.
 These prepotentials provide a natural harmonic-superspace higher-spin generalization of the spin-$\mathbf{1}$ Mezincescu prepotential  \cite{Mezincescu} and spin-$\mathbf{2}$ Gates-Siegel prepotential \cite{Gates:1981qq}. These prepotentials provide a convenient framework for constructing arbitrary abelian cubic vertices.

\subsection{Mezincescu-type prepotentials and their gauge freedom}
 The most straightforward way to introduce Mezincescu-type prepotentials is to use the analytic differential operator \eqref{eq: hat H} and represent it in the manifestly analytic form\footnote{Here we use the notation $(\mathcal{D}^+)^4 := \frac{1}{16} (\mathcal{D}^+)^2 (\bar{\mathcal{D}}^+)^2$.}:
\begin{equation}\label{eq: Mez intr}
	\hat{\mathcal{H}}^{++}_{(s)}  
	=
	(\mathcal{D}^+)^4 \left[ \Psi^{-\alpha(s-1)\dot{\alpha}(s-2)} \mathcal{D}^{-}_{\alpha}
	+
	\bar{\Psi}^{-\alpha(s-2)\dot{\alpha}(s-1) } \bar{\mathcal{D}}^{-}_{\dot{\alpha}} \right] \partial^{s-2}_{\alpha(s-2)\dot{\alpha}(s-2)}
\end{equation}	
Here, unconstrained superfields $\Psi^{-\alpha(s-1)\dot{\alpha}(s-2)}$ and $\bar{\Psi}^{-\alpha(s-2)\dot{\alpha}(s-1) } = \widetilde{\Psi^{-\alpha(s-1)\dot{\alpha}(s-2)}}$ are spin-$\mathbf{s}$ Mezincescu-type prepotentials.
Since the operator $\hat{\mathcal{H}}^{++}_{(s)} $ is  invariant under rigid $\mathcal{N}=2$ supersymmetry 
\begin{equation}
	\delta_\epsilon \hat{\mathcal{H}}^{++}_{(s)} = 0,
\end{equation}	 the prepotentials $\Psi^{-\dots}$ and $\bar{\Psi}^{-\dots}$ are also  invariant under rigid  $\mathcal{N}=2$ supersymmetry:
\begin{equation}
	\delta_\epsilon \Psi^{-\alpha(s-1)\dot{\alpha}(s-2)} = 0,
	\qquad
	\delta_\epsilon \bar{\Psi}^{-\alpha(s-2)\dot{\alpha}(s-1) } = 0.
\end{equation}
	
Equation \eqref{eq: Mez intr} establishes the relation between the Mezincescu-type and analytic prepotentials \eqref{eq: HS analytic prepotentials}. Using the representation \eqref{eq: DO cov G}, one obtains the relation between the Mezincescu prepotentials and the supersymmetry-invariant superfields $G^{++\dots}$:
\begin{equation}\label{eq: G and Psi prepot}
	\begin{split}
		G^{++\alpha(s-1)\dot{\alpha}(s-1)} &=
		\frac{i}{2} \left( (\mathcal{D}^+)^2 \bar{\mathcal{D}}^{+(\dot{\alpha}} \Psi^{-\alpha(s-1)\dot{\alpha}(s-2))}
		+
		(\bar{\mathcal{D}}^+)^2 \mathcal{D}^{+(\alpha} \bar{\Psi}^{-\alpha(s-2))\dot{\alpha}(s-1)} \right),
	\\	
	G^{++\alpha(s-1)\dot{\alpha}(s-2)+} &= (\mathcal{D}^+)^4 \Psi^{-\alpha(s-1)\dot{\alpha}(s-2)},
	\\
		G^{++\alpha(s-2)\dot{\alpha}(s-1)+} &= (\mathcal{D}^+)^4 	\bar{\Psi}^{-\alpha(s-2)\dot{\alpha}(s-1) },
		\\
		G^{++\alpha(s-2)\dot{\alpha}(s-2)} &= \frac{i}{4} \left(  (\bar{\mathcal{D}}^+)^2 \mathcal{D}^+_\alpha \Psi^{-\alpha(s-1)\dot{\alpha}(s-2)} - (\mathcal{D}^+)^2 \bar{\mathcal{D}}^+_{\dot{\alpha}} \bar{\Psi}^{-\alpha(s-2)\dot{\alpha}(s-1)} \right).
	\end{split}
\end{equation}
This construction generalizes to higher-spin case the alternative representation of the $\mathcal{N}=2$ supergravity prepotentials introduced by Zupnik \cite{Zupnik:1998td}.
The higher-spin Mezincescu prepotentials are defined modulo gauge transformations with parameters $K$ and with parameters $B$:
 \begin{subequations}\label{eq: Psi gauge and pre gauge}
\begin{equation}
		\boxed{	
	\begin{split}
	\delta_{\lambda,b} \Psi^{-}_{\alpha(s-1)\dot{\alpha}(s-2) }
		=
		\;
		&\mathcal{D}^{++} K^{(-3)}_{\alpha(s-1)\dot{\alpha}(s-2) }
		+
		\mathcal{D}^+_{(\alpha} B^{--}_{\alpha(s-2))\dot{\alpha}(s-2)}
		\\&+
		\mathcal{D}^{+\beta} B^{--}_{(\alpha(s-1)\beta)\dot{\alpha}(s-2)}
		+
		\bar{\mathcal{D}}^{+\dot{\beta}} B^{--}_{\alpha(s-1)(\dot{\alpha}(s-2)\dot{\beta})}
		\\
		&+
		\bar{\mathcal{D}}^+_{(\dot{\alpha}} B^{--}_{\alpha(s-1)\dot{\alpha}(s-3))} ,
	\end{split}
}
\end{equation}	
\begin{equation}
	\begin{split}
		\delta_{\lambda,b} \bar{\Psi}^{-}_{\alpha(s-2)\dot{\alpha}(s-1) }
		=\;
		&\mathcal{D}^{++} \bar{K}^{(-3)}_{\alpha(s-2)\dot{\alpha}(s-1) }
		-
		\bar{\mathcal{D}}^+_{(\dot{\alpha}} B^{--}_{\alpha(s-2)\dot{\alpha}(s-2))}
		\\&-
		\bar{\mathcal{D}}^{+\dot{\beta}} \bar{B}^{--}_{\alpha(s-2)(\dot{\alpha}(s-1)\dot{\beta})}
		+
		\mathcal{D}^{+\beta} B^{--}_{(\alpha(s-2)\beta)\dot{\alpha}(s-1)}
		\\&
		+\mathcal{D}^+_{(\alpha} \bar{B}^{--}_{\alpha(s-3))\dot{\alpha}(s-1)}.
		\end{split}
\end{equation}	
\end{subequations}
The transformations generated by the parameters $B$ will be referred to as pre-gauge transformations.
Superfield parameters $K$ are related to analytic $\lambda$-parameters through the relation: 
\begin{equation}\label{eq: K and lambda}
	\hat{\Lambda}_{(s)} = (\mathcal{D}^+)^4 \left( K^{(-3)\alpha(s-1)\dot{\alpha}(s-2)} \mathcal{D}^-_\alpha
	+
	\bar{K}^{(-3)\alpha(s-2)\dot{\alpha}(s-1)} \bar{\mathcal{D}}^-_{\dot{\alpha}}
	 \right) \partial^{(s-2)}_{\alpha(s-2)\dot{\alpha}(s-2)}
\end{equation}	 
and therefore generate the $\lambda$-gauge transformations \eqref{eq: GF} of the analytic prepotentials.
The pre-gauge transformations, generated by the parameters $B$, leave the analytic $h^{++}$ prepotentials invariant. All $B$ parameters are complex with the exception of  parameters $B^{--}_{\alpha(s-1)\dot{\alpha}(s-1)}$ and $B^{--}_{\alpha(s-2)\dot{\alpha}(s-2)}$, which are real with respect to the tilde-conjugation:
\begin{equation}
	\widetilde{B^{--}_{\alpha(s-1)\dot{\alpha}(s-1)}}  = B^{--}_{\alpha(s-1)\dot{\alpha}(s-1)}, 
	\qquad
	\widetilde{B^{--}_{\alpha(s-2)\dot{\alpha}(s-2)}}  = B^{--}_{\alpha(s-2)\dot{\alpha}(s-2)}. 
\end{equation}	

It is also instructive to analyze the decomposition of the Mezincescu prepotential into analytic superfields. Let us consider its decomposition into analytic superfields:
\begin{equation}\label{eq: Mez exp}
	\begin{split}
\Psi^{-\alpha(s-1)\dot{\alpha}(s-2)}
	=
	&
	-(\theta^-)^4 h^{++\alpha(s-1)\dot{\alpha}(s-2)+}
	\\&
	+
	\frac{i}{2} (\theta^-)^2 \bar{\theta}^-_{\dot{\alpha}} \left(h^{++\alpha(s-1)\dot{\alpha}(s-1)} + i g^{++\alpha(s-1)\dot{\alpha}(s-1)} \right)
	\\&
	-i \,\frac{s-1}{s} \, (\bar{\theta}^-)^2 \theta^{-(\alpha}
	\left(  h^{++\alpha(s-2)) \dot{\alpha}(s-2)}
	+
	i g^{++\alpha(s-2)) \dot{\alpha}(s-2)} \right)
	\\&
	+
	 (\bar{\theta}^-)^2 \theta^-_\alpha g^{++\alpha(s)\dot{\alpha}(s-2)}
	+
	(\theta^-)^2 g^{+\alpha(s-1)\dot{\alpha}(s-2)} + \dots
	\end{split}
\end{equation}	
Here $h^{++\dots}$ denote the original analytic prepotentials \eqref{eq: HS analytic prepotentials}. This identification is made by using relations \eqref{eq: G and Psi prepot} and \eqref{eq: G++ def}. 
The superfields $g^{\dots}$ denote additional analytic superfields appearing in the Mezincescu prepotential. Under supersymmetry transformations, various analytic superfields in the decomposition transform through each other. In particular, the original supersymmetry transformations \eqref{eq: susy prepotentials} are recovered up to terms involving the $g^{\dots}$ superfields.

All analytic superfields $g^{\dots}$  can be gauged away using pre-gauge freedom (parameters $B$ in \eqref{eq: Psi gauge and pre gauge}). For example, under pre-gauge freedom:
\begin{subequations}
\begin{equation}
	(\theta^-)^2 \bar{\theta}^-_{\dot{\alpha}} g^{++\alpha(s-1)\dot{\alpha}(s-1)} \sim \bar{\mathcal{D}}^+_{\dot{\alpha}} B^{--\alpha(s-1)\dot{\alpha}(s-1)} ,
	\quad
	B^{--}_{\alpha(s-1)\dot{\alpha}(s-1)} \sim (\theta^-)^4 g^{++}_{\alpha(s-1)\dot{\alpha}(s-1)},
\end{equation}	
\begin{equation}
	(\theta^-)^2 g^{+\alpha(s-1)\dot{\alpha}(s-2)} \sim \bar{\mathcal{D}}^{+}_{\dot{\alpha}} B^{--\alpha(s-1)\dot{\alpha}(s-1)},
	\quad
	B^{--}_{\alpha(s-1)\dot{\alpha}(s-1)}
	\sim 
	(\theta^-)^2 \bar{\theta}^-_{(\dot{\alpha}} g^+_{\dot{\alpha}(s-2)) \alpha(s-1)}.
\end{equation}	
\end{subequations}
Therefore, the analytic $h$-prepotentials may be interpreted as a particular gauge choice for the Mezincescu prepotentials.

\subsection{Harmonic-independent gauge}

Using the $K$-gauge transformations \eqref{eq: Psi gauge and pre gauge}, one may impose the harmonic-independent gauge with a finite number of fields in the harmonic expansion:
\begin{equation}
	\Psi_{\text{HI}}^{-\alpha(s-1)\dot{\alpha}(s-2)}  = \Psi^{\alpha(s-1)\dot{\alpha}(s-2)i} u^-_i,
\end{equation}
where $\Psi^{\alpha(s-1)\dot{\alpha}(s-2)i} $ is the conventional  $\mathcal{N}=2$ superfield, defined in $\mathbb{R}^{4|8}$ superspace.
Residual harmonic-independent gauge transformation of this superfield is given by:
\begin{equation}
	\begin{split}
	\delta_\xi \Psi_{\alpha(s-1)\dot{\alpha}(s-2)}^{i} 
	=&\;
	\mathcal{D}_{(\alpha k} \xi^{(ik)}_{\alpha(s-2))\dot{\alpha}(s-2)} 
	+
	\mathcal{D}^\beta_k \xi^{(ik)}_{\alpha(s-2)(\dot{\alpha}(s-2)\beta)} 
	\\&+
	\bar{\mathcal{D}}^{\dot{\beta}}_k \xi^{(ik)}_{(\alpha(s-1))(\dot{\alpha}(s-2)\dot{\beta})} 
	+
	\bar{\mathcal{D}}_{(\dot{\alpha} k} \xi^{(ik)}_{\alpha(s-1)\dot{\alpha}(s-3))}.
	\end{split}
\end{equation}	
Unlike the analytic formulation, the Mezincescu representation allows one to work with unconstrained superfields in the conventional superspace. However, the price for working with the conventional harmonic-independent superfield prepotential is the loss
of the transparent geometric and group structure. Neither the gauge superfield $\Psi$ nor the gauge parameters $\xi$ appear to admit a natural geometric interpretation (see also discussion of the spin-$\mathbf{1}$ Mezincescu prepotential in section 7.2.4 of \cite{18}).

\newpage

\section{$\mathcal{N}=2$ higher-spin supercurrents}
\label{sec: N=2 higher-spin supercurrents}

In this section, we explore the superfield structure of $\mathcal{N}=2$ higher-spin supercurrents and corresponding cubic interactions. In particular, we present a detailed discussion of $\mathcal{N}=2$ analytic higher-spin supercurrents and analytic form of cubic vertices. In the following, we will show that the analytic form is the most convenient for analyzing the component structure of $\mathcal{N}=2$ abelian interactions.

\subsection{$\mathcal{N}=2$ principal supercurrents and cubic vertices}\label{sec: principle SC}

The key object to construct $\mathcal{N}=2$ abelian cubic higher-spin vertices is the \textit{principal $\mathcal{N}=2$ supercurrent} \cite{Zaigraev:2024ryg}, defined by the equations\footnote{In the $s=2$ case, one needs to replace third and fourth equations by $(\mathcal{D}^+)^2 \mathcal{J} \approx 0$ and $(\bar{\mathcal{D}}^+)^2 \mathcal{J} \approx 0$. This case corresponds to $\mathcal{N}=2$ supergravity supercurrent \cite{Kuzenko:1999pi, Butter:2010sc}.}:
\begin{equation}\label{eq: primary supercurrent}
	\begin{cases}
		\widetilde{\mathcal{J}}_{\alpha(s-2)\dot{\alpha}(s-2)} = \mathcal{J}_{\alpha(s-2)\dot{\alpha}(s-2)},
		\\
		\mathcal{D}^{++} \mathcal{J}_{\alpha(s-2)\dot{\alpha}(s-2)} \approx 0,
		\\
		\mathcal{D}^{+\beta}\mathcal{J}_{(\beta\alpha(s-3))\dot{\alpha}(s-2)} \approx 0,
		\\
		\bar{\mathcal{D}}^{+\dot{\beta}} \mathcal{J}_{\alpha(s-2)(\dot{\beta}\dot{\alpha}(s-3))} \approx 0.
	\end{cases}	
\end{equation}	
These equations are similar to the definition of the $\mathcal{N}=2$ \textit{primary supercurrent} in \cite{ Kuzenko:2021pqm, Kuzenko:2023vgf} (see also \cite{Howe:1981qj}), which implies special superconformal properties for supercurrent, constructed from $\mathcal{N}=2$ superconformal multiplets (massless hypermultiplet and $\mathcal{N}=2$ vector multiplet). 
However, in the present paper we consider non-conformal $\mathcal N=2$ higher-spin multiplets, for which the corresponding supercurrents are not primary in the superconformal sense. To avoid confusion with the standard terminology of $\mathcal N=2$ superconformal theories, we therefore refer to them as \textit{principal supercurrents}.

\smallskip

System of equations \eqref{eq: primary supercurrent} has several important consequences. As an example one can check that every component of principal supercurrent is a conserved current. Actually, definitions \eqref{eq: primary supercurrent} imply:
\begin{equation}
	\mathcal{D}^{++} \left( \mathcal{D}^{-\beta}\mathcal{J}_{(\beta\alpha_2\alpha_{s-2})\dot{\alpha}(s-2)}   \right) \approx 0 
	\quad
	\Rightarrow
	\quad
	\mathcal{D}^{-\beta}\mathcal{J}_{(\beta\alpha_2\alpha_{s-2})\dot{\alpha}(s-2)}\approx 0.
\end{equation}	
Acting with the analytic derivative $\bar{\mathcal{D}}^{+\dot{\alpha}}$, we obtain:
\begin{equation}\label{eq: principle supercurrent ST conservation}
	\partial^{\beta\dot{\beta}} \mathcal{J}_{(\alpha(s-3)\beta)(\dot{\alpha}(s-3)\dot{\beta})} \approx 0.
\end{equation}

Using $\mathcal{N}=2$ principal supercurrent and Mezincescu-type higher-spin prepotentials, one can construct consistent cubic vertices as \cite{Zaigraev:2024ryg}:
\begin{equation}\label{eq: Psi inter}
	\boxed{S_{int} = \int d^4xd^8\theta du \; \left( \Psi^{-\alpha(s-1)\dot{\alpha}(s-2)}\mathcal{D}^+_\alpha + \bar{\Psi}^{-\alpha(s-2)\dot{\alpha}(s-1)} \bar{\mathcal{D}}^+_{\dot{\alpha}} \right) \mathcal{J}_{\alpha(s-2)\dot{\alpha}(s-2)}.}
\end{equation}	
Consistency of the vertex at the leading order  follows from the principal supercurrent conservation equations.

In \cite{Zaigraev:2024ryg}, we constructed \textit{principal $\mathcal{N}=2$  supercurrent} with spin $\mathbf{s_1}\geq 2\mathbf{s_2}$  for an arbitrary $\mathcal{N}=2$ off-shell gauge spin-$\mathbf{s_2}$ supermultiplet\footnote{We use shortened notation for symmetrized product of derivatives $\partial^p := \partial_{\alpha(p)\dot{\alpha}(p)}^p = \partial_{(\alpha_1(\dot{\alpha}_1}\dots \partial_{\alpha_p)\dot{\alpha}_p)}$ and assume symmetrization over dotted and undotted indices.} 
\begin{equation}\label{eq: general principle supercurrent}
		\boxed{
	\begin{split}
		&\mathcal{J}_{\alpha(s_1-2)\dot{\alpha}(s_1-2)} = \;\; (i)^{s_1-2s_2} \sum_{p=0}^{s_1-2s_2}  (-1)^p \frac{\binom{s_1-2}{2s_2-2+p} \binom{s_1-2}{p}}{\binom{s_1-2}{2s_2-2}} \; \partial^p \mathcal{W}_{\alpha(2s_2-2)} \partial^{s_1-2s_2-p} \bar{\mathcal{W}}_{\dot{\alpha}(2s_2-2)}
		\\&
		- 
		(i)^{s_1-2s_2} \sum_{p=0}^{s_1-2s_2-1}   \frac{(-1)^p}{4i} \frac{\binom{s_1-2}{2s_2-1+p} \binom{s_1-2}{p}}{\binom{s_1-2}{2s_2-2}} \; \partial^p \mathcal{D}^-_\alpha \mathcal{W}_{\alpha(2s_2-2)} \partial^{s_1-2s_2-p-1} \bar{\mathcal{D}}^+_{\dot{\alpha}}\bar{\mathcal{W}}_{\dot{\alpha}(2s_2-2)}
		\\&
		+
		(i)^{s_1-2s_2} \sum_{p=0}^{s_1-2s_2-1} \frac{(-1)^p}{4i} \frac{\binom{s_1-2}{2s_2-1+p} \binom{s_1-2}{p}}{\binom{s_1-2}{2s_2-2}}  \; \partial^p \mathcal{D}^+_\alpha \mathcal{W}_{\alpha(2s_2-2)} \partial^{s_1-2s_2-p-1} \bar{\mathcal{D}}^-_{\dot{\alpha}}\bar{\mathcal{W}}_{\dot{\alpha}(2s_2-2)}
		\\&+
		(i)^{s_1-2s_2} \sum_{p=0}^{s_1-2s_2-2}   \frac{(-1)^{p}}{16} \frac{\binom{s_1-2}{2s_2+p} \binom{s_1-2}{p}}{\binom{s_1-2}{2s_2-2}}
		\partial^p \mathcal{D}^+_\alpha \mathcal{D}^-_\alpha \mathcal{W}_{\alpha(2s_2-2)} \partial^{s_1-2s_2-p-2} \bar{\mathcal{D}}^+_{\dot{\alpha}} \bar{\mathcal{D}}^-_{\dot{\alpha}} \bar{\mathcal{W}}_{\dot{\alpha}(2s_2-2)}.
	\end{split}}
\end{equation}	
Here $\mathcal{W}_{\alpha(2s_2-2)}$ and $\bar{\mathcal{W}}_{\dot{\alpha}(2s_2-2)}$ are gauge-invariant higher-spin $\mathcal{N}=2$ superfield strengths, defined in \eqref{eq: SF strenghts}.
It is important to note that the second equation in \eqref{eq: primary supercurrent}, harmonic-independence equation,  is satisfied for these supercurrents off-shell.

\subsection{Analytic structure of $\mathcal{N}=2$ principal supercurrent}

To analyze the  structure of the $\mathcal{N}=2$ principal supercurrent, we consider its $\theta^-$ decomposition into analytic superfields:
\begin{equation}\label{eq: PSC GS}
	\begin{split}
	\mathcal{J}_{\alpha(s)\dot{\alpha}(s)}
	=&\;
	\mathbb{J}_{\alpha(s)\dot{\alpha}(s)} 
	+
	\theta^{-\beta} \mathbb{J}^+_{\beta\alpha(s) \dot{\alpha}(s)} 
	+
	\bar{\theta}^{-\dot{\beta}} \bar{\mathbb{J}}^+_{\alpha(s)\dot{\beta}\dot{\alpha}(s)} 
	\\&
	+ i(\theta^-)^2 \mathbb{J}^{++}_{\alpha(s)\dot{\alpha}(s)}
	-
	i (\bar{\theta}^-)^2 \bar{\mathbb{J}}^{++}_{\alpha(s)\dot{\alpha}(s)}
	-4i \theta^{-\beta} \bar{\theta}^{-\dot{\beta}} \mathbb{J}^{++}_{\beta\alpha(s) \dot{\beta}\dot{\alpha}(s)}
	\\&
		+
	(\bar{\theta}^-)^2\theta^{-\beta} \mathbb{J}^{(+3)}_{\beta\alpha(s) \dot{\alpha}(s)}
	+
	(\theta^-)^2\bar{\theta}^{-\dot{\beta}} \bar{\mathbb{J}}^{(+3)}_{\alpha(s)\dot{\beta}\dot{\alpha}(s)} 
	+
	(\theta^-)^4 \mathbb{J}^{(+4)}_{\alpha(s)\dot{\alpha}(s)}.
	\end{split}
\end{equation}	
Here signs and coefficients are chosen for  future convenience.
All $\mathbb{J}$ superfields  in the expansion are unconstrained analytic superfields and satisfy reality conditions under tilde-conjugation \eqref{eq: analit space}:
\begin{equation}
	\begin{split}
	\widetilde{\mathbb{J}_{\alpha(s)\dot{\alpha}(s)} } = \mathbb{J}_{\alpha(s)\dot{\alpha}(s)}, 
	\quad
	&\widetilde{\mathbb{J}^{++}_{\beta\alpha(s) \dot{\beta}\dot{\alpha}(s)}} = \mathbb{J}^{++}_{\beta\alpha(s) \dot{\beta}\dot{\alpha}(s)},
	\quad
	\widetilde{\mathbb{J}^{(+4)}_{\alpha(s)\dot{\alpha}(s)}} = \mathbb{J}^{(+4)}_{\alpha(s)\dot{\alpha}(s)},
	\\
		\widetilde{\mathbb{J}^+_{\beta\alpha(s) \dot{\alpha}(s)} }
	=
	\bar{\mathbb{J}}^+_{\alpha(s)\dot{\beta}\dot{\alpha}(s)} ,
	\quad
	&\widetilde{\mathbb{J}^{++}_{\alpha(s)\dot{\alpha}(s)}} = \bar{\mathbb{J}}^{++}_{\alpha(s)\dot{\alpha}(s)},
	\quad
	\widetilde{\mathbb{J}^{(+3)}_{\beta\alpha(s) \dot{\alpha}(s)}} =
	\bar{\mathbb{J}}^{(+3)}_{\alpha(s)\dot{\beta}\dot{\alpha}(s)}. 
	\end{split}
\end{equation}	

As a next step,  we investigate what properties the analytic supercurrents $\mathbb{J}$
 satisfy, when the conservation equations are taken into account.

\textbf{1.} On-shell condition $\mathcal{D}^{+\alpha} \mathcal{J}_{\alpha(s)\dot{\alpha}(s)} \approx 0$ leads to the constraints on the analytic superfields $\mathbb{J}$:
\begin{equation}
	\begin{split}
	\mathbb{J}^+_{\beta\alpha(s) \dot{\alpha}(s)}  \approx \mathbb{J}^+_{(\beta\alpha(s)) \dot{\alpha}(s)},
	\quad
	\bar{\mathbb{J}}^+_{\alpha(s)\dot{\beta}\dot{\alpha}(s)}  \approx \bar{\mathbb{J}}^+_{\alpha(s)(\dot{\beta}\dot{\alpha}(s))},
	\quad
	 \mathbb{J}^{++}_{\alpha(s)\dot{\alpha}(s)} \approx \bar{\mathbb{J}}^{++}_{\alpha(s)\dot{\alpha}(s)} \approx 0,
	 \\
	  \mathbb{J}^{++}_{\beta\alpha(s) \dot{\beta}\dot{\alpha}(s)}
	  \approx  \mathbb{J}^{++}_{(\beta\alpha(s)) (\dot{\beta}\dot{\alpha}(s))}, 
	  \quad
	  \mathbb{J}^{(+3)}_{\beta\alpha(s) \dot{\alpha}(s)} \approx 0, 
	  \quad
	  	\bar{\mathbb{J}}^{(+3)}_{\alpha(s)\dot{\beta}\dot{\alpha}(s)} \approx 0,
	  	\quad 
	  	 \mathbb{J}^{(+4)}_{\alpha(s)\dot{\alpha}(s)} \approx 0,
	 \end{split}
\end{equation}	
These equations are \textit{algebraic} and indicate that a number of terms in the expansion \eqref{eq: PSC GS} vanish on the equations of motion.
Such terms correspond to vertices proportional to the free equations of motion and are therefore trivial up to field redefinitions. Following standard terminology, we refer to them as \textit{"fake" vertices}.

Therefore, modulo terms vanishing on the equations of motion, the general expansion \eqref{eq: PSC GS}  drastically simplifies:
\begin{equation}\label{eq: PSC GS -- sim}
	 \boxed{ \mathcal{J}_{\alpha(s)\dot{\alpha}(s)}
		\approx
		\mathbb{J}_{\alpha(s)\dot{\alpha}(s)} 
		+
		\theta^{-\beta} \mathbb{J}^+_{(\beta\alpha(s)) \dot{\alpha}(s)} 
		+
		\bar{\theta}^{-\dot{\beta}} \bar{\mathbb{J}}^+_{\alpha(s)(\dot{\beta}\dot{\alpha}(s))} 
		-
		4i \theta^{-\beta} \bar{\theta}^{-\dot{\beta}} \mathbb{J}^{++}_{(\beta\alpha(s)) (\dot{\beta}\dot{\alpha}(s))} .} 
\end{equation}	
Thus, the analytic projection isolates precisely those components of the principal supercurrent that generate nontrivial cubic interactions.

Moreover, we note that the first term of this expansion does not contribute to the interaction \eqref{eq: Psi inter}, since there one takes  $\mathcal{D}^+$ derivatives of the principal supercurrent. 
We further emphasize that the $\mathbb J$ supercurrents vanishing on-shell correspond to couplings to analytic $g$-superfields \eqref{eq: Mez exp}, which are pure gauge degrees of freedom.
As a result, we see that only  analytic supercurrents
 \begin{equation}\label{eq: bf J currents}
	\begin{split}
		&\mathbb{J}_{\alpha(s-1)\dot{\alpha}(s-1)}^{++}
		=
		+i
		(\mathcal{D}^+)^4 \left\{ \theta^-_{(\alpha} \bar{\theta}^-_{(\dot{\alpha}} \mathcal{J}_{\alpha(s-2)) \dot{\alpha}(s-2))}  \right\}
		=
		\; \frac{1}{4i} \mathcal{D}^+_{(\alpha} \bar{\mathcal{D}}^+_{(\dot{\alpha}} \mathcal{J}_{\alpha(s-2))\dot{\alpha}(s-2))}\Big|_{\theta^- = 0} ,
		\\
		&
		\mathbb{J}_{\alpha(s-1)\dot{\alpha}(s-2)}^{+}
		=
		-2	(\mathcal{D}^+)^4 \left\{ \theta^-_{(\alpha} (\bar{\theta}^-)^2 \mathcal{J}_{\alpha(s-2)) \dot{\alpha}(s-2)}  \right\}
		=
		\mathcal{D}^+_{(\alpha} \mathcal{J}_{\alpha(s-2))\dot{\alpha}(s-2)}\Big|_{\theta^- = 0},
		\\
		&
		\bar{\mathbb{J}}_{\alpha(s-2)\dot{\alpha}(s-1)}^+
		=
		+
		2
		(\mathcal{D}^+)^4 \left\{ \bar{\theta}^-_{(\dot{\alpha}} (\theta^-)^2 \mathcal{J}_{\alpha(s-2) \dot{\alpha}(s-2))}  \right\}
		=
		\bar{\mathcal{D}}^+_{(\dot{\alpha}} \mathcal{J}_{\alpha(s-2)\dot{\alpha}(s-2))}\Big|_{\theta^- = 0}
	\end{split} 
\end{equation}	
 completely determine the cubic vertices.

\smallskip

\textbf{2.} Harmonic-independence condition $\mathcal{D}^{++} 	\mathcal{J}_{\alpha(s)\dot{\alpha}(s)} \approx 0 $ implies additional  relations on $\mathbb{J}$ supercurrents: 
\begin{equation}\label{eq: Jbb conditions}
	\begin{cases}
		\mathcal{D}^{++} \mathbb{J}^{++}_{(\beta\alpha(s)) (\dot{\beta}\dot{\alpha}(s))} \approx 0,
		\\
		\mathcal{D}^{++} \mathbb{J}^+_{(\beta\alpha(s))\dot{\alpha}(s)} - 4i \bar{\theta}^{+\dot{\beta}} \mathbb{J}^{++}_{(\beta\alpha(s)) (\dot{\beta}\dot{\alpha}(s))} \approx 0,
		\\
		\mathcal{D}^{++} \bar{\mathbb{J}}^+_{\alpha(s)(\dot{\beta}\dot{\alpha}(s))} 
		+
		4i \theta^{+\beta} \mathbb{J}^{++}_{(\beta\alpha(s)) (\dot{\beta}\dot{\alpha}(s))} \approx 0,
		\\
		\mathcal{D}^{++} 	\mathbb{J}_{\alpha(s)\dot{\alpha}(s)} 
		+
		\theta^{+\beta} \mathbb{J}^+_{(\beta\alpha(s)) \dot{\alpha}(s)} 
		+
		\bar{\theta}^{+\dot{\beta}} \bar{\mathbb{J}}^+_{\alpha(s)(\dot{\beta}\dot{\alpha}(s))} \approx 0.
	\end{cases}	
\end{equation}	
It is important to emphasize that the original principal supercurrent \eqref{eq: general principle supercurrent} satisfies the harmonic-independence condition off-shell. However, after projecting onto analytic supercurrents and eliminating terms vanishing on-shell, the resulting harmonic relations become purely on-shell constraints.

As an illustration, let us consider
\begin{equation}
	\begin{split}
	\mathcal{D}^{++} 
		\mathbb{J}_{\alpha(s-1)\dot{\alpha}(s-2)}^{+}
	=
	&-4 \bar{\theta}^{+\dot{\beta}}	(\mathcal{D}^+)^4 \left\{ \theta^-_{(\alpha} \bar{\theta}^-_{\dot{\beta}} \mathcal{J}_{\alpha(s-2)) \dot{\alpha}(s-2)}  \right\}
	\\&	-2 \theta^+_{(\alpha}	(\mathcal{D}^+)^4 \left\{  (\bar{\theta}^-)^2 \mathcal{J}_{\alpha(s-2)) \dot{\alpha}(s-2)}  \right\}.
	\end{split}
\end{equation}	
Using identity
\begin{equation*}
	\bar{\theta}^{+\dot{\beta}} 
	\left\{ \theta^-_\alpha \bar{\theta}^-_{\dot{\beta}} J_{\alpha(s-2)\dot{\alpha}(s-2)} \right\}
	=
	\bar{\theta}^{+\dot{\beta}} 
	\left\{ \theta^-_\alpha \bar{\theta}^-_{(\dot{\beta}} J_{\alpha(s-2)\dot{\alpha}(s-2))} \right\}
	-
	\theta^{+}_{(\dot{\alpha}} \left\{ \theta^-_\alpha \bar{\theta}^{-\dot{\beta}} J_{\alpha(s-2)\dot{\beta}\dot{\alpha}(s-3))} \right\}
\end{equation*}	
and on-shell equations for principal supercurrent
\begin{equation*}
	\begin{split}
 &(\mathcal{D}^+)^4	\left\{ \theta^-_\alpha \bar{\theta}^{-\dot{\beta}} \mathcal{J}_{\alpha(s-2)(\dot{\beta}\dot{\alpha}(s-3))} \right\}
 \approx 0,
 \\
 &
 (\mathcal{D}^+)^4 \left\{  (\bar{\theta}^-)^2 \mathcal{J}_{\alpha(s-2)) \dot{\alpha}(s-2)}  \right\}
 \approx 0,
 \end{split}
\end{equation*}	
and taking into account \eqref{eq: bf J currents},
we find
\begin{equation}
		\mathcal{D}^{++} \mathbb{J}^+_{(\beta\alpha(s))\dot{\alpha}(s)} - 4i \bar{\theta}^{+\dot{\beta}} \mathbb{J}^{++}_{(\beta\alpha(s)) (\dot{\beta}\dot{\alpha}(s))} \approx 0.
\end{equation}	

\smallskip

\textbf{3.}  It is also useful to present explicit transformation laws for $\mathbb{J}$ supercurrents under rigid $\mathcal{N}=2$ supersymmetry (here we use transformations \eqref{eq: rig N=2 susy} and the invariance condition for principal supercurrent $\delta_\epsilon \mathcal{J}_{\alpha(s)\dot{\alpha}(s)} = 0$):
\begin{equation}\label{eq: susy mathbb J}
	\begin{split}
		&\delta_\epsilon  \mathbb{J}^{++}_{\alpha(s+1) \dot{\alpha}(s+1)}  = 0,
		\\
		&
		\delta_\epsilon \mathbb{J}^+_{\alpha(s+1)\dot{\alpha}(s)}
		=
		+ 4i \bar{\epsilon}^{-\dot{\beta}} \mathbb{J}^{++}_{\alpha(s+1)(\dot{\beta}\dot{\alpha}(s))},
		\\
		& 
		\delta_\epsilon \bar{\mathbb{J}}^+_{\alpha(s)\dot{\alpha}(s+1)} 
		=
		-4i \epsilon^{-\beta} \mathbb{J}^{++}_{(\alpha(s)\beta)\dot{\alpha}(s+1)},  
		\\
		&
		\delta_\epsilon \mathbb{J}_{\alpha(s)\dot{\alpha}(s)} = - \epsilon^{-\beta} \mathbb{J}^+_{(\beta\alpha(s))\dot{\alpha}(s)}
		-
		\bar{\epsilon}^{-\dot{\beta}} \bar{\mathbb{J}}^+_{\alpha(s)(\dot{\beta}\dot{\alpha}(s))}. 
	\end{split}	
\end{equation}	

In closing this section, we find that, upon discarding the terms that vanish on the equations of motion, $\mathcal{N}=2$ principal supercurrent reduces to the set of analytic $\mathbb{J}$ supercurrents satisfying special harmonic on-shell relations  \eqref{eq: Jbb conditions}. 

\subsection{$\mathcal{N}=2$ analytic supercurrents}

We have constructed cubic vertices in non-analytic form \eqref{eq: Psi inter} using the Mezincescu-type prepotentials.
On the other hand, using analytic higher-spin prepotentials $h^{++\alpha(s-2)\dot{\alpha}(s-2)M}$ one can construct an $\mathcal{N}=2$ higher-spin cubic superfield coupling in the manifest \textit{analytic} form:
 \begin{equation}\label{eq: int analytic}
 	\begin{split}
 	S^{A}_{int} = \int d\zeta^{(-4)}\, \Big(&
 	h^{++\alpha(s-1)\dot{\alpha}(s-1)} J^{++}_{\alpha(s-1)\dot{\alpha}(s-1)}
 	+
 	h^{++\alpha(s-1)\dot{\alpha}(s-2)+} J^{+}_{\alpha(s-1)\dot{\alpha}(s-2)}
 	\\&+
 	h^{++\alpha(s-2)\dot{\alpha}(s-1)+} \bar{J}^{+}_{\alpha(s-2)\dot{\alpha}(s-1)}
 	+ h^{++\alpha(s-2)\dot{\alpha}(s-2)} J^{++}_{\alpha(s-2)\dot{\alpha}(s-2)} \Big).
 	\end{split}
 \end{equation}	
 Since analytic prepotentials describe d.o.f. of
 $\mathcal{N}=2$ higher-spin off-shell supermultiplets, the analytic form of the cubic interaction is the most natural one.
 
 \medskip
 
For consistency of this coupling,  $J^{++}$ supercurrents  must satisfy a number of conditions:

\medskip
 
 \noindent	\textbf{I.} \textbf{Analyticity} 
 	
 	\medskip
 Due to analyticity of prepotentials $h^{++\dots}$, supercurrents $J^{++}_{\dots}$ must also satisfy analyticity condition. It is essential that the analyticity conditions must be satisfied off-shell.

 \medskip
 
  \noindent		\textbf{II.} \textbf{Harmonic-independence}
 
 \medskip 
 
Linearized higher-spin gauge invariance \eqref{eq: GF} leads to on-shell conservation \textit{"harmonic" equations}, which imply the consistency of  the cubic coupling \eqref{eq: int analytic} in the leading order: 	
 \begin{equation}\label{eq: analytic cons}
 	\begin{cases}
 		\mathcal{D}^{++} J^{++}_{\alpha(s-1)\dot{\alpha}(s-1)} \approx 0,
 		\\
 		\mathcal{D}^{++} J^{++}_{\alpha(s-2)\dot{\alpha}(s-2)} \approx 0,
 		\\
 		\mathcal{D}^{++} J^{+}_{\alpha(s-1)\dot{\alpha}(s-2)}
 		-
 		4i \bar{\theta}^{+\dot{\beta}} J^{++}_{\alpha(s-1)(\dot{\alpha}(s-2)\dot{\beta})}
 		+
 		2i
 		\theta^+_{(\alpha} J^{++}_{\alpha(s-2))\dot{\alpha}(s-2)} \approx 0,
 			\\
 		\mathcal{D}^{++} \bar{J}^{+}_{\alpha(s-2)\dot{\alpha}(s-1)} 
 		+4i \theta^{+\beta} J^{++}_{(\alpha(s-2)\beta)\dot{\alpha}(s-1)}
 		+
 		2i \bar{\theta}^+_{(\dot{\alpha}} J^{++}_{\alpha(s-2)\dot{\alpha}(s-2))} \approx 0.
 	\end{cases}
 \end{equation}
  The general component solution of these equations was analyzed in  appendix C of \cite{Buchbinder:2022vra}.

  \medskip
  
  	\noindent \textbf{III.} \textbf{Supersymmetry transformations}
  	
  	\medskip
  
  Requirement of invariance of the analytic vertex \eqref{eq: int analytic} under rigid $\mathcal{N}=2$  supersymmetry due to non-trivial transformation law of analytic $h$-prepotentials \eqref{eq: susy prepotentials} leads to the transformation laws of analytic supercurrents:
  \begin{equation}\label{eq: analyt susy transforms}
  	\begin{split}
  		&\delta_\epsilon J^{++}_{\alpha(s-1)\dot{\alpha}(s-1)} = 0,
  		\\
  		& \delta_\epsilon J^{++}_{\alpha(s-2)\dot{\alpha}(s-2)} = 0,
  		\\
  		&\delta_\epsilon J^+_{\alpha(s-1)\dot{\alpha}(s-2)} = 4i \bar{\epsilon}^{-\dot{\beta}} J^{++}_{\alpha(s-1)(\dot{\alpha}(s-2)\dot{\beta})} 
  		-2i \epsilon^-_{(\alpha} J^{++}_{\alpha(s-2))\dot{\alpha}(s-2)},
  		\\
  		&
  		\delta_\epsilon \bar{J}^+_{\alpha(s-2)\dot{\alpha}(s-1)} = -4i \epsilon^{-\beta} J^{++}_{(\alpha(s-2)\beta)\dot{\alpha}(s-1)} -2i \bar{\epsilon}^{-}_{(\dot{\alpha}} J^{++}_{\alpha(s-2)\dot{\alpha}(s-2))}.
  	\end{split}
  \end{equation}

 Comparison of equations \eqref{eq: analytic cons} with  equations  \eqref{eq: PSC GS -- sim} leads to the observation, that $\mathbb{J}$ supercurrents  provide a particular solution of \eqref{eq: analytic cons}\footnote{We reserve the notation $\mathbb J$ for analytic supercurrents obtained by projection from the principal supercurrent, while $J$ denotes a general solution of the analytic conservation equations.}:
 \begin{equation}\label{eq: part solution analit}
 	\begin{split}
 J^{++}_{\alpha(s-1)\dot{\alpha}(s-1)} = \mathbb{J}^{++}_{\alpha(s-1)\dot{\alpha}(s-1)},
 \qquad
 &J^{++}_{\alpha(s-2)\dot{\alpha}(s-2)} = 0,
 \\
 J^{+}_{\alpha(s-1)\dot{\alpha}(s-2)} = \mathbb{J}^{+}_{\alpha(s-1)\dot{\alpha}(s-2)},
 \qquad
& \bar{J}^{+}_{\alpha(s-2)\dot{\alpha}(s-1)} 
 =
 \bar{\mathbb{J}}^{+}_{\alpha(s-2)\dot{\alpha}(s-1)}. 
 \end{split}
 \end{equation}	
 Indeed, by  construction, $\mathbb{J}$ supercurrents are analytic superfields. Rigid supersymmetry transformations \eqref{eq: analyt susy transforms} are also in full agreement with transformation laws \eqref{eq: susy mathbb J} for analytic $\mathbb{J}$ supercurrents.
 
 Note that the supercurrent $J^{++}_{\alpha(s-2)\dot{\alpha}(s-2)}$ is equal to zero.  Previously, a similar phenomenon was observed  in the construction of higher-spin supercurrents for the massless hypermultiplet in \cite{Buchbinder:2022vra}\footnote{See equation (5.49) in \cite{Buchbinder:2022vra}, for massless hypermultiplet $\partial_5 q^+_a = 0$.}. It will be nontrivial for massive $\mathcal{N}=2$ higher-spin multiplets. This can be understood by noting that, in the general case, the abelian superfield vertex $(\mathbf{s_1}, \mathbf{s_2}, \mathbf{s_2})$ may contain abelian component vertices of the types $(s_1, s_2, s_2)$ and $(s_1-1, s_2, s_2)$. These component vertices contain $s_1$ and $s_1-1$ spacetime derivatives, respectively. Then, from dimensional considerations, an additional parameter of mass dimension is required for both component vertices appear in the supersymmetric vertex. The component vertex $(s_1, s_2, s_2)$ is determined by the analytic supercurrent $J^{++}_{\alpha(s_1-1)\dot{\alpha}(s_1-1)}$ and, by our construction, is always present in the superfield vertex.  The vertex $(s_1-1, s_2, s_2)$ is determines by the analytic supercurrent $J^{++}_{\alpha(s_1-2)\dot{\alpha}(s_1-2)}$. Therefore, we conclude that this supercurrent will be non-trivial only for massive higher-spin multiplets.

We emphasise that, the representations under consideration for the cubic vertex are equivalent.
 It is clear that by taking the integral over nonanalytic coordinates $\int d\theta^{-4} = (\mathcal{D}^+)^4$ and using relations \eqref{eq: G and Psi prepot}  one can reduce non-analytic cubic couplings \eqref{eq: Psi inter} to analytic form \eqref{eq: int analytic}. In the procedure of such reduction the terms vanishing on the equations of motion, must be excluded, because these terms correspond to the  "fake"\, cubic vertices\footnote{Such terms correspond to cubic vertices proportional to the free equations of motion and can therefore be removed by field redefinitions. See an example in the appendix \ref{app: conserved gauge-invariant higher-spin currents}.}. This explains the relationship between the two representations for the cubic vertices.
 
 Thus, upon passing from the principal supercurrent formulation to the analytic-supercurrent formulation, the roles of analyticity and harmonic independence are effectively interchanged:
\begin{equation*}
	\begin{split}
		&\;\;\boxed{\text{Off-shell  analyticity}} \quad \longleftrightarrow \quad \boxed{\text{Off-shell  harmonic independence}}
		\\
		&	\qquad \text{Analytic action}\quad  \quad\quad \quad \quad \qquad\;\;\; \text{Full superspace action}
	\end{split}
\end{equation*}	

Given this, the main conclusion obtained from our analysis is the following:

\begin{quote}
	\textit{
\noindent The analytic formulation isolates the physically nontrivial part of the $\mathcal{N}=2$ cubic interaction in a manifestly supersymmetric form.	}
\end{quote}
	
\subsection*{Supersymmetry algebra}

When deriving the analytic supercurrents, we discarded terms that vanish on the equations of motion. A natural question arises: if in the original representation of cubic vertex \eqref{eq: Psi inter} the supersymmetry algebra closes off-shell, what happens to it upon passing to the analytic vertex?

To answer this question, we first note that discarding fake vertices corresponds to the local free (super)field redefinitions. However, this not only removes the fake vertices but also generates additional quartic vertices which, nevertheless, occur at the next order in the coupling constant.  This means that in our cubic approximation, they can be neglected.
Moreover, we know that the analytic vertex is supersymmetrically invariant by our construction.

The supersymmetry algebra remains closed off-shell under any local superfield redefinition. Closure is a property of the field transformations, not of the Lagrangian. Therefore, even if one truncates the action to cubic order (discarding quartic and higher terms), the supersymmetry transformations themselves are still closed off-shell. 

\subsection{Example: $(\mathbf{2}, \mathbf{1}, \mathbf{1})$ vertex}

The simplest example of a principal $\mathcal N=2$ supercurrent is the well-known $\mathcal N=2$ vector-multiplet supercurrent \cite{Howe:1981qj, Kuzenko:1999pi, Butter:2010sc}:
\begin{equation}\label{eq: spin 1 supercurrent}
	\mathcal{J} = \mathcal{W} \bar{\mathcal{W}}.
\end{equation}	
In accordance with the general consideration of the section \ref{sec: principle SC}, supercurrent $\mathcal{J}$ satisfies:
\begin{equation}
	\mathcal{D}^{++} \mathcal{J} = 0,
	\qquad
	\left( \mathcal{D}^+ \right)^2 \mathcal{J} \approx 0,
	\qquad
	\left( \bar{\mathcal{D}}^+ \right)^2 \mathcal{J} \approx 0.
\end{equation}	
This supercurrent 
generates coupling of $\mathcal{N}=2$ vector multiplet to  $\mathcal{N}=2$ supergravity.

Corresponding analytic supercurrents are given by:
\begin{subequations}
\begin{equation}
	\begin{split}
	\mathbb{J}^{++}_{\alpha\dot{\alpha}} 
	&=
	+ i
	(\mathcal{D}^+)^4
	\left\{ \theta^-_\alpha \bar{\theta}^-_{\dot{\alpha}} \mathcal{J} \right\}
= 	\frac{1}{4i }
\mathcal{D}^+_\alpha \bar{\mathcal{D}}^+_{\dot{\alpha}} \mathcal{J} \Big|_{\theta^- = 0}, 
	\end{split}
\end{equation}
\begin{equation}
	\begin{split}
	\mathbb{J}^+_\alpha 
	&= 
	- 2 (\mathcal{D}^+)^4 \left\{ \theta^-_\alpha (\bar{\theta}^-)^2 \mathcal{J}  \right\}
		=
+\mathcal{D}^+_\alpha \mathcal{J}   \Big|_{\theta^- = 0},
	\end{split}
\end{equation}	
\begin{equation}
	\mathbb{J} = (\mathcal{D}^+)^4 \left\{ (\theta^-)^4 \mathcal{J} \right\} 
	=
	\mathcal{J} \Big|_{\theta^- = 0} .
\end{equation}	
\end{subequations}
Let us  present the  component content of the analytic supercurrents from \eqref{eq: PSC GS -- sim}:
\begin{subequations}
 \begin{equation}
 	\mathbb{J} \sim 
 	\left(\phi + \theta^{+\rho} \psi^j_\rho u^-_j\right)
 	\times
 	\left(\bar{\phi} + \bar{\theta}^{+\dot{\rho}}\bar{\psi}^j_{\dot{\rho}} u^-_j \right),
 \end{equation}	
 \begin{equation}
 	\begin{split}
 	\mathbb{J}^+_\alpha
 	\sim & \; 
 	\left(\bar{\phi} + \bar{\theta}^{+\dot{\rho}}\bar{\psi}^j_{\dot{\rho}} u^-_j \right)
 	\times
 	\left( - \psi^i_\alpha u^+_i + \theta^{+\rho} C_{(\alpha\rho)} 
 -
 4i \bar{\theta}^{+\dot{\rho}} \partial_{\alpha\dot{\rho}} \phi  
 -
  4i \theta^{+\rho} \bar{\theta}^{+\dot{\rho}} \partial_{(\rho \dot{\rho}} \psi^i_{\alpha)} u^-_i \right), 
 	\end{split}
 \end{equation}	
 \begin{equation}
 	\begin{split}
 	\mathbb{J}_{\alpha\dot{\alpha}}^{++}
 	\sim &
 	\left( - \psi^i_\alpha u^+_i + \theta^{+\rho} C_{(\alpha\rho)} 
 	-
 	4i \bar{\theta}^{+\dot{\rho}} \partial_{\alpha\dot{\rho}} \phi  - 4i \theta^{+\rho} \bar{\theta}^{+\dot{\rho}} \partial_{(\rho \dot{\rho}} \psi^i_{\alpha)} u^-_i \right)
 	\\
 	& \times
 		\left( - \bar{\psi}^j_{\dot{\alpha}} u^+_j +
 		\bar{ \theta}^{+\dot{\rho}} \bar{C}_{(\dot{\alpha}\dot{\rho})} 
 	-
 	4i \theta^{+\rho} \partial_{\rho\dot{\alpha}} \bar{\phi}  - 4i \theta^{+\rho} \bar{\theta}^{+\dot{\rho}} \partial_{\rho (\dot{\rho}} \bar{\psi}^j_{\dot{\alpha})} u^-_j \right).
 	\end{split}
 \end{equation}	
 \end{subequations}
 One finds that among the components of these supercurrents there are conserved currents. For example, the current corresponding to invariance under rigid 
$ \mathcal{N}=2$ supersymmetry ($\theta^{+\beta}$ component) is given by:
 \begin{equation}\label{eq: susy current}
 	S^i_{\alpha\beta\dot{\alpha}}
 	\sim
 	C_{(\alpha\beta)} \bar{\psi}^i_{\dot{\alpha}}
 	+
 	\partial_{\alpha\dot{\alpha}} \bar{\phi} \psi^i_\beta.
 \end{equation}

The cubic vertex can be written  in terms of Mezincescu-type prepotentials:
\begin{equation}
	S_{int} = \int d^4x d^8\theta du \; \left( \Psi^{-\alpha} \mathcal{D}^+_\alpha + \bar{\Psi}^{-\dot{\alpha}} \bar{\mathcal{D}}^+_{\dot{\alpha}} \right) \mathcal{W} \bar{\mathcal{W}}
\end{equation}	
or, equivalently, in terms of analytic $h$-prepotentials:
\begin{equation}
	\begin{split}
	S^A_{int} = \int d\zeta^{(-4)}  
	\Big[ h^{++\alpha\dot{\alpha}} \cdot  i
	(\mathcal{D}^+)^4
	\left\{ \theta^-_\alpha \bar{\theta}^-_{\dot{\alpha}} \mathcal{W} \bar{\mathcal{W}} \right\}
	&- 2
	h^{++\alpha+} 	\cdot  (\mathcal{D}^+)^4 \left\{ \theta^-_\alpha (\bar{\theta}^-)^2 \mathcal{W} \bar{\mathcal{W}}  \right\}
\\	&+ 2
	 h^{++\dot{\alpha}+}  
	 \cdot
	 (\mathcal{D}^+)^4 \left\{ \bar{\theta}^-_{\dot{\alpha}} (\theta^-)^2 \mathcal{W} \bar{\mathcal{W}}  \right\}  \Big].
	 \end{split}
\end{equation}	
The comparison of the analytic and non-analytic vertices shows that the analytic vertex explicitly contains Grassmann coordinates. 
The reason for this difference is that the basic object entering the construction of hypermultiplet $\mathcal N=2$ supercurrents is the analytic superfield $q_a^+$, whereas the higher-spin supercurrents considered here are built from chiral higher-spin Weyl supertensors.
Ultimately, this is a consequence of the fact that the vector multiplet action \eqref{eq: spin 1 action}  is written as an integral over the full harmonic superspace. 
This feature is common to all abelian vertices considered in the present paper.

In conclusion, we note that the shape of the presented cubic vertex $(\mathbf{2}, \mathbf{1}, \mathbf{1})$  is  different in structure from the action of the  $\mathcal{N}=2$ vector multiplet on the background of $\mathcal{N}=2$ supergravity in harmonic approach, see \cite{Galperin:1987ek, Galperin:1987em, 18}. We will give some additional comments on this issue in the concluding  section \ref{sec: dis}.

\subsection{Example: $(\mathbf{3}, \mathbf{1}, \mathbf{1})$ vertex}

In this subsection, we will analyze in detail  the spin-$\mathbf{3}$ supercurrents, constructed from the spin-$\mathbf{1}$ field. 
This example is less well known than the  vertex $(\mathbf{2}, \mathbf{1}, \mathbf{1})$ example and illustrates the main features of the higher-spin $\mathcal{N}=2$ supercurrents. From general result \eqref{eq: general principle supercurrent}, we deduce, that 
\textit{principal $\mathcal{N}=2$ supercurrent} in this case  takes the form:
\begin{equation}
	\mathcal{J}_{\alpha\dot{\alpha}}
	=
	i \left( \mathcal{W} \partial_{\alpha\dot{\alpha}}  \bar{\mathcal{W}} -  \partial_{\alpha\dot{\alpha}}   \mathcal{W} \bar{\mathcal{W} }  \right)
	-
	\frac{1}{4} \left( \mathcal{D}^-_\alpha \mathcal{W} \bar{\mathcal{D}}^+_{\dot{\alpha}}  \bar{\mathcal{W} }  
	-
	\mathcal{D}^+_\alpha \mathcal{W} \bar{\mathcal{D}}^-_{\dot{\alpha}}  \bar{\mathcal{W} }   \right)
\end{equation}
and satisfies the following equations:
\begin{equation}
	\mathcal{D}^{++} \mathcal{J}_{\alpha\dot{\alpha}} = 0,
	\qquad
	\mathcal{D}^{+\alpha} \mathcal{J}_{\alpha\dot{\alpha}} \approx 0,
	\qquad
	\bar{\mathcal{D}}^{+\dot{\alpha}}\mathcal{J}_{\alpha\dot{\alpha}} \approx 0.
\end{equation}	

Corresponding analytic supercurrents are given by:
\begin{subequations}
	\begin{equation}
		\begin{split}
			\mathbb{J}^{++}_{(\alpha\beta)(\dot{\alpha}\dot{\beta})} 
			&=
			+ i
			(\mathcal{D}^+)^4
			\left\{ \theta^-_{(\alpha} \bar{\theta}^-_{(\dot{\alpha}} \mathcal{J}_{\beta)\dot{\beta})} \right\}
			= 	\frac{1}{4i }
			\mathcal{D}^+_{(\alpha} \bar{\mathcal{D}}^+_{(\dot{\alpha}} \mathcal{J}_{\beta)\dot{\beta})} \Big|_{\theta^- = 0}, 
		\end{split}
	\end{equation}
	\begin{equation}
		\begin{split}
			\mathbb{J}^+_{(\alpha\beta)\dot{\alpha}} 
			&= 
			- 2 (\mathcal{D}^+)^4 \left\{ \theta^-_{(\alpha} (\bar{\theta}^-)^2 \mathcal{J}_{\beta)\dot{\alpha}}  \right\}
			=
			+\mathcal{D}^+_{(\alpha} \mathcal{J}_{\beta)\dot{\alpha}}   \Big|_{\theta^- = 0},
		\end{split}
	\end{equation}	
	\begin{equation}
		\mathbb{J}_{\alpha\dot{\alpha}} = (\mathcal{D}^+)^4 \left\{ (\theta^-)^4 \mathcal{J}_{\alpha\dot{\alpha}} \right\} 
		=
	\mathcal{J}_{\alpha\dot{\alpha}} \Big|_{\theta^- = 0} .
	\end{equation}	
\end{subequations}

Component content of analytic supercurrents is given by\footnote{We define $\overset{\longleftrightarrow}{\partial_{\alpha\dot{\alpha}}}$ as $A \overset{\longleftrightarrow}{\partial_{\alpha\dot{\alpha}}} B :=  A \partial_{\alpha\dot{\alpha}}B - \partial_{\alpha\dot{\alpha}} A  B$. }:
\begin{subequations}
\begin{equation}
	\begin{split}
	\mathbb{J}_{\alpha\dot{\alpha}}
	\sim
&\;i \left(\phi + \theta^{+\beta} \psi^j_\beta u^-_j\right)
\times \overset{ \longleftrightarrow}{\partial_{\alpha\dot{\alpha}}}
\left(\bar{\phi} + \bar{\theta}^{+\dot{\beta}}\bar{\psi}^j_{\dot{\beta}} u^-_j \right)
\\&
+
\frac{1}{4} \psi^i_\alpha u^-_i
\times  
\left( - \bar{\psi}^j_{\dot{\alpha}} u^+_j +
\bar{ \theta}^{+\dot{\rho}} \bar{C}_{(\dot{\alpha}\dot{\rho})} 
-
4i \theta^{+\rho} \partial_{\rho\dot{\alpha}} \bar{\phi}  - 4i \theta^{+\rho} \bar{\theta}^{+\dot{\rho}} \partial_{\rho (\dot{\rho}} \bar{\psi}^j_{\dot{\alpha})} u^-_j \right)
\\&
+
\frac{1}{4}
\left( - \psi^i_\alpha u^+_i + \theta^{+\rho} C_{(\alpha\rho)} 
-
4i \bar{\theta}^{+\dot{\rho}} \partial_{\alpha\dot{\rho}} \phi  - 4i \theta^{+\rho} \bar{\theta}^{+\dot{\rho}} \partial_{(\rho \dot{\rho}} \psi^i_{\alpha)} u^-_i \right)
\times 
 \bar{\psi}^j_{\dot{\alpha}} u^-_j, 
\end{split}
\end{equation}	
\begin{equation}
	\begin{split}
	\mathbb{J}^+_{(\alpha\beta)\dot{\alpha}}
	\sim&
\;	i
	\left( - \psi^i_\alpha u^+_i + \theta^{+\rho} C_{(\alpha\rho)} 
	-
	4i \bar{\theta}^{+\dot{\rho}} \partial_{\alpha\dot{\rho}} \phi  - 4i \theta^{+\rho} \bar{\theta}^{+\dot{\rho}} \partial_{(\rho \dot{\rho}} \psi^i_{\alpha)} u^-_i \right)
 \overset{ \longleftrightarrow}{\partial_{\beta\dot{\alpha}}}
 \left(\bar{\phi} + \bar{\theta}^{+\dot{\rho}}\bar{\psi}^j_{\dot{\rho}} u^-_j \right)
 \\
 & + \frac{1}{4}
 \left( C_{(\alpha\beta)} + 4i \bar{\theta}^{+\dot{\rho}} \partial_{(\alpha\dot{\rho}} \psi^i_{\beta)} u^-_i \right)
\\& \qquad\qquad \times
 	\left( - \bar{\psi}^j_{\dot{\alpha}} u^+_j +
 \bar{ \theta}^{+\dot{\rho}} \bar{C}_{(\dot{\alpha}\dot{\rho})} 
 -
 4i \theta^{+\rho} \partial_{\rho\dot{\alpha}} \bar{\phi}  - 4i \theta^{+\rho} \bar{\theta}^{+\dot{\rho}} \partial_{\rho (\dot{\rho}} \bar{\psi}^j_{\dot{\alpha})} u^-_j \right)
 \\
 &
 -
 i \left( - \psi^i_\alpha u^+_i + \theta^{+\rho} C_{(\alpha\rho)} 
 -
 4i \bar{\theta}^{+\dot{\rho}} \partial_{\alpha\dot{\rho}} \phi  - 4i \theta^{+\rho} \bar{\theta}^{+\dot{\rho}} \partial_{(\rho \dot{\rho}} \psi^i_{\alpha)} u^-_i \right)
\\
&\qquad\qquad \qquad \times \partial_{\beta\dot{\alpha}} 
 \left(\bar{\phi} + \bar{\theta}^{+\dot{\beta}}\bar{\psi}^j_{\dot{\beta}} u^-_j \right),
 \\
 \end{split}
\end{equation}	
\begin{equation}
	\begin{split}
	\mathbb{J}^{++}_{(\alpha\beta)(\dot{\alpha}\dot{\beta})}
	\sim &
	\; i \left( - \psi^i_\alpha u^+_i + \theta^{+\rho} C_{(\alpha\rho)} 
	-
	4i \bar{\theta}^{+\dot{\rho}} \partial_{\alpha\dot{\rho}} \phi  - 4i \theta^{+\rho} \bar{\theta}^{+\dot{\rho}} \partial_{(\rho \dot{\rho}} \psi^i_{\alpha)} u^-_i \right)
	\\
	& \times  
	\overset{ \longleftrightarrow}{\partial_{\beta\dot{\beta}}}
	\left( - \bar{\psi}^j_{\dot{\alpha}} u^+_j +
	\bar{ \theta}^{+\dot{\rho}} \bar{C}_{(\dot{\alpha}\dot{\rho})} 
	-
	4i \theta^{+\rho} \partial_{\rho\dot{\alpha}} \bar{\phi}  - 4i \theta^{+\rho} \bar{\theta}^{+\dot{\rho}} \partial_{\rho (\dot{\rho}} \bar{\psi}^j_{\dot{\alpha})} u^-_j \right).
	\end{split}
\end{equation}	
\end{subequations}

In components of these analytic supercurrents we obtain (up to some coefficients) conserved currents:

\smallskip

\noindent \underline{\textit{Spin-3 current}}

\begin{equation}
	J_{(\alpha\beta)\rho(\dot{\alpha}\dot{\beta})\dot{\rho}}
	\sim
	i \left( C_{(\alpha\dot{\rho})}  \overset{ \longleftrightarrow}{\partial_{\beta\dot{\beta}}} \bar{C}_{(\dot{\alpha}\dot{\rho})} \right)
	+
	\partial_{(\rho \dot{\rho}} \psi^i_{\alpha)}  
	\overset{ \longleftrightarrow}{\partial_{\beta\dot{\beta}}}
	\bar{\psi}_{\dot{\alpha} i}
	-
	\psi^i_\alpha
		\overset{ \longleftrightarrow}{\partial_{\beta\dot{\beta}}}
	 \partial_{\rho (\dot{\rho}} \bar{\psi}_{\dot{\alpha}) i}
	 +
	 i \partial_{\alpha\dot{\rho}} \phi 
	 		\overset{ \longleftrightarrow}{\partial_{\beta\dot{\beta}}} \partial_{\rho\dot{\alpha}} \bar{\phi}.
\end{equation}	
This current decomposes into the standard spin-3 currents associated with the vector, fermionic, and scalar components of the $\mathcal N=2$ vector multiplet. The conserved current for the vector field is also known as \textit{zilch}; see the discussion in  appendix \ref{sec: ZILCH}.

\smallskip

\noindent \underline{\textit{Spin-5/2 current}}

\begin{equation}
	J^i_{(\alpha\beta)\rho(\dot{\alpha}\dot{\beta})}
	\sim
	i C_{(\alpha\rho)}
		\overset{ \longleftrightarrow}{\partial_{\beta\dot{\beta}}}
		 \bar{\psi}^i_{\dot{\alpha}}
	+
	\psi^i_\alpha
		\overset{ \longleftrightarrow}{\partial_{\beta\dot{\beta}}}
	 \partial_{\rho\dot{\alpha}} \bar{\phi}.
\end{equation}	
This is a generalization of the current \eqref{eq: susy current} corresponding to invariance under rigid $\mathcal{N}=2$ supersymmetry.

Cubic vertex can be written directly based on the obtained analytic supercurrents, completely analogous to the $(\mathbf{2}, \mathbf{1}, \mathbf{1})$ case, discussed above, and shares the same structural features.

\newpage

\section{Higher-spin gauge transformations for the $\mathcal{N}=2$ vector multiplet}
\label{sec: 5}

In the previous sections, we discussed the construction of $(\mathbf{s_1}, \mathbf{s_2}, \mathbf{s_2})$ cubic vertices for $\mathbf{s_1}\geq 2 \mathbf{s_2}$ and analyzed their superfield structure in different superfield representations.
In this section. we analyze the corresponding spin-$\mathbf{s_1}$ gauge transformations of the $\mathcal{N}=2$ vector multiplet ($\mathbf{s_2}=1$).  This provides a prototypical example of higher-spin gauge transformations and illustrates the main features of abelian $\mathcal{N}=2$ higher-spin gauge symmetries.

To derive the higher-spin transformations of the $\mathcal{N}=2$ vector multiplet, we employ the inverse Noether procedure (see, e.g., \textsection 7 of \cite{Konopleva:1981ew}) adapted to $\mathcal{N}=2$ harmonic superspace.
For this purpose, we will use that the general variation of the $\mathcal{N}=2$ vector multiplet action \eqref{eq: spin 1 action}  has  a simple form:
\begin{equation}\label{eq: gen spin 1 variation}
	\begin{split}
	\delta \int d^4x d^8\theta du \, V^{++} V^{--} &= 2 \int d^4x d^8\theta du \; \delta V^{++} V^{--}
	\\&=
	2\int d\zeta^{(-4)} \delta V^{++} (\mathcal{D}^+)^4 V^{--}  . 
	\end{split}
\end{equation}	
The goal is to find a variation $\delta V^{++}$ that would ensure gauge invariance at the leading order:
\begin{equation}\label{eq: nan var}
	\int d^4x d^8\theta du \; \left[2 \delta V^{++} V^{--}
	+
	\kappa_s
	\left( \delta \Psi^{-\alpha(s-1)\dot{\alpha}(s-2)}\mathcal{D}^+_\alpha + \delta \bar{\Psi}^{-\alpha(s-2)\dot{\alpha}(s-1)} \bar{\mathcal{D}}^+_{\dot{\alpha}} \right) \mathcal{J}_{\alpha(s-2)\dot{\alpha}(s-2)} \right]
	=
	0. 
\end{equation}	
Here  $\kappa_s$ is the spin-$\mathbf{s}$ coupling constant.

Similarly, higher-spin transformations can be analyzed in the analytic form:
\begin{equation}\label{eq: an var}
	\int d\zeta^{(-4)} \left[ 2\delta V^{++} (\mathcal{D}^+)^4V^{--} + \kappa_s \delta h^{++\alpha(s-2)\dot{\alpha}(s-2)M} J^{++}_{\alpha(s-2)\dot{\alpha}(s-2) M} \right] = 0.
\end{equation}	
We will analyze  higher-spin  transformations for $\mathcal{N}=2$ vector supermultiplet in both representations.

\subsection{Spin-$\mathbf{2}$ transformations}

We begin with the simplest example and demonstrate how this technique can be used to derive the coupling-induced spin-$\mathbf{2}$ gauge transformations acting on the vector multiplet.  Since this case corresponds to the interaction of the $\mathcal{N}=2$ vector multiplet with supergravity, we expect to obtain transformations that correspond to supertranslations in harmonic superspace.

\subsubsection{Non-analytic form}

Non-analytic representation of $(\mathbf{2}, \mathbf{1}, \mathbf{1})$ vertex is given by
\begin{equation}\label{eq: (2 1 1) vertex}
	\boxed{ S_{int} =  \int d^4x d^8\theta du\; \left( \Psi^{-\alpha} \mathcal{D}^+_\alpha + \bar{\Psi}^{-\dot{\alpha}} \bar{\mathcal{D}}^+_{\dot{\alpha}} \right) \mathcal{W} \bar{\mathcal{W}}. }
\end{equation}
According to eq. \eqref{eq: Psi gauge and pre gauge}, there are four types of gauge and pre-gauge transformations for spin-$\mathbf{2}$ Mezincescu-type prepotentials:

\medskip

\noindent $\bullet$ \underline{\textbf{Parameter  $K^{(-3)}_{\alpha}$}}

\smallskip
\noindent The variation generated by
 $\delta \Psi^-_{\alpha} = \mathcal{D}^{++} K^{(-3)}_{\alpha}$  vanishes identically due to the off-shell harmonic independence condition:
  \begin{equation}\label{eq: HA cond}
	\mathcal{D}^{++} \mathcal{J} = \mathcal{D}^{++}  \left(\mathcal{W} \bar{\mathcal{W}} \right)= 0,
\end{equation}
accordingly, there will be no superfield transformation of $V^{++}$ corresponding to it.

\medskip

\noindent $\bullet$  \underline{\textbf{Parameter $B^{--}$}}

\smallskip
\noindent Prepotential variation 
$
\delta \Psi^-_{\alpha} = \mathcal{D}^+_{\alpha} B^{--}
$ and
$\delta \bar{\Psi}^{-}_{\dot{\alpha}}
=
-
\bar{\mathcal{D}}^+_{\dot{\alpha}} B^{--}$ induces the variation of the action:
\begin{equation}
	\begin{split}
		\delta S_{int} = &\;-  \int d^4x d^8\theta du\; B^{--} \left[ (\mathcal{D}^+)^2 + (\bar{\mathcal{D}}^+)^2\right]  \mathcal{J}
		\\
		= &+   \int d^4x d^8\theta du\; B^{--} \left[ (\mathcal{D}^+)^2 \Big(\bar{\mathcal{W}}(\bar{\mathcal{D}}^+)^2 V^{--} \Big) + (\bar{\mathcal{D}}^+)^2 \Big(\mathcal{W}(\mathcal{D}^+)^2 V^{--} \Big)\right] 
		\\
		=  
		&  -  \int d^4x d^8\theta du\; \Big[ (\mathcal{D}^+)^2 \left( B^{--}  \bar{\mathcal{W}} \right) \Big((\bar{\mathcal{D}}^+)^2 V^{--} \Big) + (\bar{\mathcal{D}}^+)^2 \left( B^{--}  \mathcal{W} \right) \Big( (\mathcal{D}^+)^2 V^{--} \Big) \Big]
		\\
		=
		& \; 16   \int d^4x d^8 \theta du \;  (\mathcal{D}^+)^4 \left[ B^{--} \left( \mathcal{W} + \bar{\mathcal{W}} \right)  \right] V^{--}. 
	\end{split}
\end{equation}
Thus, one obtains the corresponding variation of analytic superfield:
\begin{equation}\label{eq: B-- transformations}
	\boxed{\delta V^{++} =  -8  \left( \mathcal{D}^+ \right)^4 \left[ B^{--} \left( \mathcal{W} + \bar{\mathcal{W}} \right) \right].}
\end{equation}	

\medskip	

\noindent $\bullet$ 	\underline{\textbf{Parameter $B^{--}_{(\alpha\beta)}$}}

\smallskip
\noindent Prepotential variation
$
\delta \Psi^-_{\alpha}
= \mathcal{D}^{+\beta} B^{--}_{(\alpha\beta)}
$ leads to the variation
\begin{equation}
	\delta S_{int} = \int d^4x d^8 \theta du \; B^{--(\alpha\beta)} \mathcal{D}^+_\alpha \mathcal{D}^+_{\beta} \mathcal{J} = 0.
\end{equation}	
This variation is vanishing due to $\mathcal{D}^+_{(\alpha} \mathcal{D}^+_{\beta)} = 0$.

\medskip

\noindent$\bullet$ \textbf{\underline{Parameter $B^{--}_{\alpha\dot{\alpha}}$}}

\smallskip
\noindent Prepotentials variations
$
\delta \Psi^-_{\alpha}
= 
\bar{\mathcal{D}}^{+\dot{\beta}} B^{--}_{\alpha\dot{\beta}}$ and $
\delta \bar{\Psi}^-_{\dot{\alpha}}
=
\mathcal{D}^{+\beta} B^{--}_{\beta\dot{\alpha}}
$ also leads to a trivial variation:
\begin{equation}
	\delta S_{int} = \int d^4xd^8\theta du \; B^{--\alpha\dot{\alpha}} \left( \mathcal{D}^+_\alpha \bar{\mathcal{D}}^+_{\dot{\alpha}} - \mathcal{D}^+_\alpha \bar{\mathcal{D}}^+_{\dot{\alpha}} \right) \mathcal{J}
	=
	0.
\end{equation}		

Therefore, we see that only the variation with $B^{--}$ parameter leads to the non-trivial transformation  \eqref{eq: B-- transformations} of the analytic prepotential $V^{++}$. All other variations are trivially satisfied due to  \textit{the special coupling structure} \eqref{eq: (2 1 1) vertex} and the supercurrent $\mathcal{J} = \mathcal{W} \bar{\mathcal{W}}$. We observe that the resulting transformations  \textit{do not admit an immediate geometric interpretation}. For this purpose, we shall consider transformations within the analytic formalism.

\subsubsection{Analytic form}

Analytic representation of $(\mathbf{2}, \mathbf{1}, \mathbf{1})$ vertex is given by: 
\begin{equation}\label{eq: analytic 211 vertex}
	\begin{split}
		S^A_{int} =  \int d\zeta^{(-4)}  
		\Big[ & h^{++\alpha\dot{\alpha}}  \cdot i (\mathcal{D}^+)^4 \left\{ \theta^-_\alpha \bar{\theta}^-_{\dot{\alpha}} \mathcal{W} \bar{\mathcal{W}} \right\}
		\\& -2
		h^{++\alpha+} (\mathcal{D}^+)^4 \left\{ \theta^-_\alpha (\bar{\theta}^-)^2 \mathcal{W} \bar{\mathcal{W}} \right\}
		+2
		h^{++\dot{\alpha}+}  (\mathcal{D}^+)^4 \left\{ \bar{\theta}^-_{\dot{\alpha}} (\theta^-)^2 \mathcal{W} \bar{\mathcal{W}}\right\}  \Big].
	\end{split}
\end{equation}	
 Let us consider the variation sequentially in each sector.

\medskip

\noindent $\bullet$ \textbf{\underline{Parameter $\lambda^{\alpha\dot{\alpha}}$}}

\smallskip
 
 It is possible to collect the full integration measure over  the harmonic superspace and transform the expression as:
\begin{equation}\label{eq: spin 2 case}
	\begin{split}
		\delta_{\lambda^{\alpha\dot{\alpha}}} S_{int}^A =& + i \int d\zeta^{(-4)} (\mathcal{D}^+)^4\;  \mathcal{D}^{++} \lambda^{\alpha\dot{\alpha}}  \left( \theta^-_\alpha \bar{\theta}^-_{\dot{\alpha}} \mathcal{W} \bar{\mathcal{W}} \right) 
		\\
		=& - i  \int d^4x d^8\theta du \;  \lambda^{\alpha\dot{\alpha}}  ( \theta^+_\alpha \bar{\theta}^-_{\dot{\alpha}} \mathcal{W} \bar{\mathcal{W}} + \theta^-_\alpha \bar{\theta}^+_{\dot{\alpha}} \mathcal{W} \bar{\mathcal{W}} ) 
		\\=
		& + i  \int d^4x d^8\theta du \;  
		\left[ 
		(\mathcal{D}^+)^2 \left( \lambda^{\alpha\dot{\alpha}} \theta^-_\alpha \bar{\theta}^+_{\dot{\alpha}} \bar{\mathcal{W}} \right) 
		+
		(\bar{\mathcal{D}}^+)^2  \left(  \lambda^{\alpha\dot{\alpha}}  \theta^+_\alpha \bar{\theta}^-_{\dot{\alpha}} \mathcal{W} \right)   \right] V^{--}.
	\end{split}
\end{equation}	
From here, one finds the variation of the analytic prepotential:
\begin{equation*}\label{eq: spin 2 for spin 1}
	\delta_{\lambda^{\alpha\dot{\alpha}}} V^{++} = \frac{i}{2}
(\mathcal{D}^+)^2 \left( \bar{\theta}^{+\dot{\alpha}} \theta^{-\alpha} \lambda_{\alpha\dot{\alpha}}  \mathcal{W}  \right)
	+
	\frac{i}{2} (\bar{\mathcal{D}}^+)^2 \left( \bar{\theta}^{-\dot{\alpha}} \theta^{+\alpha} \lambda_{\alpha\dot{\alpha}} \bar{\mathcal{W}} \right).
\end{equation*}
Using definitions of the superstrengths $\mathcal{W} = (\bar{\mathcal{D}}^+)^2 V^{--}$ and  	$\bar{\mathcal{W}} = (\mathcal{D}^+)^2 V^{--}$, one can rewrite this transformation in the manifestly analytic representation:
\begin{equation}\label{eq: spin 2 transf}
\boxed{	\delta_{\lambda^{\alpha\dot{\alpha}}} V^{++}  = 	- 8 
	(\mathcal{D}^+)^4 \left[ \lambda^{\alpha\dot{\alpha}}  \mathcal{D}^{++} \left(i \theta^{-}_{\alpha}  \bar{\theta}^{-}_{\dot{\alpha}} \right) V^{--} \right].}
\end{equation}	
To interpret the resulting transformation, it is convenient to represent it as
\begin{equation*}
\begin{split}
		\delta_{\lambda^{\alpha\dot{\alpha}}} V^{++} &=
			- 	8  \lambda^{\alpha\dot{\alpha}}  \underbrace{(\mathcal{D}^+)^4  \left[ i \bar{\theta}^{-}_{\dot{\alpha}} \theta^{-}_{\alpha}   \mathcal{D}^{--}  V^{++}  \right]}_{-\partial_{\alpha\dot{\alpha}} V^{++}}
		-
		8  (\mathcal{D}^{++} \lambda^{\alpha\dot{\alpha}})
		(\mathcal{D}^+)^4   \left[ i \bar{\theta}^{-}_{\dot{\alpha}} \theta^{-}_{\alpha}    V^{--}  \right]
		\\& \quad
		+\mathcal{D}^{++} \Big\{ 8	 (\mathcal{D}^+)^4   \left[ i\bar{\theta}^{-}_{\dot{\alpha}} \theta^{-}_{\alpha}  \lambda^{\alpha\dot{\alpha}} V^{--}  \right] \Big\}.
	\end{split} 
\end{equation*}	
The first term is the standard translation term.
 Second term is equal  to zero for a constant parameter $\lambda^{\alpha\dot{\alpha}} = \text{const}$ and so corresponds to a nontrivial local extension of global translations with parameter $\lambda_{\alpha\dot{\alpha}}$.  
 Last term  corresponds to spin-$\mathbf{1}$ gauge transformation \eqref{eq: spin 1 gauge} with composite analytic gauge parameter $\lambda =  (\mathcal{D}^+)^4   \left[ 	8 \kappa_2 i\bar{\theta}^{-\dot{\alpha}} \theta^{-\alpha}  \lambda_{\alpha\dot{\alpha}} V^{--}  \right] $. Omitting this term, the variation takes the form:
 \begin{equation}\label{eq: lambda alpha dot alpha}
 \boxed{	\delta_{\lambda^{\alpha\dot{\alpha}}} V^{++}
 	=
 	8 \lambda^{\alpha\dot{\alpha}} \partial_{\alpha\dot{\alpha}} V^{++}
 		-
 	8 (\mathcal{D}^{++} \lambda^{\alpha\dot{\alpha}})
 	(\mathcal{D}^+)^4   \left[ i \bar{\theta}^{-}_{\dot{\alpha}} \theta^{-}_{\alpha}    V^{--}  \right].} 
 \end{equation}	 
 We conclude that these transformations describe the analytic-superspace localization of spacetime translations:
 $$x^{\alpha\dot{\alpha}} \to x^{\alpha\dot{\alpha}} + \lambda^{\alpha\dot{\alpha}}(\zeta).$$

 It is worth examining the second term in more details. In the Wess-Zumino gauge \eqref{eq: vector WZ gauge}, we~get
 \begin{equation*}
 		(\mathcal{D}^+)^4   \left[ i \bar{\theta}^{-}_{\dot{\alpha}} \theta^{-}_{\alpha}    V_{WZ}^{--}  \right] \sim A_{\alpha\dot{\alpha}}
 \end{equation*}	
 and conclude that this term leads to a contribution ($\lambda^{\alpha\dot{\alpha}}(\zeta) = a^{\alpha\dot{\alpha}}(x) + \dots$):
 \begin{equation*}
 	\delta A_{\beta\dot{\beta}} \sim \partial_{\beta\dot{\beta}} a^{\alpha\dot{\alpha}} A_{\alpha\dot{\alpha}},
 \end{equation*}	
 which corresponds to a transformation with respect to the vector index. This is precisely the expected tensorial transformation law acting on the vector index of the gauge field.

 \medskip
 
 \noindent $\bullet$ \textbf{\underline{Parameter $\lambda^{+\alpha}$}}
 
 \smallskip

 $\lambda^{+\alpha}$ variation of analytic vertex \eqref{eq: analytic 211 vertex} has  the form\footnote{These transformation in the rigid limit ($\mathcal{D}^{++}\lambda^{+\alpha} = 0$, $\partial_{\beta\dot{\beta}} \lambda^{+\alpha} = 0$) are not a symmetry of the action! Corresponding global symmetry is given by $\delta_{\lambda^{\alpha\dot{\alpha}} } + \delta_{\lambda^{+\alpha}}$ with:
 	$\mathcal{D}^{++}\lambda_{rig}^{\alpha\dot{\alpha}} + 4i \lambda_{rig}^{+\alpha} \bar{\theta}^{+\dot{\alpha}} = 0$, see eq. \eqref{eq: rig susy}. These remarks directly carry over to higher-spin transformations as well.
 }
\begin{equation}
	\delta_{\lambda^{+\alpha}} S_{int}^A
	=
 \int d^4x d^8\theta du \; \Big\{  4 i \lambda^{+\alpha} \bar{\theta}^{+\dot{\alpha}} 
	\cdot i \left[ \theta^-_\alpha \bar{\theta}^-_{\dot{\alpha}} \mathcal{W} \bar{\mathcal{W}} \right]
	-
	2  \mathcal{D}^{++} \lambda^{+\alpha} \left [ \theta^-_\alpha (\bar{\theta}^-)^2 \mathcal{W} \bar{\mathcal{W}}  \right] \Big\}. 
\end{equation}	
After integrating by parts with respect to the harmonic derivative, we obtain:
\begin{equation}
	\begin{split}
	\delta_{\lambda^{+\alpha}} S_{int}^A 
	&=
	2 \int d^4xd^8\theta du \;  (\lambda^{+\alpha} \theta^+_\alpha) (\bar{\theta}^-)^2 \mathcal{W} \bar{\mathcal{W}}
	\\
	&
	=
	-32  \int d^4x d^8\theta du
	\;
	(\mathcal{D}^+)^4 \left[ (\lambda^+ \theta^+) (\bar{\theta}^-)^2 V^{--} \right] V^{--}.
	\end{split}
\end{equation}	
From this, one obtains the variations of the analytic prepotential:
\begin{equation}\label{eq: loc susy}
	\boxed{ \delta_{\lambda^{+\alpha}} V^{++}
	=
	16	\kappa_2  (\mathcal{D}^+)^4 \left[ (\lambda^{+\alpha} \theta^+_\alpha) (\bar{\theta}^-)^2 V^{--} \right].} 
\end{equation}	
The resulting variation can be brought to the form:
\begin{equation}\label{eq: lambda alpha +}
\begin{split}
		\delta_{\lambda^{+\alpha}} V^{++}
		=&
		+8 \kappa_2 \, \lambda^{\alpha+}\partial^-_\alpha V^{++} 
	\\&	- 
		16\kappa_2  \, (\mathcal{D}^{++} \lambda^{+\alpha}) (\mathcal{D}^+)^4 \left[  \theta^-_\alpha (\bar{\theta}^-)^2 V^{--} \right]	
		\\&
		-32 \kappa_2 
		(\mathcal{D}^+)^4 \left[  (\lambda^+\theta^-) (\bar{\theta}^+ \bar{\theta}^-) V^{--} \right]	
		\\& + \mathcal{D}^{++} \left\{  16 \kappa_2 \, (\mathcal{D}^+)^4 \left[ (\lambda^+\theta^-) (\bar{\theta}^-)^2 V^{--}  \right]\right\} .
		\end{split} 
\end{equation}	
The last term corresponds to gauge transformations and may be omitted. The first term generates translations along the Grassmann coordinate $$\theta^{+\alpha} \to \theta^{+\alpha} + \lambda^{+\alpha}(\zeta),$$  and the second term to a certain localization of supersymmetry transformations.  Moreover, we note that in the Wess–Zumino gauge \eqref{eq: vector WZ gauge}, the second term is equal to zero:
\begin{equation*}
	(\mathcal{D}^+)^4 \left[  \theta^-_\alpha (\bar{\theta}^-)^2 V^{--}_{WZ}	 \right]
	=0.
\end{equation*}
This contribution is gauge trivial, although this is not manifest in the superfield representation.

To interpret the third term, we use the fact that in the rigid limit, the supersymmetry transformations in $\epsilon^i_\alpha$ sector are given by parameters \eqref{eq: rig N=2 susy}:
\begin{equation}\label{eq: rig susy}
	\lambda^{\alpha\dot{\alpha}}_{\epsilon} = -4i \epsilon^{-\alpha} \bar{\theta}^{+\dot{\alpha}},
	\qquad
	\lambda^{+\alpha}_\epsilon 
	=
	\epsilon^{\alpha i} u^+_i,
	\qquad
	\bar{\lambda}^{+\dot{\alpha}}_\epsilon = 0.
\end{equation}	
Then, combining the spin-$\mathbf{2}$ transformations \eqref{eq: lambda alpha dot alpha} and \eqref{eq: lambda alpha +} we  find that the third term cancels with a $\mathcal{D}^{++}\lambda_\epsilon^{\alpha\dot{\alpha}}$ term and we obtain:
\begin{equation}
	\delta_\epsilon V^{++} = 8 \kappa_2 \left( \lambda^{\alpha\dot{\alpha}}_\epsilon \partial_{\alpha\dot{\alpha}} + \lambda_\epsilon^{+\alpha} \partial^-_\alpha \right) V^{++},
\end{equation}	
which coincides with the active form of the rigid $\mathcal{N}=2$ supersymmetry transformation.

We conclude that the transformations \eqref{eq: spin 2 transf} and
\eqref{eq: loc susy} correspond to the localization of rigid 
$\mathcal{N}=2$ supersymmetry.

\subsubsection*{Analytic vs. non-analytic}

A comparison between the transformations obtained in the non-analytic and analytic representations reveals several important differences. 
The transformations
\eqref{eq: lambda alpha dot alpha}
and
\eqref{eq: lambda alpha +},
derived in the analytic formulation, have a clear geometric interpretation, unlike the transformations generated by the $B$-parameters in
\eqref{eq: B-- transformations}.
Furthermore, the $K$-transformations associated with the analytic $\lambda$-parameters (see eq.~\eqref{eq: K and lambda}) vanish identically as a consequence of the harmonic independence condition \eqref{eq: HA cond}, without imposing the equations of motion.
By contrast, in the analytic formalism the harmonic shortness condition holds only on shell due to the explicit Grassmann-coordinate dependence of the supercurrents.

Two types of vertices and transformations are ultimately equivalent, although the form of the vertices and supergauge transformations differs significantly. This is related to the large gauge freedom possessed by the Mezincescu-type prepotentials and the freedom in definition of the cubic vertices. In our construction, this equivalence manifests itself through an interchange between \textbf{\textit{Grassmann analyticity}} and \textbf{\textit{harmonic independence}}. Thus, while the non-analytic formulation provides a compact algebraic description of the higher-spin couplings, the analytic formulation makes their geometric content manifest by interpreting them as local superspace transformations.

\subsection{Spin-$\mathbf{3}$ transformations: super zilch symmetries}

As a next step, it is instructive to analyze the spin-$\mathbf{3}$ example in detail. The principal current in this case has the form:
\begin{equation}
	\mathcal{J}_{\alpha\dot{\alpha}}
	=
	i \left( \mathcal{W} \partial_{\alpha\dot{\alpha}} \bar{\mathcal{W}} - \partial_{\alpha\dot{\alpha}} \mathcal{W}  \bar{\mathcal{W}}   \right)
	-
	\frac{1}{4}
	\left( \mathcal{D}^-_\alpha \mathcal{W} \bar{\mathcal{D}}^+_{\dot{\alpha}}\bar{\mathcal{W}}
	-
	\mathcal{D}^+_\alpha \mathcal{W} \bar{\mathcal{D}}^-_{\dot{\alpha}}\bar{\mathcal{W}}
	  \right).
\end{equation}	
For further analysis, we will represent it in the form:
\begin{equation}\label{eq: spin 3 with derivatives}
	\mathcal{J}_{\alpha\dot{\alpha}}
	=
	2i \left( \mathcal{W} \partial_{\alpha\dot{\alpha}} \bar{\mathcal{W}} - \partial_{\alpha\dot{\alpha}} \mathcal{W}  \bar{\mathcal{W}}   \right)
	+
	\frac{1}{4}
\bar{\mathcal{D}}^+_{\dot{\alpha}} 	\left( \mathcal{D}^-_\alpha \mathcal{W} \bar{\mathcal{W}}\right) 
	+
	\frac{1}{4}
	\mathcal{D}^+_\alpha\left(   \mathcal{W} \bar{\mathcal{D}}^-_{\dot{\alpha}}\bar{\mathcal{W}}
	\right).
\end{equation}

We will consider here transformations only in analytic form. General consideration of transformations in non-analytic form will be presented in the next section.

\smallskip

\noindent $\bullet$ \underline{\textbf{Parameter $\lambda^{(\alpha\beta)(\dot{\alpha}\dot{\beta})}$}}

\smallskip

Spin-$\mathbf{3}$ variation of $(\mathbf{3}, \mathbf{1}, \mathbf{1})$ vertex is given by:
\begin{equation}
	\begin{split}
\delta S &= \int d\zeta^{(-4)}\; \mathcal{D}^{++} \lambda^{(\alpha\beta)(\dot{\alpha}\dot{\beta})}	
\cdot i (\mathcal{D}^+)^4 \left( \theta^-_\alpha \bar{\theta}^-_{\dot{\alpha}} \mathcal{J}_{\beta\dot{\beta}} \right)
\\&=
-i
 \int d^4x d^8 \theta du \; \lambda^{(\alpha\beta)(\dot{\alpha}\dot{\beta})}	
\cdot\left( \theta^+_\alpha \bar{\theta}^-_{\dot{\alpha}}
+
\theta^-_\alpha \bar{\theta}^+_{\dot{\alpha}} \right) \mathcal{J}_{\beta\dot{\beta}} . 
\end{split}
\end{equation}	
$\mathcal{D}^+$ terms from supercurrent \eqref{eq: spin 3 with derivatives} do not contribute due to analyticity and symmetries of transformation parameter: 
\begin{equation}
	\int d^4x d^8\theta du \;  \frac{1}{4}  \lambda^{(\alpha\beta)(\dot{\alpha}\dot{\beta})}	
	\cdot\left( \theta^+_\alpha \bar{\theta}^-_{\dot{\alpha}}
	+
	\theta^-_\alpha \bar{\theta}^+_{\dot{\alpha}} \right) 	
	\left[
	\bar{\mathcal{D}}^+_{\dot{\beta}} 	\left( \mathcal{D}^-_\beta \mathcal{W} \bar{\mathcal{W}}\right) 
	+
		\mathcal{D}^+_\beta \left(  \mathcal{W} \bar{\mathcal{D}}^-_{\dot{\beta}}\bar{\mathcal{W}}\right)
	\right]
	=
	0.
\end{equation}	
As a result, the following contributions remain
\begin{equation}
	\delta S = 
	2
	\int d \zeta^{(-4)} \, \lambda^{(\alpha\beta)(\dot{\alpha}\dot{\beta})}	
	\cdot\left( \theta^+_\alpha \bar{\theta}^-_{\dot{\alpha}}
	+
	\theta^-_\alpha \bar{\theta}^+_{\dot{\alpha}} \right)  \left( \mathcal{W} \partial_{\alpha\dot{\alpha}} \bar{\mathcal{W}} - \partial_{\alpha\dot{\alpha}} \mathcal{W}  \bar{\mathcal{W}}   \right).
\end{equation}	

Proceeding as in the spin-$\mathbf{2}$ case  \eqref{eq: spin 2 case}, one finds
\begin{equation}
\boxed{ 	\begin{split}
	\delta V^{++} 
	=
	 -32  (\mathcal{D}^+)^4 
	  \Big\{ 
	 & \partial_{\alpha\dot{\alpha}}    \lambda^{(\alpha\beta)(\dot{\alpha}\dot{\beta})} [\theta^+_\beta \bar{\theta}^-_{\dot{\beta}} - \theta^-_\beta \bar{\theta}^+_{\dot{\beta}} ] V^{--} 
\\&	 +
2 \lambda^{(\alpha\beta)(\dot{\alpha}\dot{\beta})} [\theta^+_\beta \bar{\theta}^-_{\dot{\beta}} - \theta^-_\beta \bar{\theta}^+_{\dot{\beta}} ]  \partial_{\alpha\dot{\alpha}}   V^{--} 
 \Big\}.
 \end{split} }
\end{equation}	
Unlike the spin-$\mathbf{2}$ transformations \eqref{eq: spin 2 transf}, in this case one cannot extract a total harmonic derivative of Grassmann coordinates.

To understand the meaning of the obtained transformation, we consider these transformations in the rigid limit ($\partial_{\rho\dot{\rho}} \lambda^{(\alpha\beta)(\dot{\alpha}\dot{\beta})} = 0$, $\mathcal{D}^{++} \lambda^{(\alpha\beta)(\dot{\alpha}\dot{\beta})}  = 0 $), using the Wess–Zumino-type  gauge for vector superfield $V^{++}$. Only these components of $V^{--}$ will contribute to the component spin-1 field transformations (see Ref. \cite{Ivanov:2024gjo} for  $V^{--}_{WZ}$):
\begin{equation}
	V^{--}_{WZ} 
	\quad \to \quad
	8 (\theta^-)^2 \bar{\theta}^{-(\dot{\rho}} \bar{\theta}^{+\dot{\sigma})} \bar{\mathcal{F}}_{(\dot{\rho}\dot{\sigma})}
	-
	8 (\bar{\theta}^-)^2 \theta^{-(\rho} \theta^{-\sigma)} \mathcal{F}_{(\rho\sigma)}. 
\end{equation}	
Removing the Grassmann derivatives, we obtain:
\begin{equation}
	(\mathcal{D}^+)^4 \left\{ \lambda [\theta^+ \bar{\theta}^- - \theta^- \bar{\theta}^+ ]  \partial V_{WZ}^{--} \right\}
	=
	4 
	\theta^{+\beta} \bar{\theta}^{+\dot{\beta}} \left( \epsilon_{\alpha\beta} \bar{\mathcal{F}}_{(\dot{\alpha}\dot{\beta})}
	-
	\epsilon_{\dot{\alpha}\dot{\beta}} \mathcal{F}_{(\alpha\beta)}   \right),
\end{equation}	
from which we find the transformation law for the vector field
\begin{equation}
	\delta A_{\beta\dot{\beta}}
	=
	-64 i
	\lambda^{(\alpha_1\alpha_2)(\dot{\alpha}_1\dot{\alpha}_2)}  \partial_{\alpha_1\dot{\alpha}_1}
	\left( \epsilon_{\alpha_2\beta} \bar{\mathcal{F}}_{(\dot{\alpha}_2\dot{\beta})}
	-
	\epsilon_{\dot{\alpha}_2\dot{\beta}} \mathcal{F}_{(\alpha_2\beta)}   \right).
\end{equation}	
At this stage, it is convenient to switch to vector indices. Thus, one obtains:
\begin{equation}
	\delta A_b \sim i \lambda^{(a_1a_2)} \partial_{a_1} {}^\star F_{[a_2b]},
	\quad
	\text{where}
	\quad
	{}^\star F_{[ab]} : = \frac{i}{2} \epsilon_{abcd} F^{[cd]}.
\end{equation}	
The action of these transformations on the electromagnetic field strength has the form:
\begin{equation}
	\delta F_{[ab]} \sim i \lambda^{(cd)} \partial_{c}\partial_{d}{}^\star F_{[ab]}.
\end{equation}	
In this form, it is clearly seen that the corresponding vertex is P-odd\footnote{To construct $P$-even $(3,1,1)$ vertex one must to consider doublet of vector fields, see, e.g., \cite{Zinoviev:2010cr, Berends:1985xx}.}.
Such transformations are  known in the literature as \textit{zilch symmetry}.
We discussed the corresponding conserved \textit{zilch pseudotensor} in vector form and the connection with known results in appendix \ref{sec: ZILCH}.

\medskip

\noindent $\bullet$ \underline{\textbf{Parameter $\lambda^{+(\alpha\beta)\dot{\alpha}}$}}

\medskip

Vertex $(\mathbf{3}, \mathbf{1}, \mathbf{1})$ variation in this sector is given by
\begin{equation}
	\begin{split}
	\delta S_{int}^A = \int d\zeta^{(-4)}
	\Big[&
	4i \lambda^{+(\alpha\beta)(\dot{\alpha}} \bar{\theta}^{+\dot{\beta})} \cdot i (\mathcal{D}^+)^4 \left\{ \theta^-_{(\alpha} \bar{\theta}^-_{(\dot{\alpha}} \mathcal{J}_{\beta)\dot{\beta})} \right\}
	\\&
	+
	 \mathcal{D}^{++} \lambda^{+(\alpha\beta)\dot{\alpha}} \cdot -2 (\mathcal{D}^+)^4 \left\{ \theta^{-}_{(\alpha}  (\bar{\theta}^-)^2 \mathcal{J}_{\beta)\dot{\alpha}} \right\}   \Big].
	\end{split}
\end{equation}	
Using off-shell relation for analytic supercurrents 
\begin{equation}
	\begin{split}
 \mathcal{D}^{++}	(\mathcal{D}^+)^4 \left\{ \theta^{-}_{(\alpha}  (\bar{\theta}^-)^2 \mathcal{J}_{\beta)\dot{\alpha}} \right\}
 =
&\; \theta^{+}_{(\alpha}   (\mathcal{D}^+)^4 \left\{ (\bar{\theta}^-)^2 \mathcal{J}_{\beta)\dot{\alpha}} \right\}
 \\&+
2 \bar{\theta}^{+\dot{\beta}} (\mathcal{D}^+)^4
 \left\{ \theta^{-}_{(\alpha}  \bar{\theta}^-_{(\dot{\alpha}} \mathcal{J}_{\beta)\dot{\beta})} \right\}
 -
  \bar{\theta}^+_{\dot{\alpha}} 
 (\mathcal{D}^+)^4
 \left\{ \theta^{-}_{(\alpha}  \bar{\theta}^{-\dot{\beta}} \mathcal{J}_{\beta)\dot{\beta}} \right\},
 \end{split}
\end{equation}	
one can reduce variation to
\begin{equation}
	\delta S_{int}^A =  2 \int d^4x d^8\theta du \; 
	\lambda^{+(\alpha\beta)\dot{\alpha}} 
 \left[  \theta^{+}_{(\alpha}    \left\{ (\bar{\theta}^-)^2 \mathcal{J}_{\beta)\dot{\alpha}} \right\}
 -
 \bar{\theta}^+_{\dot{\alpha}} 
 \left\{ \theta^{-}_{(\alpha} \bar{\theta}^{-\dot{\beta}} \mathcal{J}_{\beta)\dot{\beta}} 
  \right\} \right].
\end{equation}	
First term here is the generalization of spin-$\mathbf{2}$ transformation with parameter $\lambda^{+\alpha}$. Second term is the new one.
We will consider the two terms separately.

\underline{First term} contributes:

\begin{equation}
	\begin{split}
	2\lambda^{+\alpha(2)\dot{\alpha}}& \theta^{+}_\alpha (\bar{\theta}^-)^2 
	\mathcal{J}_{\alpha\dot{\alpha }} 
	\\=& \;
	2	\lambda^{+\alpha(2)\dot{\alpha}} \theta^{+}_\alpha (\bar{\theta}^-)^2 
	\left( 2i\mathcal{W} \partial_{\alpha\dot{\alpha}} \bar{\mathcal{W}} - 2i \partial_{\alpha\dot{\alpha}} \mathcal{W}  \bar{\mathcal{W}}  
	+
	\frac{1}{4}
	\bar{\mathcal{D}}^+_{\dot{\alpha}} 	\left( \mathcal{D}^-_\alpha \mathcal{W} \bar{\mathcal{W}}\right)  \right)
		\\=&
	-64i (\mathcal{D}^+)^4  \left(	\lambda^{+\alpha(2)\dot{\alpha}} \theta^{+}_\alpha (\bar{\theta}^-)^2  \partial_{\alpha\dot{\alpha}} V^{--} \right)
V^{--}
\\&-64i (\mathcal{D}^+)^4   \partial_{\alpha\dot{\alpha}} \left(	\lambda^{+\alpha(2)\dot{\alpha}} \theta^{+}_\alpha (\bar{\theta}^-)^2   V^{--} \right)
V^{--}
\\&	
-8 (\mathcal{D}^+)^4 \left( \mathcal{D}^-_\alpha \lambda^{+\alpha(2) \dot{\alpha}} \theta^+_\alpha \bar{\theta}^-_{\dot{\alpha}} V^{--} \right) V^{--},
	\end{split}
\end{equation}	
where in the second line  we omit $\mathcal{D}^+ (\dots)$ terms, which do not contribute.

\underline{Second contribution} is given by:
\begin{equation}\label{eq: second contr}
	\begin{split}
	-2 
	\lambda^{+(\alpha\beta)\dot{\alpha}} &
	\bar{\theta}^+_{\dot{\alpha}} 
 \theta^{-}_{(\alpha} \bar{\theta}^{-\dot{\beta}} \mathcal{J}_{\beta)\dot{\beta}} 
\\
& =
 	-2 
 \lambda^{+(\alpha\beta)\dot{\alpha}} 
 \bar{\theta}^+_{\dot{\alpha}} 
 \theta^{-}_{(\alpha} \bar{\theta}^{-\dot{\beta}}
 	\left[ 2i\mathcal{W} \partial_{\beta\dot{\beta}} \bar{\mathcal{W}} - 2i \partial_{\beta\dot{\beta}} \mathcal{W}  \bar{\mathcal{W}}  
 +
 \frac{1}{4}
 \bar{\mathcal{D}}^+_{\dot{\beta}} 	\left( \mathcal{D}^-_\beta \mathcal{W} \bar{\mathcal{W}}\right)  \right]
 \\
 & =
 \lambda^{+(\alpha\beta)\dot{\alpha}} 
 \bar{\theta}^+_{\dot{\alpha}} 
 \theta^{-}_{\alpha} 
 \left[ - \mathcal{W} \partial^-_{\beta} \bar{\mathcal{W}} + 4i \partial_{\beta\dot{\beta}} \mathcal{W}  \bar{\mathcal{W}}  
 - \mathcal{D}^-_\beta \mathcal{W} \bar{\mathcal{W}}  \right]
 \\
 & =
 \lambda^{+(\alpha\beta)\dot{\alpha}} 
 \theta^{-}_{\alpha}  \bar{\theta}^+_{\dot{\alpha}} 
  \left[  \mathcal{W}  \partial^-_{\beta} \bar{\mathcal{W}}  - \partial^-_{\beta} \mathcal{W} \bar{\mathcal{W}} \right] 
  \\
  &=
  -
  16 (\mathcal{D}^+)^4 \partial^-_\beta \left( \lambda^{+(\alpha\beta)\dot{\alpha}} 
  \theta^{-}_{\alpha}  \bar{\theta}^+_{\dot{\alpha}}  V^{--}  \right) V^{--}
\\&\quad   +
  16 (\mathcal{D}^+)^4  \left( \lambda^{+(\alpha\beta)\dot{\alpha}} 
  \theta^{-}_{\alpha}  \bar{\theta}^+_{\dot{\alpha}}  \partial^-_\beta V^{--}  \right) V^{--}.
 \end{split}
\end{equation}	
In this term, we have extracted partial spinor derivatives $\partial^-_\beta$. It is precisely allows to bring the second contribution to the desired form.

The resulting formula for the variation of the analytic prepotential takes the form:
\begin{equation}
\boxed{ 	\begin{split}
	\delta V^{++}
	=&\;
	32 i (\mathcal{D}^+)^4 \left(
	\partial_{\alpha\dot{\alpha}} \lambda^{+\alpha(2)\dot{\alpha}} \theta^{+}_\alpha (\bar{\theta}^-)^2   V^{--}
	+
	2 \lambda^{+\alpha(2)\dot{\alpha}} \theta^{+}_\alpha (\bar{\theta}^-)^2  \partial_{\alpha\dot{\alpha}} V^{--} \right)
	\\
	& + 4  (\mathcal{D}^+)^4 \left( \mathcal{D}^-_\alpha \lambda^{+\alpha(2) \dot{\alpha}} \theta^+_\alpha \bar{\theta}^-_{\dot{\alpha}} V^{--} \right)
	\\
	&
	+ 8 (\mathcal{D}^+)^4  \left(
	\partial^-_\beta  \lambda^{+(\alpha\beta)\dot{\alpha}} 
	\theta^{-}_{\alpha}  \bar{\theta}^+_{\dot{\alpha}}  V^{--}
	-
	2 \lambda^{+(\alpha\beta)\dot{\alpha}} 
	\theta^{-}_{\alpha}  \bar{\theta}^+_{\dot{\alpha}}  \partial^-_\beta V^{--}  \right). 
	\end{split} }
\end{equation}	

For the rigid transformations
\begin{equation}
	\lambda_{\epsilon}^{(\alpha\beta)(\dot{\alpha}\dot{\beta})} = -4i \epsilon^{(\alpha\beta)(\dot{\alpha} i} \bar{\theta}^{+\dot{\beta})} u^-_i,
	\qquad
	\lambda_{\epsilon}^{+(\alpha\beta)\dot{\alpha}} 
	=
		\epsilon^{(\alpha\beta)\dot{\alpha} i} u^+_i,
	\qquad
	\bar{\lambda}_{\epsilon}^{+\alpha(\dot{\alpha}\dot{\beta})} = 0
\end{equation}	
we obtain
\begin{equation}
	\begin{split}
	\delta_{rig} V^{++} =
&	\;16  \epsilon^{\alpha(2)\dot{\alpha}+} 
	(\mathcal{D}^+)^4 \Big\{
	4i  \theta^{+}_\alpha (\bar{\theta}^-)^2  \partial_{\alpha\dot{\alpha}} V^{--} 
	-
	 \theta^-_\alpha \bar{\theta}^+_{\dot{\alpha}} \partial^-_\alpha V^{--} \Big\}
\\
&			+ 64\cdot 4i (\mathcal{D}^+)^4 
			\Big\{ 
		 \epsilon^{(\alpha\beta)(\dot{\alpha} -} \bar{\theta}^{+ \dot{\beta})} [\theta^+_\beta \bar{\theta}^-_{\dot{\beta}} - \theta^-_\beta \bar{\theta}^+_{\dot{\beta}} ]  \partial_{\alpha\dot{\alpha}}   V^{--} 
			\Big\}.
	\end{split} 
\end{equation}	
In the Wess-Zumino-type gauge these transformations reduce to
\begin{equation}
	\delta_\epsilon A_{\beta\dot{\beta}} \sim \epsilon^{\alpha\dot{\alpha}i}_\beta \partial_{\alpha(\dot{\alpha}} \bar{\psi}_{\dot{\beta}) i}.
\end{equation}	
Unlike bosonic transformations, this transformation is a straightforward generalization of the rigid supersymmetry transformations $\delta_\epsilon A_{\beta\dot{\beta}} \sim \epsilon_\beta^i \bar{\psi}_{\dot{\beta} i} $.

\subsection{Spin-$\mathbf{s}$ transformations}\label{eq: spin s transf}
In a similar manner, one can analyze the general case as well.
Our strategy is the following.
First, we rewrite the principal supercurrent in a form adapted to the analytic formulation, where all terms proportional to full $\mathcal{D}^+$ derivatives are isolated.
We then analyze all gauge and pre-gauge parameters of the non-analytic prepotential and determine which sectors generate non-trivial transformations of the vector multiplet.
Finally, we reformulate the same cubic coupling in terms of analytic supercurrents and derive the corresponding higher-spin transformations in analytic superspace.

For the case $\mathbf{s_2}=1$, principal higher-spin $\mathcal{N}=2$ supercurrent \eqref{eq: general principle supercurrent} is simplified and given by:
\begin{equation}\label{eq: general spin 1 principle supercurrent}
	\boxed{
		\begin{split}
			&\mathcal{J}_{\alpha(s-2)\dot{\alpha}(s-2)} = \;\; -i^{s} \sum_{p=0}^{s-2}  (-1)^p  \binom{s-2}{p}^2 \; \partial^p \mathcal{W} \partial^{s-p-2} \bar{\mathcal{W}}
			\\&
			+
			i^{s} \sum_{p=0}^{s-3}   \frac{(-1)^p}{4i} \binom{s-2}{p+1} \binom{s-2}{p}\; 
			\Big[\partial^p \mathcal{D}^- \mathcal{W} \partial^{s-p-3} \bar{\mathcal{D}}^+\bar{\mathcal{W}}
			-
			\partial^p \mathcal{D}^+ \mathcal{W} \partial^{s-p-3} \bar{\mathcal{D}}^-\bar{\mathcal{W}}
			\Big]
			\\&
			-
			i^{s} \sum_{p=0}^{s-4}   \frac{(-1)^{p}}{16} \binom{s-2}{p+2} \binom{s-2}{p}
			\partial^p \mathcal{D}^+ \mathcal{D}^- \mathcal{W} \partial^{s-p-4} \bar{\mathcal{D}}^+ \bar{\mathcal{D}}^- \bar{\mathcal{W}}.
		\end{split}
	}
\end{equation}	
The last term contributes only for $s\geq4$, the second line contributes only for $s\geq 3$. This representation of the supercurrent is particularly convenient for analyzing gauge transformations in the non-analytic formulation. For the gauge transformations in analytic formulation, we will present the $\mathcal{N}=2$ principal supercurrent in a different form.

In the principal higher-spin supercurrent
\eqref{eq: general spin 1 principle supercurrent},
corresponding to the $(\mathbf{s},\mathbf{1},\mathbf{1})$
coupling, there are three types of terms.
We can single out the terms that are total $\mathcal{D}^+$ derivatives by applying relations
\begin{equation}\label{eq: relation on W bar W}
	\begin{split}
		&	\mathcal{D}^-_\alpha \mathcal{W} \bar{\mathcal{D}}^+_{\dot{\alpha}} \bar{\mathcal{W}}
		=
		- \bar{\mathcal{D}}^+_{\dot{\alpha}} \left( 	\mathcal{D}^-_\alpha \mathcal{W}  \bar{\mathcal{W}} \right)
		+
		4i \partial_{\alpha\dot{\alpha}} \mathcal{W}  \bar{\mathcal{W}}	,
		\\
		&		\mathcal{D}^+_\alpha \mathcal{W} \bar{\mathcal{D}}^-_{\dot{\alpha}} \bar{\mathcal{W}}
		=+
		\mathcal{D}^+_\alpha \left( \mathcal{W} \bar{\mathcal{D}}^-_{\dot{\alpha}} \bar{\mathcal{W}} \right)
		+
		4i \mathcal{W} \partial_{\alpha\dot{\alpha}} \bar{\mathcal{W}},
		\\
		&
		\mathcal{D}^+_{(\alpha} \mathcal{D}^-_{\beta)} \mathcal{W}
		\bar{\mathcal{D}}^+_{(\dot{\alpha}}
		\bar{\mathcal{D}}^-_{\dot{\beta})} \bar{\mathcal{W}}
		=
		-	\mathcal{D}^+_{(\alpha} 	\bar{\mathcal{D}}^+_{(\dot{\alpha}}
		\left\{ \mathcal{D}^-_{\beta)} \mathcal{W}
		\bar{\mathcal{D}}^-_{\dot{\beta})} \bar{\mathcal{W}} \right\} 
		+
		4i \mathcal{D}^+_{(\alpha} \left( \partial_{\beta)(\dot{\alpha}} \mathcal{W} \bar{\mathcal{D}}^-_{\dot{\beta)}} \bar{\mathcal{W}} \right)
		\\&\qquad\qquad\qquad \qquad\qquad -
		4i \bar{\mathcal{D}}^+_{(\dot{\alpha}} \left( \mathcal{D}^-_{(\alpha} \mathcal{W} \partial_{\beta)\dot{\beta})} \bar{\mathcal{W}}  \right)
		- 
		16 \partial_{(\alpha(\dot{\alpha}} \mathcal{W} \partial_{\beta)\dot{\beta})} \bar{\mathcal{W}}.
	\end{split}
\end{equation}	
These relations allow one to separate terms that become total $\mathcal{D}^+$ derivatives after integration by parts and therefore do not contribute in the analytic formulation.
Applying these relations and Pascal's identity for binomial coefficients, one obtains: 
\begin{equation}\label{eq: general spin 1 principle sc}
	\begin{split}
	&\mathcal{J}_{\alpha(s-2)\dot{\alpha}(s-2)}
	=
	-i^s \sum_{p=0}^{s-2} (-1)^p
	\binom{s-1}{p} \binom{s-1}{p+1}
	\;
	\partial^p \mathcal{W} \partial^{s-p-2} \bar{\mathcal{W}}
	\\
	&
		- 
	i^{s} \sum_{p=0}^{s-3}   \frac{(-1)^p}{4i} \binom{s-1}{p+2} \binom{s-2}{p}\; 
	\Big[  \bar{\mathcal{D}}^+ \mathcal{D}^- \left( \partial^p  \mathcal{W} \partial^{s-p-3} \bar{\mathcal{W}} \right)
	+ \mathcal{D}^+ \bar{\mathcal{D}}^- \left( 
	\partial^{s-p-3}  \mathcal{W} \partial^{p} \bar{\mathcal{W}} \right)
	\Big]
\\
&
	+
i^{s} \sum_{p=0}^{s-4}   \frac{(-1)^{p}}{16} \binom{s-2}{p+2} \binom{s-2}{p}
\;
\mathcal{D}^+  \bar{\mathcal{D}}^+ 
\mathcal{D}^-
\bar{\mathcal{D}}^-
\left(
\partial^p  \mathcal{W} \partial^{s-p-4}  \bar{\mathcal{W}} \right).	
	\end{split}
\end{equation}	
As a result, the principal supercurrent splits into three structurally different contributions:
the purely bosonic part, the terms proportional to single spinor derivatives, and the terms proportional to full $\mathcal{D}^+\bar{\mathcal{D}}^+$ derivatives.

\subsubsection{Non-analytic form}

The variation of the cubic $(\mathbf{s}, \mathbf{1}, \mathbf{1})$ coupling is 
\begin{equation}
	\delta_{\Psi}S_{int} =  \int d^4xd^8\theta du \; \left( \delta\Psi^{-\alpha(s-1)\dot{\alpha}(s-2)}\mathcal{D}^+_\alpha + \delta\bar{\Psi}^{-\alpha(s-2)\dot{\alpha}(s-1)} \bar{\mathcal{D}}^+_{\dot{\alpha}} \right) \mathcal{J}_{\alpha(s-2)\dot{\alpha}(s-2)}.
\end{equation}		
As in the $\mathbf{s}=2$ case, we will consider all  parameters for gauge and pre-gauge freedom \eqref{eq: Psi gauge and pre gauge} consecutively: 

\newpage

\noindent $\bullet$ \textbf{\underline{Parameter  $K^{(-3)}_{\alpha(s-1)\dot{\alpha}(s-2)}$}}

Harmonic variation $\delta \Psi^-_{\alpha(s-1)\dot{\alpha}(s-2)} = \mathcal{D}^{++} K^{(-3)}_{\alpha(s-1)\dot{\alpha}(s-2)}$ vanishes trivially due to off-shell harmonic-independence condition $\mathcal{D}^{++} \mathcal{J}_{\alpha(s-2)\dot{\alpha}(s-2)} = 0$.

\medskip

\noindent $\bullet$ \textbf{\underline{Parameter $B^{--}_{\alpha(s-2)\dot{\alpha}(s-2)}$}}

Variations
$
\delta \Psi^-_{\alpha(s-1)\dot{\alpha}(s-2)} = \mathcal{D}^+_{(\alpha} B^{--}_{\alpha(s-2))\dot{\alpha}(s-2)}
$ ,
$\delta \bar{\Psi}^{-}_{\alpha(s-2)\dot{\alpha}(s-1) }
=
-
\bar{\mathcal{D}}^+_{(\dot{\alpha}} B^{--}_{\alpha(s-2)\dot{\alpha}(s-2))}$ gives the following variation of the action: 
\begin{equation}
	\delta S_{int} =  \int d^4x d^8\theta du\; B^{--\alpha(s-2)\dot{\alpha}(s-2)} \left[ (\mathcal{D}^+)^2 + (\bar{\mathcal{D}}^+)^2\right]  \mathcal{J}_{\alpha(s-2)\dot{\alpha}(s-2)},
\end{equation}
where 
\begin{equation}
	\begin{split}
		(\mathcal{D}^+)^2 \mathcal{J}_{\alpha(s-2)\dot{\alpha}(s-2)}
		=&
		-i^{s} \sum_{p=0}^{s-2}  (-1)^p  \binom{s-2}{p} \binom{s-1}{p+1} \; \partial^p 	(\mathcal{D}^+)^2 \mathcal{W} \partial^{s-p-2} \bar{\mathcal{W}}
		\\&
		-
		i^{s+1} \sum_{p=0}^{s-3}   \frac{(-1)^p}{4} \binom{s-2}{p} \binom{s-1}{p+2}\; 
		\partial^p \mathcal{D}^- (\mathcal{D}^+)^2 \mathcal{W} \partial^{s-p-3} \bar{\mathcal{D}}^+ \bar{\mathcal{W}}.
	\end{split}
\end{equation}	
Then,  using $(\mathcal{D}^+)^2 \mathcal{W}  = 16 (\mathcal{D}^+)^4 V^{--}$ one finds the variation in the desired form \eqref{eq: nan var}:
\begin{equation}
	\begin{split}
		\delta S_{int} =  
		\int d^4x d^8 du \;&\Big\{ 
		- 16  i^{s} (\mathcal{D}^+)^4 \left[  \sum_{p=0}^{s-2}   \binom{s-2}{p} \binom{s-1}{p+1} \; 	 \partial^p  \left( B^{--} \partial^{s-p-2} \bar{\mathcal{W}} \right) \right]  
		\\&
		+
		4 i^{s+1} (\mathcal{D}^+)^4  \left[ \sum_{p=0}^{s-3}   \binom{s-2}{p} \binom{s-1}{p+2}\; 
		\partial^p   \mathcal{D}^- \left( B^{--} \partial^{s-p-3} \bar{\mathcal{D}}^+ \bar{\mathcal{W}} \right) \right] 
		\\& + \text{tilde conjugated terms}  \;\; \Bigr\} \cdot V^{--}.
	\end{split}
\end{equation}	
From these variations one  finds the induced variation of the analytic $V^{++}$ prepotential:
\begin{equation}\label{eq: var 1}
\boxed{	\begin{split}
		\delta V^{++} = 
		\;
		+ i^{s} (\mathcal{D}^+)^4 \Bigg[ & +8 \sum_{p=0}^{s-2}   \binom{s-2}{p} \binom{s-1}{p+1} \; 	 \partial^p  \left( B^{--} \partial^{s-p-2} \bar{\mathcal{W}} \right) 
		\\&
		 -2i  \sum_{p=0}^{s-4}  \binom{s-2}{p} \binom{s-1}{p+2}
		\partial^p  \mathcal{D}^-  \left( B^{--} \partial^{s-p-3} \bar{\mathcal{D}}^+  \bar{\mathcal{W}}\right)\Bigg] 
		\\& + \text{ tilde conjugated terms}.
	\end{split}}
\end{equation}	
Here we suppress spinor indices and assume that all indices of $B^{--}$ are contracted with indices of derivatives.

\medskip	

\noindent $\bullet$ \textbf{\underline{Parameter $B^{--}_{(\alpha(s-1)\beta)\dot{\alpha}(s-2)}$ }}

	Variation
$
\delta \Psi^-_{\alpha(s-1)\dot{\alpha}(s-2)}
= \mathcal{D}^{+\beta} B^{--}_{(\alpha(s-1)\beta)\dot{\alpha}(s-2)}
$ is trivially satisfied due to $\mathcal{D}^+_{(\alpha} \mathcal{D}^+_{\beta)} = 0$.

\medskip

\noindent $\bullet$ \textbf{\underline{Parameter $B^{--}_{\alpha(s-1)\dot{\alpha}(s-1)}$}}

The variation 
$
\delta \Psi^-_{\alpha(s-1)\dot{\alpha}(s-2)}
= 
\bar{\mathcal{D}}^{+\dot{\beta}} B^{--}_{\alpha(s-1)(\dot{\alpha}(s-2)\dot{\beta})},
\delta \bar{\Psi}^-_{\alpha(s-2)\dot{\alpha}(s-1)}
=
\mathcal{D}^{+\beta} B^{--}_{(\alpha(s-2)\beta)\dot{\alpha}(s-1)}
$
also leads to a trivial variation of the cubic vertex:
\begin{equation}
	\delta S_{int} = \int d^4xd^8\theta du \; B^{--\alpha(s-1)\dot{\alpha}(s-1)} \left( \mathcal{D}^+_\alpha \bar{\mathcal{D}}^+_{\dot{\alpha}} - \mathcal{D}^+_\alpha \bar{\mathcal{D}}^+_{\dot{\alpha}} \right) \mathcal{J}_{\alpha(s-2)\dot{\alpha}(s-2)}
	=
	0.
\end{equation}		

\newpage

\noindent$\bullet$ \textbf{\underline{Parameter $B^{--}_{\alpha(s-1)\dot{\alpha}(s-3)}$}}

The Mezincescu-type prepotential variation
$
\delta \Psi^-_{\alpha(s-1)\dot{\alpha}(s-2)}
= 
\bar{\mathcal{D}}^+_{(\dot{\alpha}} B^{--}_{\alpha(s-1)\dot{\alpha}(s-3))} 
$ leads to:
\begin{equation}\label{eq: var alpha s-1 dot s-3}
	\delta_{\Psi}S_{int} =  \int d^4xd^8\theta du \; B^{--\alpha(s-1)\dot{\alpha}(s-3)}  \mathcal{D}^+_\alpha \bar{\mathcal{D}}^{+\dot{\alpha}}  \mathcal{J}_{\alpha(s-2)\dot{\alpha}(s-2)}.
\end{equation}		
Action of derivatives on the principal supercurrent gives: 
\begin{equation}
	\begin{split}
		\mathcal{D}^+_{(\alpha}\bar{\mathcal{D}}^{+\dot{\alpha}} &\mathcal{J}_{\alpha(s-2))\dot{\alpha}(s-2)}=
		\\=&
		- i^s \sum_{p=0}^{s-2} (-1)^p \binom{s-1}{p+1}\binom{s-3}{p} \mathcal{D}^+ \partial^p \mathcal{W} \partial_{\dot{\alpha}}^{s-p-2} \bar{\mathcal{D}}^{+\dot{\alpha}} \bar{\mathcal{W}} 
		\\&
		- i^{s+1} \sum_{p=0}^{s-3}  \frac{(-1)^p}{8} \binom{s-1}{p+2} \binom{s-2}{p} \frac{s-1}{s-2} 
		\partial^p \mathcal{D}^+ \mathcal{D}^- \mathcal{W} \partial^{s-p-3} (\bar{\mathcal{D}}^+)^2 \bar{\mathcal{W}}.
	\end{split}
\end{equation}	
Using the following relation
\begin{equation*}	
	\begin{split}
	&\partial_{\alpha\dot{\alpha}} \bar{\mathcal{D}}^{+\dot{\alpha}} \bar{\mathcal{W}}  
	=
	- \frac{1}{8i} \mathcal{D}^+_\alpha \mathcal{D}^{--} (\bar{\mathcal{D}}^{+})^2 \bar{\mathcal{W}}  
	=
	\frac{1}{8i} \mathcal{D}^-_\alpha (\bar{\mathcal{D}}^{+})^2 \bar{\mathcal{W}},  
	\end{split}
\end{equation*}	
we obtain
\begin{equation}
	\begin{split}
		\mathcal{D}^+_{(\alpha}\bar{\mathcal{D}}^{+\dot{\alpha}} &\mathcal{J}_{\alpha(s-2))\dot{\alpha}(s-2)}=
		\\=&
		+ 2 i^{s+1}  \Bigg[ \sum_{p=0}^{s-2} (-1)^p \binom{s-1}{p+1}\binom{s-3}{p}  \partial^p \mathcal{D}^+\mathcal{W} \partial^{s-p-3} \mathcal{D}^- 
		\\&\qquad\qquad
		- \sum_{p=0}^{s-3}  (-1)^p \binom{s-1}{p+2} \binom{s-2}{p}
		\cdot \frac{s-1}{s-2} \cdot
		\partial^p \mathcal{D}^+ \mathcal{D}^- \mathcal{W} \partial^{s-p-3} \Bigg]  (\mathcal{D}^+)^4 V^{--}.
	\end{split}
\end{equation}	
The variation \eqref{eq: var alpha s-1 dot s-3} can then be cast into the form:
\begin{equation}
	\begin{split}
		\delta_\Psi S_{int}
		=  
		\int d^4xd^8\theta du \; &B^{--\alpha(s-1)\dot{\alpha}(s-3)}  \mathcal{D}^+_\alpha \bar{\mathcal{D}}^{+\dot{\alpha}}  \mathcal{J}_{\alpha(s-2)\dot{\alpha}(s-2)}
		\\=
		- 2  (-1)^s i^{s+1}
		&\int  (\mathcal{D}^+)^4  \Bigg[ \sum_{p=0}^{s-2}  \binom{s-1}{p+1}\binom{s-3}{p} \partial^{s-p-3}  \mathcal{D}^- \left( B^{--}  \partial^p \mathcal{D}^+ \mathcal{W} \right) 
		\\&
	+ 
	 \sum_{p=0}^{s-3}  \binom{s-1}{p+2}\binom{s-2}{p} \cdot  \frac{s-1}{s-2} \cdot \partial^{s-p-3}  \left( B^{--}  \partial^p \mathcal{D}^+  \mathcal{D}^- \mathcal{W} \right)  \Bigg]   V^{--}.
	\end{split}
\end{equation}	

Therefore, we deduce the  analytic prepotential variation:
\begin{equation}\label{eq: var 2}
	\boxed{ \begin{split}
		\delta V^{++}
		=\;
		(-1)^s i^{s+1}
		(\mathcal{D}^+)^4  &\Bigg[ \sum_{p=0}^{s-2}  \binom{s-1}{p+1}\binom{s-3}{p} \partial^{s-p-3}  \mathcal{D}^- \left( B^{--}  \partial^p \mathcal{D}^+ \mathcal{W} \right) 
		\\&
	 	+ 
	\sum_{p=0}^{s-3}  \binom{s-1}{p+2}\binom{s-2}{p} \frac{s-1}{s-2} \partial^{s-p-3}  \left( B^{--}  \partial^p \mathcal{D}^+  \mathcal{D}^- \mathcal{W} \right) 
		\\&
		 + \text{tilde conjugated terms }.
	\end{split}}
\end{equation}	

In this subsection, we have obtained gauge transformations for $\mathcal{N}=2$ vector supermultiplet prepotential, starting from non-analytic form of coupling. Only the
$B^{--}_{\alpha(s-2)\dot{\alpha}(s-2)}$
and
$B^{--}_{\alpha(s-1)\dot{\alpha}(s-3)}$
sectors lead to the non-trivial transformations, which are given by equations \eqref{eq: var 1} and \eqref{eq: var 2}.

\subsubsection{Analytic form}

In the analytic form we represent the $(\mathbf{s}, \mathbf{1}, \mathbf{1})$ vertex using manifestly analytic $J$ supercurrents, defined in eq. \eqref{eq: bf J currents}: 
 \begin{equation}\label{eq: int analytic B}
	\begin{split}
		S^{A}_{int} = \int d\zeta^{(-4)}\, \Big(&
		h^{++\alpha(s-1)\dot{\alpha}(s-1)} J^{++}_{\alpha(s-1)\dot{\alpha}(s-1)}
		+
		h^{++\alpha(s-1)\dot{\alpha}(s-2)+} J^{+}_{\alpha(s-1)\dot{\alpha}(s-2)}
		\\&+
		h^{++\alpha(s-2)\dot{\alpha}(s-1)+}\bar{J}^{+}_{\alpha(s-2)\dot{\alpha}(s-1)}
		+ h^{++\alpha(s-2)\dot{\alpha}(s-2)} J^{++}_{\alpha(s-2)\dot{\alpha}(s-2)} \Big).
	\end{split}
\end{equation}	
One needs to bring the $\delta_\lambda h^{++}$ variation of this expression to the form \eqref{eq: gen spin 1 variation}. 

\medskip

\noindent $\bullet$ \textbf{\underline{Parameter $\lambda^{\alpha(s-1)\dot{\alpha}(s-1)}$}}

\smallskip

Using the condition of the off-shell covariant harmonic-independence of the principal supercurrent \eqref{eq: general spin 1 principle supercurrent}, we bring the variation to the form:
\begin{equation}
	\begin{split}
	\delta_\lambda S_{int}^A =& -i  \int d\zeta^{(-4)} \; \left(\mathcal{D}^+\right)^4 \mathcal{D}^{++} \lambda^{\alpha(s-1)\dot{\alpha}(s-1)} \left\{ \theta^-_\alpha \bar{\theta}_{\dot{\alpha}} \mathcal{J}_{\alpha(s-2)\dot{\alpha}(s-2)} \right\}
	\\
	=& +i \int d^4x d^8\theta du \; \lambda^{\alpha(s-1)\dot{\alpha}(s-1)} \mathcal{D}^{++} \left( \theta^-_\alpha \bar{\theta}^-_{\dot{\alpha}}  \right)  \mathcal{J}_{\alpha(s-2)\dot{\alpha}(s-2)}.
	\end{split}
\end{equation}
Full $\mathcal{D}^+$ derivatives from eq.
\eqref{eq: general spin 1 principle sc} do not contribute to the variation.
Other  terms can be transformed to the desired form \eqref{eq: gen spin 1 variation} using identities valid modulo total derivatives (here for brevity, we omit the indices):
\begin{subequations}
	\begin{equation}
		\begin{split}
			\lambda  \cdot \theta^+ \bar{\theta}^-
			\partial^p \mathcal{W} \partial^{s-p-2} \bar{\mathcal{W}} 
			=&\,
			(-1)^{p+1} \partial^p \left\{ \lambda \cdot \left( \bar{\mathcal{D}}^+ \right)^2 \left(  \theta^+ \bar{\theta}^- \partial^{s-p-2} \bar{\mathcal{W}} \right)  \right\} V^{--}
			\\
			=&\, 16  \cdot (-1)^{p+1}  \left( \mathcal{D}^+ \right)^4 \partial^p \left\{ \lambda  \cdot  \theta^+ \bar{\theta}^- \partial^{s-p-2} V^{--}  \right\} V^{--},
		\end{split}
	\end{equation}
	\begin{equation}
		\begin{split}
			\lambda  \cdot \theta^- \bar{\theta}^+
			\partial^p \mathcal{W} \partial^{s-p-2} \bar{\mathcal{W}} 
			=&\,
			(-1)^{s-p-1} \partial^{s-p-2} \left\{ \lambda \cdot \left( \bar{\mathcal{D}}^+ \right)^2 \left(  \theta^- \bar{\theta}^+ \partial^{p} \mathcal{W} \right)  \right\} V^{--}
			\\
			=&\, 16  \cdot (-1)^{s-p-1}  \left( \mathcal{D}^+ \right)^4 \partial^{s-p-2} \left\{ \lambda  \cdot \theta^- \bar{\theta}^+ \partial^{p} V^{--}  \right\}
			 V^{--}.
		\end{split}
	\end{equation}
\end{subequations}

Collecting coefficients, one deduces
\begin{equation}\label{eq: arb spin bos}
\boxed{	\delta V^{++} = - 8  i^{s+1} (\mathcal{D}^+)^4 \left[
	\sum_{p=0}^{s-2}
	\binom{s-1}{p} \binom{s-1}{p+1}
	\partial^p \Big\{ \lambda \cdot  \left(  \theta^+ \bar{\theta}^- 
	+
	(-1)^s
	\theta^- \bar{\theta}^+ 
	\right)
	\partial^{s-p-2} V^{--} \Big\}
	\right]. }
\end{equation}	

The structure of the transformations generated by
\eqref{eq: arb spin bos}
depends crucially on the parity of the spin.
For even values of $s$, the transformations reduce to higher-spin extensions of spacetime translations.
By contrast, odd-spin transformations are parity odd and involve the dual Maxwell field strength.
We now discuss these two cases separately.

\medskip

\noindent $\bullet$  In the case of \underline{even spins $s$}, the resulting formula can be further simplified by representing the $\theta$-terms as a total harmonic derivative:
$$
[\theta^+ \bar{\theta}^- +\theta^- \bar{\theta}^+  ] = \mathcal{D}^{++}  [\theta^- \bar{\theta}^-  ]. 
$$
Upon separating the terms that reduce to gauge transformations, the terms inside the $\{\}$-brackets acquire the form:
\begin{equation}
	\Big\{ \lambda   \left[\theta^+ \bar{\theta}^- +  \theta^- \bar{\theta}^+ \right] \partial^{q}V^{--} \Big\}
	\quad
	 \to
	   \quad
	  - \Big\{ \lambda \left[\theta^- \bar{\theta}^-  \right] \mathcal{D}^{--} \partial^{q} V^{++} \Big\}
	  - \Big\{ \mathcal{D}^{++}\lambda  \left[\theta^- \bar{\theta}^-  \right] \partial^{q} V^{--} \Big\}.
\end{equation}	
The action of the spinor derivatives then yields:
\begin{equation}
(\mathcal{D}^+)^4 \Big\{ \lambda \cdot  \left[\theta^- \bar{\theta}^-  \right] \mathcal{D}^{--} \partial^{q} V^{++} \Big\}
=
-i \lambda \cdot  \partial^{q+1} V^{++}. 
\end{equation}	
After these simplifications, the resulting transformation takes a  form:
\begin{equation}
	\begin{split}
	\delta V^{++} =& + 8 i^{s}  
	\sum_{p=0}^{s-2}
	\binom{s-1}{p} \binom{s-1}{p+1}
	\partial^p \left\{ \lambda \cdot 
	\partial^{s-p-1} V^{++} \right\}
	\\
	& 
	+ 8  i^{s+1} (\mathcal{D}^+)^4 \left[
	\sum_{p=0}^{s-2}
	\binom{s-1}{p} \binom{s-1}{p+1}
	\partial^p \left\{ \mathcal{D}^{++} \lambda \cdot \left(  \theta^- \bar{\theta}^- 
	\right)
	\partial^{s-p-2} V^{--} \right\}
	\right].
	\end{split}
\end{equation}	
 For  rigid transformations with parameters satisfying $\partial \lambda = 0$ and $\mathcal{D}^{++} \lambda = 0$ these transformations reduce to the form:
\begin{equation}
	\delta_{rig} V^{++} = 8  i^s \binom{2s-2}{s-2} 
 \lambda^{\alpha(s-1)\dot{\alpha}(s-1)} 	\cdot \partial^{s-1}_{\alpha(s-1)\dot{\alpha}(s-1)}  V^{++}.
\end{equation}	
Thus, the even-spin transformations are naturally interpreted as gauged higher-spin extensions of spacetime translations.

\medskip

\noindent $\bullet$ 
Since the combination
$\theta^+\bar\theta^- - \theta^-\bar\theta^+$
cannot be represented as a total harmonic derivative,
the \underline{odd spin-$s$ transformations} do not reduce to higher-spin translations.
Instead, they correspond to parity-odd transformations involving the dual field strength:
\begin{equation}
		\delta A_b \sim i \lambda^{(a_1\dots a_{s-1})} \partial_{a_1} \dots \partial_{a_{s-2}} {}^\star F_{[a_{s-1}b]}.
\end{equation}	
The corresponding vertices are  parity odd, similarly to the spin-$\mathbf{3}$ case. Transformation law \eqref{eq: arb spin bos} then can be interpreted as superfield generalisation of combination of electromagnetic duality and higher-spin translation.

\medskip

\noindent $\bullet$ \textbf{\underline{Parameter $\lambda^{\alpha(s-1)\dot{\alpha}(s-2)}$}}

\smallskip

\begin{equation}\label{eq: spin s ferm var}
	\begin{split}
	\delta_\lambda S_{int}^A
	=
	\int d\zeta^{(-4)} \Big\{& 4i \lambda^{+\alpha(s-1)(\dot{\alpha}(s-2)} \bar{\theta}^{+\dot{\alpha})} \cdot  i (\mathcal{D}^+)^4 \left( \theta^-_\alpha \bar{\theta}^-_{\dot{\alpha}} \mathcal{J}_{\alpha(s-2) \dot{\alpha}(s-2)} \right)
	\\&
	+ \mathcal{D}^{++}  \lambda^{\alpha(s-1)\dot{\alpha}(s-2)} \cdot -2  (\mathcal{D}^+)^4 \left( \theta^-_\alpha (\bar{\theta}^-)^2 \mathcal{J}_{\alpha(s-2) \dot{\alpha}(s-2)} \right)   \Big\}. 
	\end{split}
\end{equation}	
After integration by parts, we obtain (here we assume symmetry over all upper and lower indices):
\begin{equation}
	\begin{split}
	\delta_\lambda S_{int}^A 
	=
	2 \int d^4xd^8\theta du\; \lambda^{+\alpha(s-1)\dot{\alpha}(s-2)}  \Big[  &\theta^+_\alpha \left\{ (\bar{\theta}^-)^2 \mathcal{J}_{\alpha(s-2) \dot{\alpha}(s-2)}  \right\}
	\\&-
	\bar{\theta}^+_{\dot{\alpha}}
	\left\{
	\theta^-_{\alpha} \bar{\theta}^{-\dot{\beta}}  \mathcal{J}_{\alpha(s-2) \dot{\alpha}(s-3)\dot{\beta}}  \right\}
	\Big] .
	\end{split}
\end{equation}
Terms, presented in \eqref{eq: general spin 1 principle sc}  as total $\mathcal{D}^+$ derivatives, do not contribute to the variation, but  $\bar{\mathcal{D}}^+$ contribute. 

The fermionic sector is more subtle because the terms containing
$\partial \mathcal W \, \bar{\mathcal W}$
and
$\bar{\mathcal D}^+\mathcal D^-(\mathcal W \bar{\mathcal W})$
must be rearranged simultaneously in order to reconstruct variations proportional to $V^{--}$. The required recombination relies on complete symmetrization over dotted indices. To perform this, we will consider terms separately.

\underline{First term} in \eqref{eq: spin s ferm var} give 2 contributions:
\begin{subequations}
\begin{equation}
	\begin{split}
	2  \lambda^{+\alpha(s-1)\dot{\alpha}(s-2)}   &\theta^+_\alpha  (\bar{\theta}^-)^2 
	\partial^p \mathcal{W} \partial^{s-p-2} \bar{\mathcal{W}}
	\\
	&=
	-
	32 \cdot (-1)^p \cdot
	(\mathcal{D}^{+})^4
	\partial^p 
	\left(  \lambda^{+\alpha(s-1)\dot{\alpha}(s-2)}   \theta^+_\alpha  (\bar{\theta}^-)^2 
	\partial^{s-p-2} V^{--}	
	\right),
	\end{split}
\end{equation}	
\begin{equation}
	\begin{split}
		2  \lambda^{+\alpha(s-1)\dot{\alpha}(s-2)}   &\theta^+_\alpha  (\bar{\theta}^-)^2 
		\bar{\mathcal{D}}^+ \mathcal{D}^-
		\left( 
		\partial^p \mathcal{W} \partial^{s-p-3} \bar{\mathcal{W}} \right)
		\\
		&=
		+
		64 \cdot  (-1)^p  \cdot
		(\mathcal{D}^{+})^4
		\partial^p 
		\left(  \mathcal{D}^-_\alpha \lambda^{+\alpha(s-1)\dot{\alpha}(s-2)}   \theta^+_\alpha  \bar{\theta}^-_{\dot{\alpha}} 
		\partial^{s-p-3} V^{--}	
		\right).
	\end{split}
\end{equation}	
\end{subequations}

\underline{Second term} in \eqref{eq: spin s ferm var} is more tricky.
As in the spin-$\mathbf{3}$ case, we need combine  $\partial^{p+1} \mathcal{W} \partial^{s-p-3} \bar{\mathcal{W}}$ and 
$\bar{\mathcal{D}}^+ \mathcal{D}^-
\left(\partial^{p} \mathcal{W} \partial^{s-p-3} \bar{\mathcal{W}} \right)$ terms to present variation to the desired form.

Using complete symmetrization over the $s-2$ dotted indices, one may distribute one derivative between the two factors with relative weights  $\frac{p+1}{s-2}$
and 
$\frac{s-p-3}{s-2}$:
\begin{equation}
	\begin{split}
		-2 \lambda^{+\alpha(s-1)\dot{\alpha}(s-2)}
		&\bar{\theta}^+_{\dot{\alpha}} \theta^-_\alpha \bar{\theta}^{-\dot{\beta}}
		 \partial^{p+1} \mathcal{W} \partial^{s-p-3} \bar{\mathcal{W}} 
		\\
		=&
		-2\cdot  \frac{p+1}{s-2} \cdot
		\lambda^{+\alpha(s-1)\dot{\alpha}(s-2)}
		\bar{\theta}^+_{\dot{\alpha}} \theta^-_\alpha
	 \bar{\theta}^{-\dot{\beta}}
		\partial^{p+1}_{\dots \dot{\beta}\dots} \mathcal{W} \partial^{s-p-3} \bar{\mathcal{W}} 
		\\
		&
			-2\cdot  \frac{s-3-p}{s-2} \cdot
		\lambda^{+\alpha(s-1)\dot{\alpha}(s-2)}
		\bar{\theta}^+_{\dot{\alpha}} \theta^-_\alpha
		\bar{\theta}^{-\dot{\beta}}
		\partial^{p+1} \mathcal{W} \partial_{\dots \dot{\beta}\dots}^{s-p-3} \bar{\mathcal{W}} 
		\\
		=&
			-2\cdot \frac{1}{4i} \cdot  \frac{p+1}{s-2} \cdot
		\lambda^{+\alpha(s-1)\dot{\alpha}(s-2)}
		\bar{\theta}^+_{\dot{\alpha}} \theta^-_\alpha
	(\mathcal{D}^- + \partial^- )
		\partial^p \mathcal{W} \partial^{s-p-3} \bar{\mathcal{W}} 
		\\
		&
		-2\cdot   \frac{1}{4i} \cdot   \frac{s-3-p}{s-2} \cdot
		\lambda^{+\alpha(s-1)\dot{\alpha}(s-2)}
		\bar{\theta}^+_{\dot{\alpha}} \theta^-_\alpha
		\partial^{p+1} \mathcal{W} \partial^{s-p-4} \partial^-  \bar{\mathcal{W}}. 
	\end{split}
\end{equation}	

The term with spinor derivatives gives
\begin{equation}
	\begin{split}
	-2 \cdot \frac{1}{4i} \cdot \lambda^{+\alpha(s-1)\dot{\alpha}(s-2)}
	&\bar{\theta}^+_{\dot{\alpha}} \theta^-_\alpha \bar{\theta}^{-\dot{\beta}}
	\bar{\mathcal{D}}^+  \mathcal{D}^-
	\left( \partial^p \mathcal{W} \partial^{s-p-3} \bar{\mathcal{W}} \right)
\\&
=
-2\cdot \frac{1}{4i} \cdot  \frac{s-1}{s-2} \cdot
\lambda^{+\alpha(s-1)\dot{\alpha}(s-2)}
\bar{\theta}^+_{\dot{\alpha}} \theta^-_\alpha
  \mathcal{D}^-
\left( \partial^p \mathcal{W} \partial^{s-p-3} \bar{\mathcal{W}} \right).
	\end{split}
\end{equation}	

Using identity  
\begin{equation*}
	\binom{s-1}{p+1}  \cdot (p+1)
	=
	(s-1) \cdot 
\binom{s-2}{p} ,
\end{equation*}	
 we reduce the sum of these terms to the desired form:
\begin{subequations}
\begin{equation}
	\begin{split}
	\lambda^{+\alpha(s-1)\dot{\alpha}(s-2)}
&	\bar{\theta}^+_{\dot{\alpha}} \theta^-_\alpha 
	\partial^-_\alpha 
	\partial^p \mathcal{W} 
	\partial^{s-p-3} \bar{\mathcal{W}}
	\\&=
	- 16 \cdot  (-1)^{s-p}
	(\bar{\mathcal{D}}^+)^4
	\partial^{s-p-3} \left( 	\lambda^{+\alpha(s-1)\dot{\alpha}(s-2)}
	\bar{\theta}^+_{\dot{\alpha}} \theta^-_\alpha 
	\partial^-_\alpha 
	\partial^p V^{--}  \right)
	V^{--},
	\end{split}
\end{equation}	
\begin{equation}
	\begin{split}
	\lambda^{+\alpha(s-1)\dot{\alpha}(s-2)}
	&\bar{\theta}^+_{\dot{\alpha}} \theta^-_\alpha 
	\partial^p \mathcal{W} 
	\partial^{s-p-3} 	\partial^-_\alpha  \bar{\mathcal{W}}
	\\&=
	+ 16 \cdot (-1)^{s-p}
	(\mathcal{D}^+)^4
	\partial^{s-p-3} 	\partial^-_\alpha  \left( 	\lambda^{+\alpha(s-1)\dot{\alpha}(s-2)}
	\bar{\theta}^+_{\dot{\alpha}} \theta^-_\alpha 
	\partial^p V^{--}  \right)
	V^{--}.
	\end{split}
\end{equation}	
\end{subequations}
After that we obtain:
\begin{equation}\label{eq: arb spin ferm}
	\boxed{ 
		\begin{split}
	\delta V^{++} = &- 16 i^s (\mathcal{D}^{+})^4 \Bigg\{ \sum_{p=0}^{s-2} \binom{s-1}{p} \binom{s-1}{p+1} 
	\partial^p 
	\left(  \lambda^{+\alpha(s-1)\dot{\alpha}(s-2)}   \theta^+_\alpha  (\bar{\theta}^-)^2 
	\partial^{s-p-2} V^{--}	
	\right)
	\\
	&-2 
	 \sum_{p=0}^{s-3}   \frac{1}{4i} \binom{s-1}{p+2} \binom{s-2}{p}\; 	
	\partial^p 
	\left(  \mathcal{D}^-_\alpha \lambda^{+\alpha(s-1)\dot{\alpha}(s-2)}   \theta^+_\alpha  \bar{\theta}^-_{\dot{\alpha}} 
	\partial^{s-p-3} V^{--}	
	\right)
	\\&
	+ (-1)^s 
	\sum_{p=0}^{s-3}
	\frac{1}{4i} \binom{s-1}{p+2} \binom{s-2}{p} \frac{s-1}{s-2}
	\partial^{s-p-3} \left( 	\lambda^{+\alpha(s-1)\dot{\alpha}(s-2)}
	\bar{\theta}^+_{\dot{\alpha}} \theta^-_\alpha 
	\partial^-_\alpha 
	\partial^p V^{--}  \right)
	\\
	&
	- (-1)^s 
	\sum_{p=0}^{s-3}
		\frac{1}{4i}
\binom{s-1}{p+2} \binom{s-2}{p} \frac{s-1}{s-2}
	\partial^{s-p-3} 	\partial^-_\alpha  \left( 	\lambda^{+\alpha(s-1)\dot{\alpha}(s-2)}
	\bar{\theta}^+_{\dot{\alpha}} \theta^-_\alpha 
	\partial^p V^{--}  \right) \Bigg\}.
	\end{split} 
}
\end{equation}	

In the rigid case, the transformation simplifies to
\begin{equation}
	\begin{split}
	\delta_{rig} V^{++}
	=
	-8 i^s \lambda^{+\alpha(s-1)\dot{\alpha}(s-2)} \Bigg[ 2\cdot
	&\binom{2s-2}{s-2} \theta^+_\alpha (\bar{\theta}^-)^2 \partial^{s-2} V^{--}
	\\&
	- i (-1)^s \frac{s-1}{s-2} 
	\binom{2s-3}{s-3}
	\bar{\theta}^+_{\dot{\alpha}} \theta^-_\alpha \partial^-_\alpha \partial^{s-3} V^{--}
	\Bigg],
	\end{split}
\end{equation}	
and combining with contribution \eqref{eq: arb spin bos} with parameter $\lambda^{\alpha(s-1)\dot{\alpha}(s-1)} = -4i \epsilon^{\alpha(s-1)(\dot{\alpha}(s-2)-} \bar{\theta}^{+\dot{\alpha})}$ one deduces the higher-spin supersymmetry transformation of type:
\begin{equation}
	\delta_\epsilon A_{\beta\dot{\beta}} \sim \epsilon_{\beta}^{\alpha(s-2)\dot{\alpha}(s-2) i} \partial^{s-2}_{\alpha(s-2)(\dot{\alpha}(s-2)} \bar{\psi}_{\dot{\beta})i}. 
\end{equation}	
The obtained higher-spin supersymmetry transformation correctly reduces to the standard $\mathcal{N}=2$ supersymmetry transformation in the $s=2$ case.

\medskip

To summarize, the cubic $(\mathbf s,\mathbf1,\mathbf1)$ couplings induce two qualitatively distinct classes of higher-spin transformations.
For even spins, they generate higher-spin extensions of translational symmetries, whereas odd spins produce parity-odd transformations involving the dual Maxwell field strength.
The fermionic parameters similarly induce higher-spin extensions of $\mathcal N=2$ supersymmetry transformations, reducing to the ordinary $\mathcal N=2$ supersymmetry transformations at $s=2$.

\newpage
\section{Component structure on Bel--Robinson diagonal}
\label{sec: 6}

All component currents contained in the $\mathcal N=2$ principal supercurrent are conserved as a consequence of \eqref{eq: principle supercurrent ST conservation}.
 In this section, we analyze how these component currents appear in the corresponding component cubic vertices. 
 We explicitly derive several component higher-spin currents and analyze the component structure of the corresponding superfield cubic couplings on the $\mathcal{N}=2$ Bel-Robinson diagonal, i.e., for vertices $(\mathbf{s}= 2\mathbf{s_2}, \mathbf{s_2}, \mathbf{s_2})$. We focus on the couplings of the spin-$s$ component field with various component fields from the spin-$s_2$ supermultiplet.
The remaining sectors can be analyzed in a completely analogous way. Finally, we illustrate the general construction on the example of the $(\mathbf 2,\mathbf 1,\mathbf 1)$ vertex.

The analytic form of cubic vertex \eqref{eq: int analytic} is more useful for the exploration of field content due to the particularly simple form of the Wess-Zumino-type gauge.
In Wess-Zumino gauge \eqref{eq: WZ gauge} in the \underline{spin-$s$ sector} there are contributions from components of analytic prepotentials:
\begin{equation}
	\begin{split}
	&h^{++\alpha(s-1)\dot{\alpha}(s-1)} = -4i \theta^{+}_{\beta}\bar{\theta}^{+}_{\dot{\beta}}  \Phi_{}^{(\alpha(s-1)\beta)(\dot{\alpha}(s-1)\dot{\beta})} 
	-
	4i \left(\frac{s-1}{s} \right)^2 \theta^{+(\alpha} \bar{\theta}^{+(\dot{\alpha}} \Phi^{\alpha(s-2))\dot{\alpha}(s-2))},
\\&
	h^{++\alpha(s-1)\dot{\alpha}(s-2)+} = -2i (\theta^+)^2 \bar{\theta}^{+}_{\dot{\beta}} P^{\dot{\beta}\alpha(s-1)\dot{\alpha}(s-2)},
\\&
	h^{++\alpha(s-2)\dot{\alpha}(s-1)+} = \widetilde{	h^{++\alpha(s-1)\dot{\alpha}(s-2)+}}.
	\end{split}
\end{equation}	
Here $P$ field  is constructed from spin-$s$ fields:
\begin{equation}\label{eq: P fields}
	\begin{split}
		P_{\dot{\beta}\alpha(s-1)\dot{\alpha}(s-2)}
		=&\,
		\frac{s-1}{2} \partial^{\rho\dot{\rho}} \Phi_{(\alpha(s-1)\rho)(\dot{\alpha}(s-2)\dot{\beta}\dot{\rho})} 
		-
		\frac{(s+1)(s-1)^2}{2s^2} 	\partial_{(\alpha(\dot{\beta}} \Phi_{\alpha(s-2))\dot{\alpha}(s-2))}
		\\&
		-
		\epsilon_{\dot{\beta}(\dot{\alpha}} \frac{(s-1)(s-2)}{s^2}  	\partial_{(\alpha}^{\dot{\rho}} 	\Phi_{\alpha(s-2))(\dot{\alpha}(s-3)\dot{\rho}))} .
	\end{split}
\end{equation}	
This expression can be derived in the same way as in the detailed analysis of the spin-$2$ and spin-$3$ cases performed in \cite{Buchbinder:2021ite}.
 The field $P$ plays the role of the trace part of the higher-spin Lorentz connection and is responsible for the appearance of trace contributions in the corresponding conserved currents.

The structure of the principal $\mathcal{N}=2$ supercurrent has the simplest form in the $\mathbf{s}=2\mathbf{s_2}$ case:
\begin{equation}
	\mathcal{J}_{\alpha(2s_2-2)\dot{\alpha}(2s_2-2)} = \mathcal{W}_{\alpha(2s_2-2)} \bar{\mathcal{W}}_{\dot{\alpha}(2s_2-2)}.
\end{equation}
Due to this factorized form, the analysis of the component reduction considerably simplifies. In particular, different spin sectors become separated already at the level of the supercurrent expansion.
We begin with the bosonic sector and afterwards discuss fermionic contributions.

\subsection{Bosonic sector}

At the component level, the \textit{bosonic sector} of the principal supercurrent
$\mathcal J_{\alpha(2s_2-2)\dot\alpha(2s_2-2)}$
takes the form\footnote{See eq. \eqref{eq: W on-shell} for  the component content of higher-spin  $\mathcal{N}=2$ Weyl supertensor $\mathcal{W}_{\alpha(2s_2-2)}$. Here we also present all equations up to the unessential numerical coefficients.}:
\begin{equation}\label{eq: BR diag current}
	\begin{split}
	\mathcal{J}_{\alpha(2s_2-2)\dot{\alpha}(2s_2-2)} \sim\; & \;\; C_{\alpha(2s_2-2)} \bar{C}_{\dot{\alpha}(2s_2-2)}
	\\
	& + 4i \theta^{+\beta} \bar{\theta}^{-\dot{\beta}} C_{\alpha(2s_2-2)} \partial_{\beta\dot{\beta}} \bar{C}_{\dot{\alpha}(2s_2-2)}
		\\
	& + 4i \theta^{-\beta} \bar{\theta}^{+\dot{\beta}}  \partial_{\beta\dot{\beta}} C_{\alpha(2s_2-2)} \bar{C}_{\dot{\alpha}(2s_2-2)}
	\\ 
	&+ \theta^{+\beta}\theta^{-\gamma} \bar{\theta}^{+\dot{\beta}} \bar{\theta}^{-\dot{\gamma}} C_{(\alpha(2s_2-2)\beta\gamma)} \bar{C}_{(\dot{\alpha}(2s_2-2)\dot{\beta}\dot{\gamma})}
	\\
	& 
	-16 \, \theta^{-\beta} \bar{\theta}^{+\dot{\beta}} \theta^{+\gamma} \bar{\theta}^{-\dot{\gamma}} \partial_{\beta\dot{\beta}} C_{\alpha(2s_2-2)}  \partial_{\gamma\dot{\gamma}} \bar{C}_{\dot{\alpha}(2s_2-2)}.
	\end{split}
\end{equation}	
Here we retain only terms bilinear in the spin-$s_2$ and in the spin-$(s_2-1)$ fields, since the remaining terms do not contribute to the interaction with the spin-$s$ field.
Tensor $C_{\alpha(2s_2)}\bar{C}_{\dot{\alpha}(2s_2)}$ is the higher-spin generalization of the linearized Bel-Robinson tensor. On the component equations of motion these tensors satisfy conservation equations, e.g.,
\begin{equation}\label{eq: BR cons}
	\begin{split}
& \partial^{\alpha\dot{\alpha}} \left(	C_{\alpha(2s_2-2)} \bar{C}_{\dot{\alpha}(2s_2-2)} \right)
	\approx 0,
	\\
	&\partial^{\alpha{\dot{\alpha}}}
	\left( \partial_{\beta\dot{\beta}} C_{\alpha(2s_2-2)}  \partial_{\gamma\dot{\gamma}} \bar{C}_{\dot{\alpha}(2s_2-2)} \right) \approx 0.
	\end{split}
\end{equation}	
These conservation laws are the higher-spin analogs of the conservation of the ordinary Bel--Robinson tensor in the linearized gravity.

\smallskip

\noindent $\bullet$ In the  \underline{spin-$s_2$ sector}, there is only one non-trivial analytic supercurrent:
\begin{equation}
	J^{++}_{\alpha(s-1)\dot{\alpha}(s-1)}
	=
	\frac{1}{4i} \theta^{+\beta} \bar{\theta}^{+\dot{\beta}}  C_{(\alpha(s-1)\beta)}\bar{C}_{(\dot{\alpha}(s-1)\dot{\beta})} ,
\end{equation}	
and one obtains for the $\mathcal{N}=2$ vertex in $(s=2s_2,s_2,s_2)$ sector:
\begin{equation}
	\begin{split}
	S_{(\mathbf{s}, \mathbf{s_2}, \mathbf{s_2})}\Big|_{(s, s_2, s_2)}
	=&
	-
	\frac{1}{4}
	\int d^4x \; \Phi^{\alpha(s)\dot{\alpha}(s)} C_{\alpha(2s_2)} \bar{C}_{\dot{\alpha}(2s_2)}. 
	\end{split}
\end{equation}	
This reproduces the diagonal higher-spin Bel--Robinson coupling. For $s_2=1$, this expression reduces to the standard linearized gravitational coupling to the Maxwell Bel--Robinson tensor, providing a consistency check of the general construction.

\medskip

\noindent $\bullet$ In the \underline{spin-$(s_2-1)$ sector}, there are three analytic supercurrents:
\begin{equation}
	\begin{split}
	&J^{++}_{\alpha(s-1)\dot{\alpha}(s-1)}
	=
	4i\, \theta^{+\beta} \bar{\theta}^{+\dot{\beta}} 
	\,
	\partial_{(\alpha\dot{\beta}} C_{\alpha(2s_2-2))}  \partial_{\beta(\dot{\alpha}} \bar{C}_{\dot{\alpha}(2s_2-2))},
\\
&
	J^+_{\alpha(s-1)\dot{\alpha}(s-2)} = 
	4i \bar{\theta}^{+\dot{\beta}} \partial_{(\alpha\dot{\beta}} C_{\alpha(2s_2-2))} \bar{C}_{\dot{\alpha}(2s_2-2)},
\\
&	\bar{J}^+_{\alpha(s-2)\dot{\alpha}(s-1)} = 
	- 4i \theta^{+\beta} C_{\alpha(2s_2-2)} \partial_{\beta(\dot{\alpha}} \bar{C}_{\dot{\alpha}(2s_2-2)}.
	\end{split}
\end{equation}

In the $(s, s_2-1, s_2-1)$  component sector one finds the contribution:
\begin{equation}
	\begin{split}
		S_{(\mathbf{s}, \mathbf{s_2}, \mathbf{s_2})}\Big|_{(s, s_2-1, s_2-1)}
		=&
		\int d^4x \;
		\left(  \Phi^{\alpha(s)\dot{\alpha}(s)}\,
		J_{\alpha(s)\dot{\alpha}(s)} + \Phi^{\alpha(s-2)\dot{\alpha}(s-2)}\,
		T_{\alpha(s-2)\dot{\alpha}(s-2)}   \right),
	\end{split}
\end{equation}	
where the corresponding component higher-spin current and current trace are given by
\begin{equation}
	\begin{split}
	&J_{\alpha(s)\dot{\alpha}(s)} 
	=
	- 2 \Big[ (s-1) \partial^2_{\alpha(2)\dot{\alpha}(2)} C_{\alpha(2s_2-2)}\bar{C}_{\dot{\alpha}(2s_2-2)} 
\\&\qquad \qquad\qquad\qquad	+
	2(s-2) \partial_{\alpha\dot{\alpha}} C_{\alpha(2s_2-2)} \partial_{\alpha\dot{\alpha}}  \bar{C}_{\dot{\alpha}(2s_2-2)} 
\\&\qquad \qquad\qquad\qquad\quad	+
	(s-1)  C_{\alpha(2s_2-2)} \partial^2_{\alpha(2)\dot{\alpha}(2)} \bar{C}_{\dot{\alpha}(2s_2-2)} \Big],
\\
	&T_{\alpha(s-2)\dot{\alpha}(s-2)}
	=
	8 \frac{s-1}{s^2}
	\partial^{\beta\dot{\beta}} C_{\alpha(2s_2-2)} \cdot \partial_{\beta\dot{\beta}} \bar{C}_{\dot{\alpha}(2s_2-2)}.
	\end{split}
\end{equation}	
They satisfy the conservation equation:
\begin{equation}
	\partial^{\beta\dot{\beta}} J_{(\alpha(s-1)\beta)(\dot{\alpha}(s-1)\dot{\beta})} + \partial_{(\alpha(\dot{\alpha}} T_{\alpha(s-2))\dot{\alpha}(s-2))}
	=
	0.
\end{equation}	
This current provides an example of the higher-spin currents, presented in appendix  \ref{app: conserved gauge-invariant higher-spin currents}. The current trace appears due to terms coming from the $P$-field. Note that, unlike the 
$\mathcal{N}=1$ case (see, e.g., \cite{Buchbinder:2018wzq}), we have no explicit supertrace in the interaction ansatz.

\subsection{Fermionic sector}

The fermionic field content of $\mathcal{N}=2$ principal supercurrent in the \textit{harmonic-independent sector} is given by
\begin{equation}
	\begin{split}
	\mathcal{J}_{\alpha(2s_2-2)\dot{\alpha}(2s_2-2)}
	\sim\;& \left[ \theta^{+\beta}\bar{\theta}^{-\dot{\beta}} - \theta^{-\beta}\bar{\theta}^{+\dot{\beta}}\right]
	\hat{C}^i_{(\beta\alpha(2s_2-2))} \hat{\bar{C}}_{(\dot{\beta}\dot{\alpha}(2s_2-2))i}
	\\& + 4i \theta^{-\beta}\theta^{+\rho}\bar{\theta}^{+\dot{\rho}} \bar{\theta}^{-\dot{\beta}}
	\Big( \hat{C}^i_{(\beta\alpha(2s_2-2))} \partial_{\rho\dot{\rho}} \hat{\bar{C}}_{(\dot{\beta}\dot{\alpha}(2s_2-2))i}
	\\&\qquad\qquad\qquad\qquad\quad-
	\partial_{\rho\dot{\rho}}
	\hat{C}^i_{(\beta\alpha(2s_2-2))}  \hat{\bar{C}}_{(\dot{\beta}\dot{\alpha}(2s_2-2))i} \Big)
	\\&
	+
	(\bar{\theta}^-)^2 \theta^{+\rho} \bar{\theta}^{+\dot{\rho}} \hat{C}^i_{\rho\alpha(2s-2)} \check{C}_{\dot{\rho} \dot{\alpha}(2s-2) i}
\\&	-
	(\theta^-)^2 \theta^{+\rho} \bar{\theta}^{+\dot{\rho}}
	\check{\bar{C}}^i_{\rho\alpha(2s-2)} \hat{\bar{C}}_{\dot{\rho} \dot{\alpha}(2s-2) i}
	\\
	&
	+2i (\theta^-)^2 (\theta^+)^2 \bar{\theta}^{+(\dot{\gamma}} \bar{\theta}^{-\dot{\rho})} 
	\check{\bar{C}}^i_{\rho\alpha(2s-2)} \partial^\rho_{\dot{\gamma}} \hat{\bar{C}}_{\dot{\rho}\dot{\alpha}(2s-2) i}
	\\
	&
	+2i (\bar{\theta}^-)^2 (\bar{\theta}^+)^2 \theta^{+(\gamma} \theta^{-\rho)} 
	\partial_\gamma^{\dot{\rho}}
	\hat{C}^i_{\rho\alpha(2s-2)} \check{C}_{\dot{\rho}\dot{\alpha}(2s-2) i}
	\\&
	 +\left[ (\bar{\theta}^+)^2 (\theta^-)^2 \theta^{+\rho} \bar{\theta}^{-\dot{\rho}}  
	-
	(\theta^+)^2 (\bar{\theta}^-)^2 \theta^{-\rho} \bar{\theta}^{+\dot{\rho}}
	\right]
	\check{\bar{C}}^i_{\rho\alpha(2s-2)}
	\check{C}_{\dot{\rho}\dot{\alpha}(2s-2) i}. 
	\end{split}
\end{equation}	
One can verify that all component contributions satisfy the corresponding conservation law.

According to eq. \eqref{eq: bf J currents}, the on-shell analytic supercurrents in \underline{spin-$(s_2-1/2)$ sector} are given by:
\begin{equation}
	\begin{split}
	&J^{++}_{\alpha(2s_2-1)\dot{\alpha}(2s_2-1)} 
	=
	- 
	\theta^{+\rho}\bar{\theta}^{+\dot{\rho}} 
	\Big( \hat{C}^i_{\alpha(2s_2-1)} \partial_{\rho\dot{\rho}} \hat{\bar{C}}_{\dot{\alpha}(2s_2-1)i}
	-
	\partial_{\rho\dot{\rho}}
	\hat{C}^i_{\alpha(2s_2-1)}  \hat{\bar{C}}_{\dot{\alpha}(2s_2-1)i} \Big), 
\\
&	J^+_{\alpha(2s_2-1)\dot{\alpha}(2s_2-2)} = - \bar{\theta}^{+\dot{\beta}}
	\hat{C}^i_{\alpha(2s_2-1)} \hat{\bar{C}}_{(\dot{\beta}\dot{\alpha}(2s_2-2))i}, 
\\
&	\bar{J}^+_{\alpha(2s_2-2)\dot{\alpha}(2s_2-1)} = 
	- \theta^{+\beta}  	\hat{C}^i_{(\beta\alpha(2s_2-2))} \hat{\bar{C}}_{\dot{\alpha}(2s_2-1)i}.
		\end{split}
\end{equation}	
A remarkable feature of the fermionic sector is that the field strengths $\check C$ completely decouple from the on-shell analytic supercurrents. As a result, the cubic vertices depend only on the gauge-invariant Weyl-like tensors $\hat C$.

Therefore, one obtains the following component interaction:
\begin{equation}
		S_{(\mathbf{s}, \mathbf{s_2}, \mathbf{s_2})}\Big|_{(s, s_2-\frac{1}{2}, s_2-\frac{1}{2})}
		=
		\int d^4x \, \Phi^{\alpha(s)\dot{\alpha}(s)}
		J_{\alpha(s)\dot{\alpha}(s)},
\end{equation}	
where conserved current is given by
\begin{equation}
		J_{\alpha(s)\dot{\alpha}(s)} =  i \left( \hat{C}^{i}_{\alpha(2s_2-1)} \partial_{\alpha\dot{\alpha}} \hat{\bar{C}}_{\dot{\alpha}(2s_2-1)i} - \partial_{\alpha\dot{\alpha}} \hat{C}^{i}_{\alpha(2s_2-1)}  \hat{\bar{C}}_{\dot{\alpha}(2s_2-1)i} \right).
\end{equation}	

\medskip

We  explicitly demonstrated how the superfield interaction reproduces the expected component higher-spin currents and cubic couplings in different spin sectors.
In particular, the spin-$s$ contribution originating from the $P$-fields \eqref{eq: P fields} generates nontrivial trace contributions to the conserved currents in the  $(s, s_2-1, s_2-1)$ sector. In the hypermultiplet case such terms lead to "fake"\, interactions, demonstrated in \cite{Buchbinder:2022vra}.

\smallskip

To summarize, the component reduction of the analytic superfield vertex  reproduces conserved higher-spin currents and cubic interactions in all physical spin sectors of the $\mathcal N=2$ higher-spin multiplet. In particular, the structure of the resulting vertices naturally separates into bosonic spin-$s_2$ and spin-$(s_2-1)$ sectors together with the fermionic spin-$(s_2-\frac12)$ sector.

\subsection*{Example: component content of $(\mathbf{2}, \mathbf{1}, \mathbf{1})$ vertex}

As a nontrivial consistency check of the general construction, let us consider the special case $\mathbf{s} = \mathbf{2}$, $\mathbf{s_2} = \mathbf{1}$. This case corresponds to the linearized interactions of $\mathcal{N}=2$ supergravity with the $\mathcal{N}=2$ vector multiplet. 
At the component level, the spin-$2$ sector decomposes into the $(2,1,1)$, $(2,\frac12,\frac12)$ and $(2,0,0)$ cubic vertices:
\begin{subequations}
\begin{equation}
	S_{(2,1,1)} = - \frac{1}{4} \int d^4x\, \Phi^{\alpha(2)\dot{\alpha}(2)} C_{\alpha(2)}\bar{C}_{\dot{\alpha}(2)},
\end{equation}	
\begin{equation}
	S_{(2, \frac{1}{2}, \frac{1}{2})} =i \int d^4x \,
	\Phi^{\alpha(2)\dot{\alpha}(2)} \left(\psi^i_\alpha \partial_{\alpha\dot{\alpha}} \bar{\psi}_{\dot{\alpha}i} - \partial_{\alpha\dot{\alpha}} \psi^i_\alpha  \bar{\psi}_{\dot{\alpha}i} \right).
\end{equation}	
\begin{equation}
	S_{(2,0,0)}  = \int d^4x \left(\Phi^{\alpha(2)\dot{\alpha}(2)} \cdot (-2) [\partial_{\alpha\dot{\alpha}}\partial_{\alpha\dot{\alpha}}\phi  \bar{\phi} + \phi  \partial_{\alpha\dot{\alpha}}\partial_{\alpha\dot{\alpha}} \bar{\phi} ] + \Phi \cdot 2 \partial_{\alpha\dot{\alpha}} \phi  \partial^{\alpha\dot{\alpha}} \bar{\phi}        \right).
\end{equation}	
\end{subequations}
Here $\Phi^{\alpha(2)\dot{\alpha}(2)}$ and $\Phi$ describe the graviton sector. The tensors $C_{\alpha(2)}$ and $\bar C_{\dot\alpha(2)}$ correspond to the self-dual and anti-self-dual parts of the Maxwell field strength. The fermionic Weyl-like tensor reduces to the spin-$\frac12$ fermion $\psi^i_\alpha$, while the spin-$(s_2-1)$ Weyl tensor reduces to the complex scalar $\phi$.
The resulting conserved spin-2 currents coincide with the standard stress-energy tensors of the physical fields of the $\mathcal{N}=2$ vector multiplet \eqref{eq: N=2 vector}, confirming consistency with ordinary $\mathcal{N}=2$ supergravity couplings.


\section{Discussion}\label{sec: dis}

In this work, we have analyzed Berends--Burgers--van Dam cubic $(\mathbf{s_1}, \mathbf{s_2}, \mathbf{s_2}),$ $\mathbf{s_1}\geq 2\mathbf{s_2}$ interactions for $\mathcal{N}=2$ higher-spin multiplets:
\begin{equation}\label{eq: final vertex}
	S_{int} = \int d^4x\,d^8\theta\, du \; \left( \Psi^{-\alpha(s_1-1)\dot{\alpha}(s_1-2)}\mathcal{D}^+_\alpha + \bar{\Psi}^{-\alpha(s_1-2)\dot{\alpha}(s_1-1)} \bar{\mathcal{D}}^+_{\dot{\alpha}} \right) \mathcal{J}_{\alpha(s_1-2)\dot{\alpha}(s_1-2)},
\end{equation}	
where $\Psi^-_{\alpha(s_1-1)\dot{\alpha}(s_1-2)}$ and $\bar{\Psi}^-_{\alpha(s_1-2)\dot{\alpha}(s_1-1)}$ are the Mezincescu-type $\mathcal{N}=2$ higher-spin prepotentials, while $\mathcal{J}_{\alpha(s_1-2)\dot{\alpha}(s_1-2)}$ is the corresponding  gauge-invariant $\mathcal{N}=2$ principal supercurrent. 

\medskip

The main results can be summarized as follows:

\begin{itemize}
\item 
The abelian superfield $(\mathbf{s_1}, \mathbf{s_2}, \mathbf{s}_2)$ vertex
admits an equivalent analytic representation in terms of
$\mathcal N=2$ analytic supercurrents:
 \begin{equation}\label{eq: int analytic 0}
	\begin{split}
		S^{A}_{int} = \int d\zeta^{(-4)}\, \Big(&
		h^{++\alpha(s-1)\dot{\alpha}(s-1)} J^{++}_{\alpha(s-1)\dot{\alpha}(s-1)}
		+
		h^{++\alpha(s-1)\dot{\alpha}(s-2)+} J^{+}_{\alpha(s-1)\dot{\alpha}(s-2)}
		\\&+
		h^{++\alpha(s-2)\dot{\alpha}(s-1)+} \bar{J}^{+}_{\alpha(s-2)\dot{\alpha}(s-1)}
		+ h^{++\alpha(s-2)\dot{\alpha}(s-2)} J^{++}_{\alpha(s-2)\dot{\alpha}(s-2)} \Big).
	\end{split}
\end{equation}	
Corresponding analytic $J$ supercurrents are given by 
 \begin{equation}
	\begin{split}
		&J_{\alpha(s-1)\dot{\alpha}(s-1)}^{++}
		=
		+i
		(\mathcal{D}^+)^4 \left\{ \theta^-_{(\alpha} \bar{\theta}^-_{(\dot{\alpha}} \mathcal{J}_{\alpha(s-2)) \dot{\alpha}(s-2))}  \right\}
		=
		\; \frac{1}{4i} \mathcal{D}^+_{(\alpha} \bar{\mathcal{D}}^+_{(\dot{\alpha}} \mathcal{J}_{\alpha(s-2))\dot{\alpha}(s-2))}\Big|_{\theta^- = 0} ,
		\\
		&
		J_{\alpha(s-1)\dot{\alpha}(s-2)}^{+}
		=
		-2	(\mathcal{D}^+)^4 \left\{ \theta^-_{(\alpha} (\bar{\theta}^-)^2 \mathcal{J}_{\alpha(s-2)) \dot{\alpha}(s-2)}  \right\}
		=
		\mathcal{D}^+_{(\alpha} \mathcal{J}_{\alpha(s-2))\dot{\alpha}(s-2)}\Big|_{\theta^- = 0},
		\\
		&
		\bar{J}_{\alpha(s-2)\dot{\alpha}(s-1)}^+
		=
		+
		2
		(\mathcal{D}^+)^4 \left\{ \bar{\theta}^-_{(\dot{\alpha}} (\theta^-)^2 \mathcal{J}_{\alpha(s-2) \dot{\alpha}(s-2))}  \right\}
		=
		\bar{\mathcal{D}}^+_{(\dot{\alpha}} \mathcal{J}_{\alpha(s-2)\dot{\alpha}(s-2))}\Big|_{\theta^- = 0},
		\\
		&
		J^{++}_{\alpha(s-2)\dot{\alpha}(s-2)} = 0
	\end{split} 
\end{equation}	
and coincide with the analytic components of the principal $\mathcal{N}=2$ higher-spin
supercurrent that remain non-vanishing on shell.
 
 \item We explicitly derived arbitrary-spin $\mathbf{s}\geq \mathbf{2}$ superfield gauge transformations for $\mathcal{N}=2$ vector multiplet in both analytic (eqs. \eqref{eq: arb spin bos} and \eqref{eq: arb spin ferm}) and non-analytic (eqs. \eqref{eq: var 1} and \eqref{eq: var 2}) forms.  
 The non-analytic formulation is technically more compact, although it obscures the geometric structure of the transformations, in contrast to the analytic ones. A distinctive feature of the analytic gauge transformations is that the transformation law of the analytic prepotential $V^{++}$ depends on $V^{--}$.
 This feature implies the harmonic non-locality and complicates the analysis of the closure of the higher-spin algebra.

 \item Using the analytic form of the BBvD vertex, we investigated the component structure of the interactions \eqref{eq: final vertex} on the Bel-Robinson diagonal ($\mathbf{s_1}=2\mathbf{s_2}$), providing a nontrivial consistency check of the construction. 
 A notable feature is that in the component sector $(s_1, s_2-1, s_2-1)$ the corresponding higher-spin current possesses a non-vanishing trace, whereas in other sectors $(s_1, s_2, s_2)$, $(s_1, s_2-\frac{1}{2}, s_2-\frac{1}{2})$ give rise to traceless higher-spin currents.  Another notable property is the absence of contributions from the spinor field strength $\check{C}^i_{\alpha(2s_1-1)}$.

\end{itemize}

\newpage

\medskip

We conclude by outlining several directions for future investigation.

\medskip

\noindent $\bullet$ \textbf{Maxwell multiplet on $\mathcal{N}=2$ supergravity background}

\smallskip

\noindent Among the $\mathcal{N}=2$ cubic interactions constructed in this paper, there is the well-known vertex $(\mathbf{2}, \mathbf{1}, \mathbf{1})$. This cubic interaction  can be obtained as the cubic part of the full $\mathcal N=2$ vector multiplet action on $\mathcal{N}=2$ supergravity background \cite{18, Galperin:1987em, Galperin:1987ek, Ivanov:2022vwc}:
\begin{equation}\label{eq: EM NL}
	S = \int d^4x d^8 \theta du\; E \; V^{++}V^{--},
\end{equation}	
where $E$ is the full superspace integration measure and $V^{++}$ and $V^{--}$ satisfy the zero-curvature equation on supergravity background:
\begin{equation}
	\left( \mathcal{D}^{++} + \kappa_2\hat{\mathcal{H}}^{++}_{(2)} \right) V^{--} =  	\left( \mathcal{D}^{--} + \kappa_2\hat{\mathcal{H}}^{--}_{(2)} \right) V^{++}. 
\end{equation}	
Unfortunately, the straightforward linearization of the action \eqref{eq: EM NL} is a technically involved procedure that does not allow for a straightforward definition of $\mathcal{N}=2$ supercurrents.
It would be interesting to reproduce the obtained vertex $(\mathbf{2}, \mathbf{1}, \mathbf{1})$ and corresponding supercurrent \eqref{eq: spin 1 supercurrent}  in this approach.

One may conjecture that an arbitrary $(\mathbf{s_1}, \mathbf{s_2}, \mathbf{s_2})$ vertex  ($\mathbf{s}_1\geq 2\mathbf{s}_2$) can be rewritten in similar form. For example, the vertex $(\mathbf{s},\mathbf{1},\mathbf{1})$ may admit the structure:
 \begin{equation}
 	S_{(\mathbf{s}, \mathbf{1}, \mathbf{1})} = \int d^4xd^8\theta du \; E_{(s)} V^{++} V^{--},
 \end{equation}	
 where $E_{(s)}$ is a spin-$s$ superspace integration measure and $V^{--}$ is defined as the solution of the zero-curvature equation on higher-spin background:
 \begin{equation}
 	\left( \mathcal{D}^{++} + \kappa_s \hat{\mathcal{H}}^{++}_{(s)} \right) V^{--} =  	\left( \mathcal{D}^{--} + \kappa_s \hat{\mathcal{H}}^{--}_{(s)} \right) V^{++}. 
 \end{equation}	
 Moreover, this assumption can provide useful insight into the structure of the quartic (or higher order) higher-spin vertices.

\medskip

\noindent $\bullet$ \textbf{More general $\mathcal{N}=2$ higher-spin vertices}

\smallskip

\noindent A natural problem is to construct more general $\mathcal{N}=2$ supercurrents that correspond to other abelian interactions. A more difficult and, arguably, more important task is the construction of $\mathcal{N}=2$ non‑abelian vertices. We expect the appearance of harmonic non‑localities, as in the cubic vertex of Yang-Mills theory \cite{Zupnik:1987vm}:
\begin{equation}
	S_{(\mathbf{1},\mathbf{1}, \mathbf{1} )} 
	\sim
	\int d^4x d^8\theta du du_1 du_2
	\frac{V^{++}(u) V^{++}(u_1) V^{++}(u_2) }{ (u^+u_1^+)(u_1^+u_2^+)(u_2^+ u^+)}. 
\end{equation}
This example suggests that one can expect that $\mathcal{N}=2$ superfield non-abelian vertices will have a substantially simpler structure than their component counterparts.

\medskip

\noindent $\bullet$ \textbf{Supersymmetric Born-Infeld cubic vertices}

\smallskip

\noindent Direct attempts to supersymmetrize type-I  Born-Infeld  interactions \eqref{eq: type 1} for  highest spin $s$ from spin-$\mathbf{s}$ supermultiplet do not appear to produce the desired component structures. In particular, using chiral higher-spin Weyl supertensors \eqref{eq: SF strenghts} one cannot obtain component terms of the desired BI cubic form. For example, the Born-Infeld vertex $(\mathbf{2},\mathbf{2},\mathbf{2})$\footnote{See also discussion in \cite{Ivanov:2024gjo}.}
\begin{equation}
	\begin{split}
	S_{(\mathbf{2},\mathbf{2},\mathbf{2})}
	=
&	\int d^4x d^4\theta\; \mathcal{W}_{\alpha}^{\;\;\beta} 
	\mathcal{W}_{\beta}^{\;\;\gamma} 
	\mathcal{W}_{\gamma}^{\;\;\alpha} 
\\&	\sim
	\int d^4x\; \left( C^{\alpha\beta\gamma\delta} C_{\alpha\beta\gamma\rho} C^\rho_{\delta} 
	+
C^{\alpha\beta\gamma\delta} \hat{C}^i_{\alpha\beta\rho} \hat{C}^\rho_{\gamma\delta i}
+
\hat{C}^{\alpha\beta\gamma i} \check{\bar{C}}_{\alpha\beta \delta i} C_\gamma^\delta
	\right) 
	\end{split}
\end{equation}	
contains component interactions of the types $(2,2,1)$, $(2,\frac{3}{2},\frac{3}{2})$ and $(1,\frac{3}{2},\frac{3}{2})$  Born-Infeld-type vertices, but does not contain the Born-Infeld vertex $(2,2,2)$.
 This phenomenon is reminiscent of the absence of the cubic  supersymmetric invariant $C_{\alpha\beta}^{\quad\gamma\delta} C_{\gamma\delta}^{\quad\rho\sigma} C_{\rho\sigma}^{\quad\alpha\beta}$  in supergravity \cite{Ferrara:1977mv}, where it corresponds to the two‑loop finiteness of supergravity.  
 
This observation also appears closely related  to the Metsaev classification of $\mathcal{N}=1$  higher-spin cubic interactions  \cite{Metsaev:2019dqt}. According to this classification, using only integer $(s,s-\frac{1}{2})$ or half-integer $(s+\frac{1}{2},s)$  multiplets, it is impossible to construct vertices containing the maximum number of derivatives. However, such interactions can be constructed using both types of multiplets simultaneously. This suggests that such vertices may exist only for $\mathcal{N}=2$ multiplets with the highest half-integer spin.
In terms of $\mathcal{N}=2$ supersymmetry, this implies that such vertices exist only for $\mathcal{N}=2$ multiplets having the highest half-integer spin $(s+\frac{1}{2}, 2\times s, s-\frac{1}{2} )$. At present, the $\mathcal{N}=2$ superfield description of such supermultiplets is unknown\footnote{In the paper \cite{Ivanov:2024bsb}, a theory of the $\mathcal{N}=2$ conformal gravitino in harmonic superspace is constructed, which can be generalized to an arbitrary spin value. However, the construction of non-conformal multiplets via conformal compensators is unknown.}.


\medskip

\noindent $\bullet$ \textbf{Fradkin-Vasiliev mechanism in AdS}

\smallskip

\noindent Famously, higher-spin fields on (A)dS background admit quasi-minimal cubic coupling \cite{Fradkin:1986qy, Fradkin:1987ks}.  It consists of the minimal coupling plus a finite tail of non-minimal cubic vertices dressed by the cosmological constant.  Therefore, it is natural to study how  the Fradkin-Vasiliev mechanism is implemented in harmonic superspace language.  In view of recent progress in the description of 
$\mathcal{N}=2$ harmonic AdS$_4$ superspace \cite{Ivanov:2025jdp, Gargett:2025xcg}, this problem appears particularly interesting.

\medskip

\noindent $\bullet$ \textbf{Unfolded dynamics approach}

\smallskip

\noindent Recently,  Misuna constructed an unfolded system that describes an on-shell free massless hypermultiplet \cite{Misuna:2026bhy}. The unfolded dynamics approach \cite{Vasiliev:2005zu} is an efficient framework
 for constructing consistent interactions of massless higher-spin fields (see, e.g., for the recent developments  \cite{Didenko:2023vna, Didenko:2024zpd}). We hope that extending these results to $\mathcal{N}=2$ higher-spin harmonic superfields may lead to an efficient approach for constructing and analyzing $\mathcal{N}=2$ superfield higher-spin interactions.

\medskip

Our obtained results indicate that harmonic superspace
provides a natural framework for the description of
$\mathcal N=2$ higher-spin interactions,
combining manifest supersymmetry with a  compact
superfield structure.
We expect that this approach might be particularly useful
for the systematic study of non-abelian interactions
and AdS deformations.

\acknowledgments

The author is grateful to Yu. Zinoviev for discussions of cubic vertices, to E. Ivanov for numerous fruitful discussions and continued support, and to the anonymous JHEP referee for questions and comments that significantly improved the text. Special thanks go to Polina Petriakova for her help with the text preparation.
This work  was  supported by the grant  \verb|#| 25-1-1-10-4 from
the Foundation for the Advancement of Theoretical Physics and
Mathematics ``BASIS''.

\appendix

\section{Higher-spin fields in the spinor notation: free lagrangians and field strengths}\label{app A}

In this appendix, we review the spinor formulation of the Fronsdal and Fang--Fronsdal higher-spin theories employed throughout this paper. We use the standard spinor notation, e.g.,  $$\alpha(s):= (\alpha_1\dots\alpha_s),\qquad x_{\alpha\dot{\alpha}}:= \sigma^m_{\alpha\dot{\alpha}}x_m \qquad   
\partial_{\alpha\dot{\alpha}} := \frac{1}{2}\sigma^m_{\alpha\dot{\alpha}}\partial_m,
\qquad
 \partial_{\alpha\dot{\beta}} \partial^{\dot{\beta}\beta} 
 = \frac{1}{4} \delta_\alpha^\beta \Box.$$ 
Our main goal is to introduce the field strengths that are the basic building blocks for gauge-invariant higher-spin currents; see, e.g., \eqref{eq: N=0 current}. A discussion of the relation between the spinor formalism and the vector one can be found, for example, in chapter 6.9 of \cite{BK}. 

This appendix is organized as follows. First, we review the Fronsdal formulation for integer spins and introduce the corresponding gauge-invariant curvatures and Weyl tensors. Further, we  discuss the analogous Fang--Fronsdal formulation for half-integer spins. Finally, we explain why only the Weyl-like curvatures contribute to nontrivial cubic interactions.

\smallskip

\noindent \textit{\underline{Integer higher spin}} $s$ is described by the bosonic fields:
\begin{equation}
	\left\{\Phi^{\alpha(s)\dot{\alpha}(s)}, \Phi^{\alpha(s-2)\dot{\alpha}(s-2)} \right\},
\end{equation}	
defined up to the gauge transformations:
\begin{equation}\label{eq: spin s gauge}
	\delta_\xi  \Phi^{\alpha(s)\dot{\alpha}(s)}=  \partial^{(\alpha(\dot{\alpha}} \xi^{\alpha(s-1))\dot{\alpha}(s-1))},
	\qquad
		\delta_\xi  \Phi^{\alpha(s-2)\dot{\alpha}(s-2)}=  \partial_{\alpha\dot{\alpha}} \xi^{\alpha(s-1)\dot{\alpha}(s-1)}.
\end{equation}	

The Fronsdal gauge-invariant action can be written in the form:
\begin{equation}
	S_{(s)} = (-1)^s \int d^4x \left( \Phi^{\alpha(s)\dot{\alpha}(s)} \mathcal{R}_{\alpha(s)\dot{\alpha}(s)} - \frac{(s-1)^2}{s^2} \frac{2s-1}{s} \Phi^{\alpha(s-2)\dot{\alpha}(s-2)} \mathcal{R}_{\alpha(s-2)\dot{\alpha}(s-2)}  \right). 
\end{equation}	
Here, we introduce the linearized gauge-invariant higher-spin curvatures: 
\begin{subequations}
\begin{equation}\label{eq: curv 1}
	\begin{split}
	\mathcal{R}_{\alpha(s)\dot{\alpha}(s)}
	:=\;&
	\Box \Phi_{\alpha(s)\dot{\alpha}(s)} -2s\, \partial_{\alpha\dot{\alpha}} \partial^{\beta\dot{\beta}}  \Phi_{(\alpha(s-1)\beta)(\dot{\alpha}(s-1)\dot{\beta})} 
	\\&
	+
	\frac{2(s-1)^2}{s} \partial_{\alpha\dot{\alpha}}  \partial_{\alpha\dot{\alpha}} \Phi_{\alpha(s-2)\dot{\alpha}(s-2)},
	\end{split}
\end{equation}	
\begin{equation}\label{eq: curv 2}
	\begin{split}
	\mathcal{R}_{\alpha(s-2)\dot{\alpha}(s-2)}
		:=\;&
		\Box \Phi_{\alpha(s-2)\dot{\alpha}(s-2)}
		+
		\frac{2(s-2)^2}{2s-1} \partial_{\alpha\dot{\alpha}} \partial^{\beta\dot{\beta}}\Phi_{(\alpha(s-3)\beta)(\dot{\alpha}(s-3)\dot{\beta})}
		\\&
		-
		\frac{2s^2}{2s-1}  
		\partial^{\beta\dot{\beta}}\partial^{\gamma\dot{\gamma}} \Phi_{(\alpha(s-2)\beta\gamma)(\dot{\alpha}(s-2)\dot{\beta}\dot{\gamma})}.
		\end{split}
\end{equation}
\end{subequations}
These curvatures generalize the irreducible components of the linearized Ricci tensor and scalar curvature of the spin-2 field, and correspond to Fronsdal--de Wit--Freedman higher-spin curvatures~\cite{deWit:1979sib}  presented in the spinor notation.
The equations of motion for spin-$s$ theory can be written in terms of the linearized curvatures:
\begin{equation}\label{eom s}
	\begin{cases}
			\mathcal{R}_{\alpha(s)\dot{\alpha}(s)} \approx 0,
			\\
				\mathcal{R}_{\alpha(s-2)\dot{\alpha}(s-2)} \approx 0.
	\end{cases}	
\end{equation}	

In addition to the Fronsdal curvatures, one can construct the gauge-invariant higher-spin Weyl tensor: 
\begin{equation}\label{eq: bos C}
	C_{(\alpha_1\dots\alpha_{2s})} := \partial_{(\alpha_1}^{\;\;\dot{\beta}_1} \dots \partial_{\alpha_s}^{\;\;\dot{\beta}_s} \Phi_{\alpha_{s+1}\dots \alpha_{2s})(\dot{\beta}_1 \dots \dot{\beta}_s)},
\end{equation}	
which generalizes the linearized self-dual Weyl tensor to higher-spin case. For spin $s$, the tensor contains $s$ derivatives. This tensor satisfies the off-shell equality:
\begin{equation}
	\partial_{\dot{\alpha}}^{\;\;\alpha_{2s}} 	C_{(\alpha_1\dots\alpha_{2s})}
	=
	- \frac{1}{4} \partial_{(\alpha_1}^{\;\;\dot{\beta}_1} \dots \partial_{\alpha_{s-1}}^{\;\;\dot{\beta}_{s-1}} \mathcal{R}_{\alpha_{s}\dots \alpha_{2s-1})(\dot{\beta}_1 \dots \dot{\beta}_{s-1}\dot{\alpha})}.
\end{equation} 
It follows that, on the equations of motion, the higher-spin Weyl tensor does not vanish, but instead satisfies the following equation: 
\begin{equation}\label{eq: on-shell bos}
	\partial_{\dot{\alpha}}^{\;\;\alpha_1} 	C_{(\alpha_1\dots\alpha_{2s})}
	  \approx 0
	\qquad
	\Rightarrow
	\qquad
	\Box C_{(\alpha_1\dots\alpha_{2s})}  \approx 0 .
\end{equation} 
Thus, the higher-spin Weyl tensor describes the propagating on-shell degrees of freedom.

\smallskip

\noindent \textit{\underline{Half-integer higher spin}} $s-\frac{1}{2}$ is described by the complex spinor fields:
\begin{equation}
	\left\{ \psi^{\alpha(s)\dot{\alpha}(s-1)}, \psi^{\alpha(s-2)\dot{\alpha}(s-1)}, \psi^{\alpha(s-2)\dot{\alpha}(s-3)} \right\}
\end{equation}	
with gauge transformations:
\begin{equation}
	\begin{split}
\delta \psi_{\alpha(s)\dot{\alpha}(s-1)} &=  \partial_{\alpha\dot{\alpha}} \xi_{\alpha(s-1)\dot{\alpha}(s-2)},
\\
\delta \psi_{\alpha(s-2)\dot{\alpha}(s-1)} &=  \partial_{\dot{\alpha}}^\alpha \xi_{\alpha(s-1)\dot{\alpha}(s-2)}, 
\\
\delta \psi_{\alpha(s-2) \dot{\alpha}(s-3)} &=
\partial^{\alpha\dot{\alpha}}\xi_{\alpha(s-1)\dot{\alpha}(s-2)}. 
	\end{split}
\end{equation}

The Fang-Fronsdal gauge-invariant action is given by:
\begin{equation}
	\begin{split}
		S_{(s-\frac{1}{2})}
		=
		i(-1)^s \int d^4x \; \Bigr(& \bar{\psi}^{\alpha(s-1)\dot{\alpha}(s)} \mathcal{R}_{\alpha(s-1)\dot{\alpha}(s)}
		+
		\frac{s-1}{s}
		\bar{\psi}^{\alpha(s-1)\dot{\alpha}(s-2)} \mathcal{R}_{\alpha(s-1)\dot{\alpha}(s-2)}
		\\
		&
		+
		\frac{s-2}{s-1}
		\bar{\psi}^{\alpha(s-3)\dot{\alpha}(s-2)}
		\mathcal{R}_{\alpha(s-3)\dot{\alpha}(s-2)}
		\Bigr),
	\end{split}
\end{equation}
where the fermionic higher-spin curvatures are defined by (we assume full symmetrization over spinor indices): 
\begin{subequations}
	\begin{equation}
		\mathcal{R}_{\alpha(s-1)\dot{\alpha}(s)} := \partial_{\dot{\alpha}}^\alpha \psi_{\alpha(s)\dot{\alpha}(s-1)} - \frac{s-1}{s} \partial_{\alpha\dot{\alpha}} \psi_{\alpha(s-2)\dot{\alpha}(s-1)},
	\end{equation}
	\begin{equation}
		\begin{split}
		\mathcal{R}_{\alpha(s-1)\dot{\alpha}(s-2)} :=&\;
		 \partial^{\alpha\dot{\alpha}} \psi_{\alpha(s)\dot{\alpha}(s-1)}
		+
		\frac{2s-1}{s(s-1)}
		\partial_\alpha^{\dot{\alpha}} \psi_{\alpha(s-2)\dot{\alpha}(s-1)}
		\\
		&
		-
		\frac{s(s-2)}{(s-1)^2} \partial_{\alpha\dot{\alpha}} \psi_{\alpha(s-2)\dot{\alpha}(s-3)},
		\end{split}
	\end{equation}
	\begin{equation}
		\mathcal{R}_{\alpha(s-3)\dot{\alpha}(s-2)}
		:=
		\partial^{\alpha\dot{\alpha}} \psi_{\alpha(s-2)\dot{\alpha}(s-1)}
		-
		\frac{s-2}{s-1} \partial_{\dot{\alpha}}^\alpha \psi_{\alpha(s-2)\dot{\alpha}(s-3)}.
	\end{equation}
\end{subequations} 
These curvatures satisfy the differential identity:
\begin{equation}
	\partial^{\alpha\dot{\alpha}}  \mathcal{R}_{\alpha(s-1)\dot{\alpha}(s)}
	-
	\frac{s-1}{s}
	\partial^{\;\alpha}_{\dot{\alpha}} \mathcal{R}_{\alpha(s-1)\dot{\alpha}(s-2)}
	+ 
	\frac{s-2}{s-1}
	\partial_{\alpha\dot{\alpha}}  \mathcal{R}_{\alpha(s-3)\dot{\alpha}(s-2)} = 0.
\end{equation}

The equations of motion for spin $s-\frac{1}{2}$ are given by:
\begin{equation}\label{eom s-1/2}
	\begin{cases}
			\mathcal{R}_{\alpha(s-1)\dot{\alpha}(s)}  \approx 0,
			\\
				\mathcal{R}_{\alpha(s-1)\dot{\alpha}(s-2)} \approx 0,
				\\  
					\mathcal{R}_{\alpha(s-3)\dot{\alpha}(s-2)} \approx 0.
	\end{cases}	
\end{equation}	

In analogy with the bosonic case, one can  introduce two complex gauge-invariant Weyl-like tensors, which do not vanish on the equations of motions:
\begin{equation}\label{eq: ferm C}
	\begin{split}
	&\hat{C}_{(\alpha_1\dots \alpha_{2s-1})} := \partial_{(\alpha_1}^{\;\;\dot{\beta}_1} \dots \partial_{\alpha_{s-1}}^{\;\;\dot{\beta}_{s-1}} \psi_{\alpha_s \dots \alpha_{2s-1})(\dot{\beta}_1\dots\dot{\beta}_{s-1})},
	\\
	&\check{C}_{(\dot{\alpha}_1\ldots \dot{\alpha}_{2s-1})}
	:=
	\partial_{(\dot{\alpha}_1}^{\;\;\beta_1} \dots \partial^{\;\;\beta_s}_{\dot{\alpha}_s}
	\psi_{(\beta_1\dots\beta_s) \dot{\alpha}_{s+1}\ldots \dot{\alpha}_{2s-1})},
	\end{split}
\end{equation}	
which satisfy on-shell conditions analogous to the bosonic ones \eqref{eq: on-shell bos}:
\begin{equation}
	\begin{split}
	\partial_{\dot{\alpha}}^{\alpha_1} \hat{C}_{(\alpha_1\dots\alpha_{2s-1})}
	=&\,
	\frac{1}{s} (\partial^{s-1})^{\dot{\beta}(s-1)}_{\alpha(s-1)} \mathcal{R}_{\alpha(s-1) (\dot{\beta}(s-1)\dot{\alpha})}
	\\&+
	\frac{(s-1)^2}{s(2s-1)} 
	\partial_{(\alpha\dot{\alpha}} 
	(\partial^{s-2})^{\dot{\beta}(s-2)}_{\alpha(s-2)} 
	\mathcal{R}_{\alpha(s-1))\dot{\beta}(s-2)},
	\end{split}
\end{equation}	
\begin{equation}
	\begin{split}
	\partial_{\alpha}^{\dot{\alpha}_1} \check{C}_{(\dot{\alpha}_1\ldots \dot{\alpha}_{2s-1})}
	=
	&-
	\frac{1}{4}  (\partial^{s-2})^{\beta(s-2)}_{\dot{\alpha}(s-2)} \Box \mathcal{R}_{(\alpha \beta(s-2))\dot{\alpha}(s)}
\\&	+
	\frac{s-1}{2s-1} \partial_{\alpha\dot{\alpha}} 
	(\partial^{s-1})^{\beta(s-1)}_{\dot{\alpha}(s-1)} \mathcal{R}_{\beta(s-1)\dot{\alpha}(s-2)}.
	\end{split}
\end{equation}	

On shell, the only nonvanishing gauge-invariant local tensors are the Weyl-like spin-tensors\footnote{More precisely, any local gauge invariant is equivalent on shell to a function of the Weyl tensors and their derivatives, see, e.g., \cite{Bekaert:2005ka}.} \eqref{eq: bos C} and \eqref{eq: ferm C}. 
Therefore, only these tensors can contribute to nontrivial cubic vertices such as \eqref{eq: type 1} and \eqref{eq: type 2}. Vertices constructed solely from the curvatures $\mathcal R$ are proportional to the free equations of motion and can be removed by local field redefinitions. Therefore, such vertices correspond to "fake" interactions.

\section{On the conserved gauge-invariant higher-spin currents}
\label{app: conserved gauge-invariant higher-spin currents}

The goal of this appendix is to classify all conserved gauge-invariant currents constructed from higher-spin Weyl tensors and coupled to a spin-$s$ Fronsdal field.
We show that the conservation equations admit several independent solutions, but only one of them corresponds to a nontrivial cubic interaction, while the remaining ones are "fake" vertices removable by field redefinitions.

\smallskip

We consider the most general current coupling to a spin-$s$ Fronsdal field:
\begin{equation}\label{eq: com coupling}
 S_{int} 
 =
  \int d^4 x \left( \Phi^{\alpha(s)\dot{\alpha}(s)} J_{\alpha(s)\dot{\alpha}(s)} + \Phi^{\alpha(s-2)\dot{\alpha}(s-2)} T_{\alpha(s-2)\dot{\alpha}(s-2)} \right).
\end{equation}
Our aim is to explicitly construct a manifestly gauge-invariant higher-spin current $J$ and the corresponding current trace   $T$ in terms of spin-$s_2$ Fronsdal higher-spin fields.
The requirement of  gauge invariance of the cubic vertex \eqref{eq: com coupling} under \eqref{eq: spin s gauge}  leads to the conservation equation for the current:
\begin{equation}\label{eq: conservation}
	\partial^{\beta\dot{\beta}} J_{(\alpha(s-1)\beta)(\dot{\alpha}(s-1)\dot{\beta})} + \partial_{(\alpha(\dot{\alpha}} T_{\alpha(s-2))\dot{\alpha}(s-2))}
	\approx
	0.
\end{equation}

The most general gauge-invariant ansatz, bilinear in the Weyl-like spin-$s_2$ tensor $C_{\alpha(2s_2)}$ with the minimal number of space-time derivatives compatible with gauge invariance and index structure for $J_{\alpha(s)\dot{\alpha}(s)}$ is given by:
\begin{equation}\label{eq: J}
	J_{\alpha(s)\dot{\alpha}(s)}
	=
	\sum_{p=0}^{s-2s_2} 
	\alpha_p \; \partial^p C_{\alpha(2s_2)} \partial^{s-2s_2-p} \bar{C}_{\dot{\alpha}(2s_2)},
	\qquad
	\alpha_p^* = \alpha_{s-2s_2-p}.
\end{equation}	  
Here, we assume symmetrization over all dotted and undotted spinor indices. Derivatives are distributed over the Weyl tensors in all inequivalent completely symmetrized ways.

For the current trace $T_{\alpha(s-2)\dot{\alpha}(s-2)}$ we must include the same number of derivatives as in $J$, so the general ansatz has the form:
\begin{equation}\label{eq: T}
	\begin{split}
	T_{\alpha(s-2)\dot{\alpha}(s-2)} = \sum^{s-2s_2-2}_{p=0} \beta_p& \; (\partial^{p+1} C)_{(\alpha(2s_2+p)\beta)(\dot{\alpha}(p)\dot{\beta})} 
	\\
	&\times (\partial^{s-2s_2-p-1} \bar{C})^{\beta\dot{\beta}}_{\alpha(s-2s_2-p-2)\dot{\alpha}(s-p-2)} .
	\end{split}
\end{equation}	
In both factors, we assume full symmetrization over the dotted and undotted spinor indices, otherwise these terms vanish on-shell due to \eqref{eq: on-shell bos}.
The reality condition $T^*_{\dots} = T_{\dots}$ implies the relation:
\begin{equation}\label{eq: relity beta}
	\beta^*_p = \beta_{s-2s_2-p-2}.
\end{equation}	
Let us note that a nontrivial current trace $T$ exists only for $s\geq 2s_2+2$.

\medskip 

In total, the general ansatz contains  $s-2s_2+1$  independent real $\alpha$-parameters and $s-2s_2-1$  independent real $\beta$-parameters.
Conservation equation \eqref{eq: conservation}, after substitution of ansatz \eqref{eq: J} and \eqref{eq: T}  leads to a system of  recurrence equations ($p=0,\dots,s-2s_2-1$):
\begin{equation}\label{eq: alpha and beta}
	\alpha_p \cdot  \frac{s-2s_2-p}{s} \cdot  \frac{s-p}{s} + \alpha_{p+1} \cdot  \frac{2s_2+p+1}{s} \cdot  \frac{p+1}{s}
	+
	\beta_p + \beta_{p-1} = 0.
\end{equation}	
In this equation, we impose the boundary conditions $\alpha_p=0$ for $p<0$ or $p>N$, and $\beta_p=0$ for $p<0$
or  $p>N-2$.

The recurrence relations do not determine all coefficients uniquely.
Different choices of the trace coefficients $\beta_p$ lead to different conserved currents. 

\smallskip 

\noindent $\bullet$ Setting all trace coefficients to zero,
$\beta_q = 0$ ($q=0,\dots , s-2s_2-2$), one obtains the unique traceless conserved current:
\begin{equation}\label{eq: hom spl}
		\alpha^{(0)}_p = c  (i)^{s-2s_2} (-1)^p \frac{\binom{s}{2s_2+p}\binom{s}{p}}{\binom{s}{2s_2}},
		\qquad
		p=0,\dots s-2s_2,
		\qquad
		c\in \mathbb{R}.
\end{equation}	
The resulting traceless conserved current takes the form ($\partial^{\alpha\dot{\alpha}} J_{\alpha(s)\dot{\alpha}(s)} \approx 0$)\footnote{In components, $\mathcal{N}=1$ higher-spin supercurrents, constructed in \cite{Buchbinder:2018wzq}, contain higher-spin current \eqref{eq: N=0 current} (see eq. 22a). It coincides with the current \eqref{eq: N=0 current} due to the relation: $$\frac{\binom{s_1}{2s_2+p}\binom{s_1}{p}}{\binom{s_1}{2s_2}} = \frac{\binom{s_1-2s_2}{p}\binom{s_1}{p}}{\binom{2s_2+p}{2s_2}}.$$} is given by
\begin{equation}\label{eq: N=0 current}
	\boxed{J_{\alpha(s)\dot{\alpha}(s)} = c (i)^{s-2s_2} \sum_{p=0}^{s-2s_2} (-1)^p \frac{\binom{s}{2s_2+p}\binom{s}{p}}{\binom{s}{2s_2}} \partial^p C_{\alpha(2s_2)} \partial^{s-2s_2-p} \bar{C}_{\dot{\alpha}(2s_2)}.}
\end{equation}	
This current is the unique traceless conserved spin-$s$ current constructed from the spin-$s_2$ Weyl tensors with the minimal number of derivatives.

\smallskip

The remaining solutions necessarily involve nonvanishing current traces. As we discuss below, currents with nonvanishing traces are closely related to ``fake'' cubic vertices proportional to the free equations of motion.

\smallskip

The general solution of the recurrence relation in the presence of "sources" $\beta_p$ has the form:
\begin{equation}\label{eq: general solution}
	\boxed{ \alpha_p = \alpha_p^{(0)} \left[ \frac{\alpha_0}{\alpha_0^{(0)}} + s^2 \sum_{j=0}^{p-1} \frac{\beta_j + \beta_{j-1}}{(s-2s_2-j)(s-j) \cdot \alpha_j^{(0)}} \right],
	\qquad
	p=1,2,\dots, s-2s_2. }
\end{equation}
Here $\alpha_p^{(0)}$ are  defined by \eqref{eq: hom spl}.
Reallity conditions $\alpha_p^* = \alpha_{s-2s_2-p}$ and $\beta^*_p = \beta_{s-2s_2-p-2}$ give equation on $\alpha_0$:
\begin{equation}\label{eq: alpha 0 cond}
\alpha_0 - (-1)^{s-2s_2} \alpha_0^* = - \alpha_0^{(0)} \cdot s^2 \sum_{j=0}^{s-2s_2-1} \frac{\beta_j + \beta_{j-1}}{(s-2s_2-j)(s-j) \cdot \alpha_j^{(0)}},
	\qquad
\alpha_0^{(0)}
=
c(i)^{s-2s_2}.
\end{equation}
We note an interesting property of the denominator of the resulting sum:
\begin{equation}
	\begin{split}
	D_j:
	&=
	(s-2s_2-j)(s-j) \cdot \alpha_j^{(0)}
	\overset{\eqref{eq: alpha and beta}}{=}
	- (2s_2+j+1) (j+1) \alpha_{j+1}^{(0)}
\\&	=
	-
	(-1)^{s-2s_2}  (2s_2+j+1) (j+1) \alpha_{s-2s_2-j-1}^{(0)}
	=
	-
	(-1)^{s-2s_2} D_{s-2s_2-j-1}.
	\end{split}
\end{equation}

Using the formulas \eqref{eq: general solution}, \eqref{eq: alpha 0 cond} and substituting the given coefficients $\beta_p$ (which determine the current trace) that satisfy the reality conditions \eqref{eq: relity beta}, we can straightforwardly construct conserved currents. So we conclude that the conservation equations for spin-$s$ currents admit $s-2s_2$ linearly independent solutions, constructed from the spin-$s_2$ Weyl-like tensor. However, as we show below, only one of them corresponds to a nontrivial cubic interaction.

\newpage 
Let us provide an  simple subclass of these higher-spin currents.

\smallskip

\noindent  $\bullet$ 
For example, if we fix 
$$
\beta_p= (-1)^p \beta_0 \neq 0,
\qquad
\beta_0 = b_0 i^{s-2s_2} 
\quad\text{with}
\quad
 b_0 \in \mathbb{R}
$$ 
then from eq. \eqref{eq: gen alpha 0} we obtain
\begin{equation}\label{eq: gen alpha 0}
	\begin{split}
	\alpha_0 - (-1)^{s-2s_2} \alpha_0^*
&	=
	-
	\alpha_0^{(0)} \cdot s^2 \left[ \frac{\beta_0}{D_0} + \frac{\beta_{s-2s_2-2}}{D_{s-2s_2-1}} \right]
\\&	=
	-
	\alpha_0^{(0)}
	\frac{s}{s-2s_2}
	\left[ \frac{\beta_0}{\alpha^{(0)}_0} 
	-
	\left(  \frac{\beta_0}{\alpha_0^{(0)} } \right)^* \right]
	=
	0.
	\end{split}
\end{equation}	
So there are two cases:
\begin{equation*}
	\begin{split}
	&\text{even} \qquad  s-2s_2 
	\qquad	\qquad
	\text{Im}\,  \alpha_0 = 0,
	\\
	&\text{odd} \;\;\;\;\;\quad  s-2s_2 
		\qquad	\qquad
	\text{Re}\,  \alpha_0 = 0.
	\end{split}
\end{equation*}	
This implies that we can in both cases fix $\alpha_0 = 0$.
Then the solution \eqref{eq: general solution} simplifies to:
\begin{equation}
	\alpha_p = (-1)^p \beta_0 \cdot s^2 \cdot \frac{(s-2s_2-1)!}{(s-2s_2-p)!} \frac{(s-1)!}{(s-p)!}
	\frac{1}{p!} \frac{(2s_2)!}{(2s_2+p)!}
		=
			(-1)^{p} \beta_0 \frac{\binom{s}{2s_2+p} \binom{s}{p}} {\binom{s-1}{2s_2}}.
 \end{equation}	
 Here $p=1, \dots s-2s_2-1$. 

\medskip

\noindent  $\bullet$  This example admits a fairly simple generalization. The solution of eq. \eqref{eq: alpha and beta} for 
$$\alpha_0 = \alpha_1 = \dots = \alpha_{k} = 0,
\qquad k \in \left\{ 0,1, \dots, \left[\frac{s-2s_2-2}{2}\right]\right\},
\qquad
s - 2s_2 \geq 2$$
$$\beta_0 = \beta_1 = \dots = \beta_{k-1} = 0\quad  \text{and}\quad \beta_{p} = (-1)^{p-k} \beta_k,\qquad p = k, \dots,  s-2s_2-k-2$$
 is given by: 
\begin{multline}
	\alpha^{(k)}_p = (-1)^{p-k} \beta_k \cdot s^2 \cdot  
	\frac{(s-2s_2-k-1)!}{(s-2s_2-p)!} \frac{(s-k-1)!}{(s-p)!}
	\frac{k!}{p!} \frac{(2s_2+k)!}{(2s_2+p)!}
\\	=
	(-1)^{p-k} \beta_k \frac{\binom{s}{2s_2+p} \binom{s}{p}}{\binom{s-1}{k} \binom{s-1}{2s_2+k}}.
\end{multline}
Here $p=k+1, \dots, s-2s_2-k-1$ and $\beta_k = b_k i^{s-2s_2-2k}$ with $b_k\in \mathbb{R}$. 

\smallskip

The corresponding higher-spin currents and current traces for $k \in \left\{ 0,1, \dots, [\frac{s-2s_2-2}{2}]\right\}$ are given by:
\begin{subequations}
\begin{equation}
	J^{(k)}_{\alpha(s)\dot{\alpha}(s)}
	=
	b_k  i^{s-2s_2-2k}
	\sum_{p=k+1}^{s-2s_2-k-1}
		(-1)^{p-k}  \frac{\binom{s}{2s_2+p} \binom{s}{p}}{\binom{s-1}{k} \binom{s-1}{2s_2+k}} \partial^p C_{\alpha(2s_2)} \partial^{s-2s_2-p} \bar{C}_{\dot{\alpha}(2s_2)},
\end{equation}	
\begin{equation}
	\begin{split}
	T^{(k)}_{\alpha(s-2)\dot{\alpha}(s-2)} 
	=
		b_k  i^{s-2s_2-2k}
		\sum_{p=k}^{s-2s_2-k-2}
		(-1)^{p-k}
	&	(\partial^{p+1} C)_{(\alpha(2s_2+p)\beta)(\dot{\alpha}(p)\dot{\beta})}
		\\
	&	\times (\partial^{s-2s_2-p-1} \bar{C})^{\beta\dot{\beta}}_{\alpha(s-2s_2-p-2)\dot{\alpha}(s-p-2)}.
	\end{split}
\end{equation}	
\end{subequations}
The higher-spin currents with nontrivial trace constructed in this way have either purely real coefficients for even values of $s-2s_2$ or purely imaginary coefficients for odd values $s-2s_2$. There are $[\frac{s-2s_2}{2}]$ such currents in total.

\newpage 

As illustrative examples, we present the first few higher-spin currents for the electromagnetic (spin-1) field, whose $\mathcal{N}=2$ supersymmetric generalization is studied in section \ref{sec: 5}.

\medskip

\noindent $\bullet$  \underline{Spin $s=2$ current}:
\begin{equation}
	J_{\alpha(2)\dot{\alpha}(2)} = C_{\alpha(2)} \bar{C}_{\dot{\alpha}(2)}.
\end{equation}	
This current corresponds to the traceless electromagnetic stress tensor.

\smallskip

\noindent $\bullet$  \underline{Spin $s=3$ current}:
\begin{equation}\label{eq: zilch in spinor}
	J_{\alpha(3)\dot{\alpha}(3)} =   i \left( C_{\alpha(2)} \partial \bar{C}_{\dot{\alpha}(2)} - \partial C_{\alpha(2)} \bar{C}_{\dot{\alpha}(2)}  \right).
\end{equation}	
This current corresponds to the spinorial form of the traceless symmetric part of the  \textit{Lipkin  zilch pseudotensor} \cite{Lipkin, Kibble, Zilch}. The symmetry transformation that corresponds to the conservation of this current is discussed in appendix \ref{sec: ZILCH}.

\smallskip

\noindent $\bullet$  \underline{Spin $s=4$ current}:

\smallskip

In this case, there are two linearly independent conserved currents:
\begin{equation}
	J_{\alpha(4)\dot{\alpha}(4)}
	=
	C_{\alpha(2)} \partial^2 \bar{C}_{\dot{\alpha}(2)} - \frac{8}{3} \partial  C_{\alpha(2)} \partial \bar{C}_{\dot{\alpha}(2)}  + \partial^2 C_{\alpha(2)}  \bar{C}_{\dot{\alpha}(2)},  
\end{equation}	
\begin{equation}
	J^{(0)}_{\alpha(4)\dot{\alpha}(4)}
	=
 \frac{16}{3} \partial C_{\alpha(2)} \partial \bar{C}_{\dot{\alpha}(2)},
	\qquad
	T^{(0)}_{\alpha(2)\dot{\alpha}(2)}
	=
	-  (\partial C)_{(\alpha(2)\beta)\dot{\beta}} (\partial \bar{C})^{\beta\dot{\beta}}_{\dot{\alpha}(2)}.
\end{equation}	
These currents are electromagnetic analogues of the gravitational Bel--Robinson tensor and are known in the literature as \textit{the Chevreton superenergy tensor} \cite{Bergqvist:2003an, Chevreton}.

\smallskip

\noindent $\bullet$  \underline{Spin $s=5$ current}:

\medskip 
In this case, there is traceless current 
\begin{equation}\label{eq: spin 5-1}
	\begin{split}
	J_{\alpha(5)\dot{\alpha}(5)} =
	&\;
	i
	\left(  C_{\alpha(2)} \partial^3 \bar{C}_{\dot{\alpha}(2)}
	 - 5 \partial C_{\alpha(2)} \partial^2 \bar{C}_{\dot{\alpha}(2)}
	 +
	 5  \partial^2 C_{\alpha(2)} \partial \bar{C}_{\dot{\alpha}(2)} 
	 -
	  \partial^3 C_{\alpha(2)}  \bar{C}_{\dot{\alpha}(2)}
	   \right),
	\end{split}
\end{equation}	
and two independent conserved currents with nontrivial trace:
\begin{subequations}\label{eq: spin 5-2}
\begin{equation}
		J^{(0)}_{\alpha(5)\dot{\alpha}(5)} =
			i \cdot \frac{25}{3}\left( \partial C_{\alpha(2)} \partial^2 \bar{C}_{\dot{\alpha}(2)} - \partial^2 C_{\alpha(2)} \partial \bar{C}_{\dot{\alpha}(2)} \right),
\end{equation}	
\begin{equation}
	T^{(0)}_{\alpha(3)\dot{\alpha}(3)}
	=	
-i \cdot \left( (\partial C)_{(\alpha(2)\beta)\dot{\beta}} (\partial^{2}\bar{C})^{\beta\dot{\beta}}_{\alpha\dot{\alpha}(3)} - (\partial^2 C)_{(\alpha(3)\beta)(\dot{\alpha}\dot{\beta})} (\partial \bar{C})^{\beta\dot{\beta}}_{\dot{\alpha}(2)}   \right);
\end{equation}	
\end{subequations}
\begin{subequations}\label{eq: spin 5-3}
\begin{equation}
	\begin{split}
		J^{(1)}_{\alpha(5)\dot{\alpha}(5)} =
		&\;
		-\frac{25}{24}
		\left(  C_{\alpha(2)} \partial^3 \bar{C}_{\dot{\alpha}(2)}
		+3 \partial C_{\alpha(2)} \partial^2 \bar{C}_{\dot{\alpha}(2)}
		+
		3  \partial^2 C_{\alpha(2)} \partial \bar{C}_{\dot{\alpha}(2)} 
		+
		\partial^3 C_{\alpha(2)}  \bar{C}_{\dot{\alpha}(2)}
		\right),
	\end{split}
\end{equation}	
\begin{equation}
	T^{(1)}_{\alpha(3)\dot{\alpha}(3)} 
	=
	 (\partial C)_{(\alpha(2)\beta)\dot{\beta}} (\partial^{2}\bar{C})^{\beta\dot{\beta}}_{\alpha\dot{\alpha}(3)} + (\partial^2 C)_{(\alpha(3)\beta)(\dot{\alpha}\dot{\beta})} (\partial \bar{C})^{\beta\dot{\beta}}_{\dot{\alpha}(2)}.  
\end{equation}	
\end{subequations}
Note that the current $J^{(1)}_{\alpha(5)\dot{\alpha}(5)}$ corresponds to the parity-invariant $(5,1,1)$ vertex.

\medskip

For higher values of $s$, the number of independent currents increases, and in the electromagnetic case ($s_2=1$) the total number of linearly independent currents is $s-2$ for $s\geq 4$.

\subsection*{"Fake"\, vertices}

As we have shown above, the general solution of the conservation equations for currents \eqref{eq: conservation} has $s-2s_2$ independent solutions for $s\geq 2s_2-2$. 
We demonstrate explicitly in the example of the $(4,2,2)$ vertex that one linear combination of the conserved currents corresponds to a ``fake'' cubic vertex removable by the field redefinition.

Let us consider the cubic vertex
\begin{equation}
	\mathcal{L}^{fake}_{(4,1,1)}
	=	\mathcal{R}_{(s=4)}^{\alpha(2)\dot{\alpha}(2)}
	C_{\alpha(2)} \bar{C}_{\dot{\alpha}(2)},
\end{equation}	
which is proportional to the free equations of motion and corresponds to a "fake" interaction.
Let us express this vertex in terms of conserved currents.
After integrating by parts (using definition \eqref{eq: curv 2} for $s=4 $), one finds three terms on the spin-$1$ shell:
	\begin{equation}
		\begin{split}
	&	\Phi^{\alpha(2)\dot{\alpha}(2)} \Box \left( 	C_{\alpha(2)} \bar{C}_{\dot{\alpha}(2)} \right)
		\approx 
		4\Phi^{\alpha(2)\dot{\alpha}(2)} \; \partial^{\beta\dot{\beta}}	C_{\alpha(2)}  \partial_{\beta\dot{\beta}} \bar{C}_{\dot{\alpha}(2)}
		=
		-4 \Phi^{\alpha(2)\dot{\alpha}(2)} \; T^{(0)}_{\alpha(2)\dot{\alpha}(2) }
		, 
\\
	&	\Phi^{(\alpha\beta)(\dot{\alpha}\dot{\beta})} \partial_{\beta\dot{\beta}} \partial^{\alpha\dot{\alpha}} \left( 	C_{\alpha(2)} \bar{C}_{\dot{\alpha}(2)} \right)
		\approx 0,
\\&
	\Phi^{\alpha(4)\dot{\alpha}(4)} 
		\partial^2_{\alpha(2)\dot{\alpha}(2)} \left( 	C_{\alpha(2)} \bar{C}_{\dot{\alpha}(2)} \right) 
		=
		\Phi^{\alpha(4)\dot{\alpha}(4)}  
		\left( J_{\alpha(4)\dot{\alpha}(4)} 
		+
		\frac{7}{8} J_{\alpha(4)\dot{\alpha}(4)}^{(0)}  \right).
		\end{split}
	\end{equation}	
Consequently, we can express this vertex (on  the spin-$1$ shell) as:
\begin{equation}
	\begin{split}
		\mathcal{L}^{fake}_{(4,2,2)}
		&=	\mathcal{R}^{\alpha(2)\dot{\alpha}(2)}
		C_{\alpha(2)} \bar{C}_{\dot{\alpha}(2)}
		\\&	\approx
		- \frac{32}{7}  
		\Phi^{\alpha(4)\dot{\alpha}(4)}  
		J_{\alpha(4)\dot{\alpha}(4)} 
		-
		4
		\left( 	\Phi^{\alpha(4)\dot{\alpha}(4)}  
		J_{\alpha(4)\dot{\alpha}(4)}^{(0)}  
		+
		\Phi^{\alpha(2)\dot{\alpha}(2)}  T_{{\alpha(2)\dot{\alpha}(2)} }
		\right).
	\end{split}
\end{equation}	
We see that such a combination of spin-$4$ currents corresponds to a "fake"\, cubic $(4,1,1)$ interaction. 

Analogously, one can construct "fake"  $(5,1,1)$ interactions:
\begin{equation}
	\begin{split}
&	\mathcal{L}^{fake-1}_{(5,1,1)} = \mathcal{R}^{\alpha(3)\dot{\alpha}(3)}_{(s=5)} J_{\alpha(3)\dot{\alpha}(3)}
\overset{\eqref{eq: zilch in spinor}}{=}
\mathcal{R}^{\alpha(3)\dot{\alpha}(3)}_{(s=5)} i  \left( C_{\alpha(2)} \partial \bar{C}_{\dot{\alpha}(2)} - \partial C_{\alpha(2)} \bar{C}_{\dot{\alpha}(2)}  \right)
,
\\
	&	\mathcal{L}^{fake-2}_{(5,1,1)} = \mathcal{R}^{\alpha(3)\dot{\alpha}(3)}_{(s=5)} \left( C_{\alpha(2)} \partial \bar{C}_{\dot{\alpha}(2)} + \partial C_{\alpha(2)} \bar{C}_{\dot{\alpha}(2)}  \right).
		\end{split}
\end{equation}
This means that out of two parity-breaking currents \eqref{eq: spin 5-1}, \eqref{eq: spin 5-2} and one parity-invariant current \eqref{eq: spin 5-3}, only one nontrivial parity-brealing vertex remains.

 Therefore, when constructing consistent cubic vertices, one may restrict attention to a single current, for example the traceless higher-spin current \eqref{eq: N=0 current}. An analogous statement also holds for arbitrary-spin abelian interactions: \textit{only a single nontrivial abelian vertex with minimal number of derivatives remains for all spin values}.


\section{Zilch transformations and their higher-spin generalizations}\label{sec: ZILCH}

In this appendix, we derive the zilch symmetry transformation directly from the zilch current by means of the inverse Noether procedure, whose $\mathcal{N}=2$ superfield version is applied in section \ref{sec: 5}. In contrast to the standard duality-symmetric formulations commonly used in the literature, the derivation can be performed entirely within the conventional Maxwell framework. We then discuss the higher-spin generalization of these transformations and their relation to parity properties of higher-spin couplings.

Let us consider the zilch current in the Kibble form \cite{Kibble}\footnote{We define the dual tensor by ${}^\star F_{[ab]} : = \frac{i}{2} \epsilon_{abcd} F^{[cd]}$.}:
\begin{equation}\label{eq: zilch current}
	z_{\mu\nu|\rho} = 2i F_{\mu\sigma} \overset{\text{\tiny $\leftrightarrow$ }}{\partial_\rho} {}^\star F^{\sigma}_{\;\;\nu} 
	=
 i \left( F_{\mu\sigma}   \partial_\rho {}^\star F^{\sigma}_{\;\;\nu} 
	-
	\partial_\rho F_{\mu\sigma}    {}^\star F^{\sigma}_{\;\;\nu}\right).
\end{equation}	
 This form is the most convenient for our purposes, since in this form the conservation law is the most easily verified:
\begin{equation}
		\partial^\rho  z_{\mu\nu|\rho}
		=
		 i \left( F_{\mu\sigma}   \Box {}^\star F^{\sigma}_{\;\;\nu} 
		-
		\Box F_{\mu\sigma}    {}^\star F^{\sigma}_{\;\;\nu}\right)  \approx 0.
\end{equation}	
The conserved spin-3 current \eqref{eq: zilch in spinor} is obtained by taking the completely symmetric traceless part of
\begin{equation}
	Z_{(\mu\nu\rho)} =\frac{1}{3} \left(	z_{\mu\nu|\rho}  
	+
		z_{\rho \mu|\nu} 
		+
			z_{\nu\rho |\mu}  \right).
\end{equation}
Such a current was considered, e.g., in the work  \cite{Krivsky:1994np}. Since the current explicitly involves the Levi-Civita tensor, the corresponding $(3,1,1)$ cubic vertex is parity odd.

The general variation of the electromagnetic field Lagrangian has the form (hereafter, we consider equalities up to a total derivative):
\begin{equation}\label{eq: EM variation}
	\mathcal{L}_{EM} = -\frac{1}{4} F^{\mu\nu} F_{\mu\nu}
	\qquad
	\Rightarrow
	\qquad
	\delta 	\mathcal{L}_{EM} 
	=
	 \delta A_{\sigma} \partial_\rho F^{\rho\sigma}.
\end{equation}

According to Noether's theorem, the variation of the Lagrangian with respect to the symmetry transformations corresponding to the current \eqref{eq: zilch current} has the form:
\begin{equation}
	\delta \mathcal{L}_{EM} = \frac{1}{2}  \xi^{\mu\nu} \partial^{\rho} z_{\mu\nu|\rho}
	 =
	 \xi^{\mu\nu} \cdot \frac{i}{2} \left(  F_{\mu\sigma}   \Box {}^\star F^{\sigma}_{\;\;\nu} 
	 -
	 \Box F_{\mu\sigma}    {}^\star F^{\sigma}_{\;\;\nu} \right).
\end{equation}	
Indeed, integrating by parts and using the rigidity condition $\partial_\rho \xi^{\mu\nu}=0$, the invariance of the Maxwell action immediately follows.

In order to find the transformation law for the vector field, we need to bring this variation to the form
\eqref{eq: EM variation}.
Using off-shell identities
\begin{equation}
	\Box F_{\mu\nu} = 2 \partial_{[\mu} \partial^\sigma F_{\sigma \nu]},
	\qquad
	\Box {}^\star F^{\rho\sigma} 
	=
	\frac{i}{2} \epsilon^{\rho\sigma\mu\nu} \Box F_{\mu\nu}
	= 	i \epsilon^{\rho\sigma\mu\nu} \partial_{[\mu} \partial^\kappa F_{\kappa \nu]},
\end{equation}	
up to total derivatives one finds:
\begin{subequations}
\begin{equation}
	\begin{split}
	\xi^{\mu\nu} \cdot  \Box F_{\mu\sigma}    {}^\star F^{\sigma}_{\;\;\nu}
	 &=
	 \xi^{\mu\nu} 
	 \cdot
	 2 \partial_{[\mu} \partial^\rho F_{\rho \sigma]}
	 {}^\star F^{\sigma}_{\;\;\nu}
	 =
	- 2  \xi^{\mu\nu} 
	 \cdot
	 \partial^\rho F_{\rho [\sigma}
	  \partial_{\mu]} {}^\star F^{\sigma}_{\;\;\nu}
	\\&  =
	  -   \xi^{\mu\nu} 
	  \cdot
	  \partial_{\mu} {}^\star F_{\sigma\nu}
	   \partial_\rho F^{\rho \sigma},
	  \end{split}
\end{equation}	
\begin{equation}
	\begin{split}
	 \xi^{\mu\nu} \cdot  F_{\mu}^{\;\;\sigma}   \Box {}^\star F_{\sigma\nu} 
	 &=
	\frac{i}{2} 
	  \xi^{\mu\nu} \cdot  F_{\mu}^{\;\;\sigma}  
	  \epsilon_{\sigma\nu\kappa \delta} \Box F^{\kappa\delta}
	  =
	  i
	 \xi^{\mu\nu} \cdot  F_{\mu}^{\;\;\sigma}  
	 \epsilon_{\sigma\nu\kappa \delta} \partial^\kappa \partial_\rho F^{\rho\delta}
	  \\
	  &=
	  -
	  i
	 \xi^{\mu\nu} \cdot  \underbrace{\partial^\kappa F_{\mu}^{\;\;\sigma} }_{-\frac{1}{2}\partial_\mu F^{\kappa\sigma}} 
	 \epsilon_{\sigma\nu\kappa \delta}  \partial_\rho F^{\rho\delta}
	 =
	\frac{i}{2}
	 \xi^{\mu\nu} \cdot  \partial_\mu F^{\kappa\sigma}
	 \epsilon_{\sigma\nu\kappa \delta}  \partial_\rho F^{\rho\delta}
	 \\
	 &=
	  \xi^{\mu\nu} \cdot  \partial_\mu {}^\star F_{\nu\sigma} \partial_{\rho} F^{\rho\sigma}.
	  \end{split}
\end{equation}	
\end{subequations}
Comparing with \eqref{eq: EM variation}, one finds the symmetry transformation:
\begin{equation}
	\boxed{\delta A_\sigma = i  \xi^{\mu\nu} \partial_\mu {}^\star F_{\nu\sigma}.}
\end{equation}	
In the literature, such transformations are known as \textit{zilch symmetry} transformations. 
Note that to the best of our knowledge, this approach to the derivation of the transformation law has never been employed in the literature.  Moreover, discussions of zilch transformations are often formulated in a framework that is invariant under electromagnetic duality, see  \cite{Zilch} and references therein.  As the derivation presented above shows, there is no need for such a formalism (see also \cite{Letsios:2022bid}).

The structure of the transformation becomes more transparent after introducing the dual vector potential
 ${}^\star F_{\nu\sigma} := i \left( \partial_{\nu} B_{\sigma} - \partial_{\sigma} B_{\nu} \right)$. We can rewrite the zilch transformation law\footnote{Up to gauge transformations $\delta_{gauge} A_\sigma = \partial_\sigma \left(\xi^{\mu\nu} \partial_{\mu} B_{\nu}  \right)$.} as follows:
\begin{equation}
	\delta A_\sigma = -  \xi^{\mu\nu} \partial_\mu \partial_\nu  B_{\sigma}
	\qquad
	\Rightarrow
	\qquad
	\delta F_{\rho\sigma} = i  \xi^{\mu\nu} \partial_\mu \partial_\nu  {}^\star F_{\rho\sigma}.
\end{equation}
The transformation combines electromagnetic duality and spin-3 translation-like transformation.

This structure immediately suggests a higher-spin generalization.
 It is easy to construct generalizations of the translation and zilch transformations to the higher-spin case (see, e.g., \cite{Zilch}):
\begin{equation}
	\begin{split}
	&\delta_k^{(1)} A_\sigma = \xi^{\mu_1\dots \mu_k} \partial_{\mu_1} \dots  \partial_{\mu_k} A_{\sigma}
	\quad
	\Rightarrow
	\quad
	\delta_k^{(1)}  F_{\rho\sigma} 
	=
	\xi^{\mu_1\dots \mu_k} \partial_{\mu_1} \dots  \partial_{\mu_k} F_{\rho\sigma},
	\\
	&\delta_k^{(2)} A_\sigma = \xi^{\mu_1\dots \mu_k} \partial_{\mu_1} \dots  \partial_{\mu_k} B_{\sigma}
	\quad
	\Rightarrow
	\quad
	\delta_k^{(2)} F_{\rho\sigma} = -i \xi^{\mu_1\dots \mu_k} \partial_{\mu_1} \dots  \partial_{\mu_k} {}^\star F_{\rho\sigma}.
	\end{split}
\end{equation}	
Both classes of transformations leave  the Maxwell equations invariant: 
$$\partial_\mu F^{\mu\nu} \approx 0,
\qquad  \partial_\mu {}^\star F^{\mu\nu} = 0.$$ 
However, one can verify that the invariance of the Maxwell action is preserved only for $$\delta^{(1)}_k \;\;\text{ with} \;\; k=1,3,5,\dots \quad  \text{and for} \quad  \delta^{(2)}_k \;\;  \text{with}\;\; k=0, 2, 4, \dots$$  This pattern is consistent with the structure of cubic higher-spin vertices: transformations of type $\delta_k^{(1)}$ generate couplings to even-spin gauge fields, whereas transformations of type $\delta_k^{(2)}$, involving electromagnetic duality, generate couplings to odd-spin fields.

Let us emphasize, that for a real scalar field it is impossible to construct abelian cubic interactions with odd spins, since there are no non-trivial conserved currents. Such interactions can be constructed only for a complex scalar field, since for such a field one can introduce an electric charge. For vector fields, however, such odd-spin couplings do exist. Their existence is closely related to the interplay between higher-spin translation symmetries and electromagnetic duality. We note that in the harmonic superspace approach, these differences appear automatically, see  section \ref{eq: spin s transf}.

Similar subtleties arise for  abelian interactions of arbitrarily high-spin fields with odd spins. For example, for the (5,2,2) vertex is associated with a transformation mixing the graviton with its dual field:
\begin{equation}
	\delta h_{\mu\nu} = \xi^{\rho_1\rho_2\rho_3\rho_4} \partial_{\rho_1} \partial_{\rho_2} \partial_{\rho_3} \partial_{\rho_4} 
	\tilde{h}_{\mu\nu}.
\end{equation}	
A detailed discussion of the definition and some applications of EM duality for gravity and higher-spin fields can be found in the review \cite{Danehkar:2018yjp}.

\section{On the structure of gauge-invariant $\mathcal{N}=2$ superfield higher-spin Weyl tensors}
\label{app: structure of strength}

In this appendix, we discuss the general structure of gauge-invariant higher-spin superfield Weyl-like tensor in harmonic superspace and explain how the corresponding chiral strengths arise from the composite "half-analytic" superfield $\mathcal{H}^{++}_{\dots}$.

Our goal is to identify a composite harmonic superfield whose
chirality properties naturally generate the gauge-invariant higher-spin
strength.
Using the $G^{++}$ potentials introduced in \eqref{eq: G++ def}, one may define the composite $\mathcal N=2$ superfield:
\begin{equation}\label{eq: H++}
	\mathcal{H}^{++}_{\alpha(2s-2)}
	:=
	\partial_{(\alpha_1}^{\;\;\dot{\beta}_1} \dots  	\partial_{\alpha_{s-2}}^{\;\;\dot{\beta}_{s-2}} \left(
	\mathcal{D}^-_{\alpha_{s-1}} G^{+++}_{\alpha_s\dots\alpha_{2s-2})(\dot{\beta}_1\dots\dot{\beta}_{s-2})}
	+
	\partial_{\alpha_{s-1}}^{\;\;\dot{\beta}_{s-1}} G^{++}_{\alpha_s\dots\alpha_{2s-2})(\dot{\beta}_1\dots\dot{\beta}_{s-1})}\right).
\end{equation}	
This is the natural  higher-spin generalization of superfield $\mathcal{H}^{++}_{(\alpha\beta)}$ introduced in \cite{IZ} in the spin-$\mathbf{2}$ case.
This object satisfies the "half-analyticity" condition\footnote{"Half-analyticity" condition plays an important role in the construction of $\mathcal{N}=2$ conformal theories in harmonic approach \cite{Ivanov:2024bsb, Ivanov:2025gvs} and allows generalization to curved $\mathcal{N}=2$ harmonic superspace \cite{Galperin:1987em}.}:
\begin{equation}\label{eq: H-A condition}
		\bar{\mathcal{D}}^+_{\dot{\alpha}}\mathcal{H}^{++}_{\alpha(2s-2)} = 0,
\end{equation}	
which uniquely fixes the relative coefficient between two contributions  in \eqref{eq: H++}.

From this object one can construct its negatively-charged counterpart as the solution of the zero-curvature equation:
\begin{equation}\label{eq: ZC hcal}
	\mathcal{D}^{++} \mathcal{H}^{--}_{\alpha(2s-2)} =  	\mathcal{D}^{--} \mathcal{H}^{++}_{\alpha(2s-2)}. 
\end{equation}	

A convenient representative of the solution can be written explicitly in terms of the negatively charged potentials $G^{--}$, defined by equations \eqref{zero-curv}: 
 \begin{equation}\label{eq: mathcal H --}
 	\begin{split}
 		\mathcal{H}^{--}_{\alpha(2s-2)}
 	=
 	\partial_{(\alpha_1}^{\;\;\dot{\beta}_1}& \dots  	\partial_{\alpha_{s-2}}^{\;\;\dot{\beta}_{s-2}} \Bigr(
 		\mathcal{D}^+_{\alpha_{s-1}} G^{---}_{\alpha_s\dots\alpha_{2s-2})(\dot{\beta}_1\dots\dot{\beta}_{s-2})}
 		\\&	+
 	\mathcal{D}^-_{\alpha_{s-1}}
  G^{--+}_{\alpha_s\dots\alpha_{2s-2})(\dot{\beta}_1\dots\dot{\beta}_{s-2})}
 +
 	\partial_{\alpha_{s-1}}^{\;\;\dot{\beta}_{s-1}} G^{--}_{\alpha_s\dots\alpha_{2s-2})(\dot{\beta}_1\dots\dot{\beta}_{s-1})}\Bigr).
 	\end{split}
 \end{equation}	
 
 The gauge transformations of $\mathcal{H}^{\pm\pm}_{\alpha(2s-2)}$ under \eqref{eq:G++ gauge} and \eqref{Lambda gauge} take the form:
 \begin{equation}
 	\delta_\lambda \mathcal{H}^{\pm\pm}_{\alpha(2s-2)} 
 	=
 	\mathcal{D}^{\pm\pm} \Lambda_{\alpha(2s-2)},
 \end{equation}	
 where we introduced the composite gauge superparameter 
 \begin{equation}
 	\Lambda_{\alpha(2s-2)} = 	\partial_{(\alpha_1}^{\;\;\dot{\beta}_1} \dots  	\partial_{\alpha_{s-2}}^{\;\;\dot{\beta}_{s-2}} \left(
 	\mathcal{D}^-_{\alpha_{s-1}} \Lambda^{+}_{\alpha_s\dots\alpha_{2s-2})(\dot{\beta}_1\dots\dot{\beta}_{s-2})}
 	+
 	\partial_{\alpha_{s-1}}^{\;\;\dot{\beta}_{s-1}} \Lambda^{}_{\alpha_s\dots\alpha_{2s-2})(\dot{\beta}_1\dots\dot{\beta}_{s-1})}\right),
 \end{equation}	
which also satisfies the "half-analyticity" condition:
 \begin{equation}
 	\bar{\mathcal{D}}^+_{\dot{\alpha}} \Lambda_{\alpha(2s-2)} = 0.
 \end{equation}
 
 The "half-analyticity" condition is crucial because it guarantees gauge invariance of the chiral $\mathcal{N}=2$ superfield  Weyl-like tensor:
 \begin{equation}\label{eq: app W}
 	\mathcal{W}_{\alpha(2s-2)} :=\left(\bar{\mathcal{D}}^+\right)^2 \mathcal{H}^{--}_{\alpha(2s-2)},
 	\qquad
 	\delta_\lambda 	\mathcal{W}_{\alpha(2s-2)} = 0.
 \end{equation}	
The superfield strength also possesses several important properties. Firstly, it is covariantly independent of the  harmonic variables:
 \begin{equation}\label{eq: W harm indep}
 			\mathcal{D}^{++} \mathcal{W}_{\alpha(2s-2)} = 0,
 			\qquad
 				\mathcal{D}^{--} \mathcal{W}_{\alpha(2s-2)} = 0.
 \end{equation}	
 The first equality follows directly from the zero curvature equation \eqref{eq: ZC hcal} and the "half-analyticity" condition \eqref{eq: H-A condition}. The second equality follows immediately in the harmonic superspace central basis.
 Moreover, by construction $\mathcal{W}_{\alpha(2s-2)}$ satisfies chirality conditions:
\begin{equation}
	\bar{\mathcal{D}}^+_{\dot{\alpha}} \mathcal{W}_{\alpha(2s-2)}  = 0,
	\qquad
		\bar{\mathcal{D}}^-_{\dot{\alpha}} \mathcal{W}_{\alpha(2s-2)}  = 0.
\end{equation}	
Analogously, one can construct the conjugate superfield strength $\bar{\mathcal{W}}_{\dot{\alpha}(2s-2)} = \widetilde{\mathcal{W}_{\alpha(2s-2)}}$. It also has analogous properties.

Comparing the structure of the higher-spin superfield strengths \eqref{eq: app W} with the spin-$1$ case \eqref{eq: W spin 1}, one observes that   $\mathcal{H}^{++}_{\alpha(2s-2)}$ plays a role analogous to that of the prepotential $V^{++}$. This analogy suggests that
$\mathcal H^{++}_{\alpha(2s-2)}$
may provide the natural starting point for a harmonic-superspace
construction of non-abelian higher-spin interactions, in close analogy
with the role of $V^{++}$ in $\mathcal N=2$ SYM theory. 

\section{Component content and on-shell equations of $\mathcal{N}=2$ higher-spin Weyl supertensor}
\label{eq: cons N=2 str}

In this appendix, we derive the  on-shell conditions:
\begin{equation}\label{eq: sf strength}
	\mathcal{D}^{+\alpha} \mathcal{W}_{\alpha(2s-2)} \approx 0,
	\qquad
	\mathcal{D}^{-\alpha} \mathcal{W}_{\alpha(2s-2)} \approx 0.
\end{equation}	
A direct derivation of these equations from the superfield equations of motion \eqref{eq: super EOM} is rather cumbersome. A more elegant way is to use component analysis in Wess-Zumino-type gauge. Indeed, the requirement of gauge invariance essentialy fixes (up to numerical coefficients) on-shell form of the higher-spin superfield strength\footnote{To fix the numerical coefficients one needs explicitly solve zero curvature equations \eqref{zero-curv} in the Wess-Zumino-type gauge. Concrete example of such procedure was presented in spin-$\mathbf{2}$ case in \cite{Ivanov:2024gjo}.} \eqref{eq: SF strenghts}:
\begin{equation}\label{eq: W on-shell}
	\begin{split}
	\mathcal{W}_{\alpha(2s-2)} \sim \;&
	C_{\alpha(2s-2)}
	+
	\theta^{+\beta} \hat{C}^i _{(\beta\alpha(2s-2))} u^-_i
	-
	\theta^{-\beta} \hat{C}^i _{(\beta\alpha(2s-2))} u^+_i
\\&	+
	\theta^{+\beta} \theta^{-\gamma} C_{(\alpha(2s-2)\beta\gamma)}
	+  4i \theta^{-\beta} \bar{\theta}^{+\dot{\beta}} \partial_{\beta\dot{\beta}} C_{\alpha(2s-2)}
\\
&	- 
	4i \theta^{-(\beta} \theta^{+\gamma)} \bar{\theta}^{+\dot{\gamma}}  \partial_{\gamma\dot{\gamma}} \hat{C}^i _{(\beta\alpha(2s-2))} u^-_i
\\&	+
	(\theta^+)^2  \theta^{-\rho}  \check{\bar{C}}^i_{(\rho\alpha(2s-2))} u^-_i 
	-
		(\theta^-)^2  \theta^{+\rho}  \check{\bar{C}}^i_{(\rho\alpha(2s-2))} u^+_i 
	.
	\end{split}
\end{equation}	
Here $C_{\alpha(2s)}$, $C_{\alpha(2s-2)}$, and $C^i_{\alpha(2s-1)}$ denote the Weyl-like tensors associated with the spin-$s$, spin-$(s-1)$, and spin-$(s-\frac12)$ fields, respectively.
In deriving \eqref{eq: W on-shell} we also used dimensional analysis\footnote{Dimensions of field strengths are  $[\mathcal{W}_{\alpha(2s-2)}]= s$,  $[C_{\alpha(2s)}] = s+1$,  $[C_{\alpha(2s-1)}] =s+\frac{1}{2}$.}, the requirement of  covariant harmonic-independence $\mathcal{D}^{++}\mathcal{W}_{\alpha(2s-2)}=0$, and the chirality property $\bar{\mathcal{D}}^{\pm}_{\dot{\beta}} \mathcal{W}_{\alpha(2s-2)} = 0$. 

All higher-order terms in the component expansion involve derivatives of the Weyl-like tensors and therefore vanish on-shell as a consequence of \eqref{eq: on-shell bos}.
It immediately follows from the expansion \eqref{eq: W on-shell}  that equations \eqref{eq: sf strength} are satisfied on-shell. 

The expansion \eqref{eq: W on-shell} also implies several useful on-shell identities:
\begin{subequations}\label{eq: on-shell W cond}
\begin{equation}
	\bar{\mathcal{D}}^+_{\dot{\alpha}} 	\mathcal{D}^{-\alpha} \mathcal{W}_{\alpha(2s-2)} \approx 0
	\quad
	\Rightarrow
	\quad
	\partial_{\dot{\alpha}}^{\;\alpha} \mathcal{W}_{\alpha(2s-2)} \approx 0
	\quad
	\Rightarrow
	\quad
	\Box \mathcal{W}_{\alpha(2s-2)} \approx 0,
\end{equation}	
\begin{equation}\label{eq: C3b}
	(\mathcal{D}^+)^2 	\mathcal{W}_{\alpha(2s-2)} \approx 0,
	\qquad
	(\mathcal{D}^-)^2 	\mathcal{W}_{\alpha(2s-2)} \approx 0,
\end{equation}	
\begin{equation}
		\mathcal{D}^+_{\beta}  \mathcal{W}_{\alpha(2s-2)} \approx 
			\mathcal{D}^+_{(\beta}  \mathcal{W}_{\alpha(2s-2))},
			\qquad
				\mathcal{D}^-_{\beta}  \mathcal{W}_{\alpha(2s-2)} \approx 
			\mathcal{D}^-_{(\beta}  \mathcal{W}_{\alpha(2s-2))},
\end{equation}	
\begin{equation}
	\partial^p_{\alpha(p)\dot{\alpha}(p)} \mathcal{W}_{\alpha(2s-2)} \approx  	\partial^p_{(\alpha(p)\dot{\alpha}(p)} \mathcal{W}_{\alpha(2s-2))},
	\quad
		\partial^p_{\alpha(p)\dot{\alpha}(p)} \bar{\mathcal{W}}_{\dot{\alpha}(2s-2)} \approx  	\partial^p_{\alpha(p)(\dot{\alpha}(p)} \mathcal{W}_{\dot{\alpha}(2s-2))}.
\end{equation}	
\end{subequations}
These identities play a crucial role in the derivation of conserved $\mathcal N=2$ higher-spin supercurrents.


\end{document}